\documentclass[sn-apa]{sn-jnl}
\usepackage{graphicx}%
\usepackage{amsmath,amssymb,amsfonts}%
\usepackage[title]{appendix}%
\usepackage{xcolor}%
\usepackage{manyfoot}%

\newcommand{\ftilde}{\ensuremath{\tilde{f}}}
\newcommand{\taustar}{\ensuremath{\overset{*}{\tau}}}
\newcommand{\mat}[1]{\ensuremath{\mathbf{#1}}}
\newcommand{\ftildebold}{\ensuremath{\tilde{\mathbf{f}}}}

\newcommand{\stim}[1]{\textsc{#1}}
\newcommand{\stimone}{\stim{x}}
\newcommand{\stimtwo}{\stim{y}}
\newcommand{\stimthree}{\stim{z}}

\newcommand{\baseone}{\ensuremath{x}}
\newcommand{\basetwo}{\ensuremath{y}}
\newcommand{\basethree}{\ensuremath{z}}

\renewcommand{\L}[1]{\ensuremath{\mathcal{L}\left\{#1\right\}}}

\newcommand{\Z}[1]{\ensuremath{\mathcal{Z}\left\{#1\right\}}}

\newcommand{\PastLapl}[1]{\ensuremath{F^-_{#1}}}
\newcommand{\FutLapl}[1]{\ensuremath{F^+_{#1}}}
\newcommand{\PastLaplbold}[1]{\ensuremath{\mathbf{F}^-_{#1}}}
\newcommand{\FutLaplbold}[1]{\ensuremath{\mathbf{F}^+_{#1}}}

\newcommand{\MXY}{\ensuremath{M_\basetwo^{\ \baseone}(s)}}
\newcommand{\barMXY}{\ensuremath{\bar{M}_\basetwo^{\ \baseone}(s)}}
\newcommand{\barMYX}{\ensuremath{\bar{M}_\baseone^{\ \basetwo}(s)}}

\newcommand{\barMXYbare}{\ensuremath{\bar{M}_\basetwo^{\ \baseone}}}
\newcommand{\barMYXbare}{\ensuremath{\bar{M}_\baseone^{\ \basetwo}}}

\newcommand{\MXYbare}{\ensuremath{M_\basetwo^{\ \baseone}}}
\newcommand{\MYZ}{\ensuremath{M_\basethree^{\ \basetwo}(s)}}
\newcommand{\MYZbare}{\ensuremath{M_\basethree^{\ \basetwo}}}
\newcommand{\MXZ}{\ensuremath{M_\basethree^{\ \baseone}(s)}}
\newcommand{\MXZbare}{\ensuremath{M_\basethree^{\ \baseone}}}
\newcommand{\MYX}{\ensuremath{M_\baseone^{\ \basetwo}(s)}}

\newcommand{\smin}{\ensuremath{s_{\textnormal{min}}}}
\newcommand{\smax}{\ensuremath{s_{\textnormal{max}}}}

\newcommand{\learn}{\rho}

\renewcommand{\not}[1]{\ensuremath{\tilde{#1}}} 

\begin{document}
\title[Temporal relationships]{Learning temporal
relationships between symbols with Laplace Neural Manifolds} 


\author*[1]{\fnm{Marc W.} \sur{Howard}}\email{marc777@bu.edu}

\author[1]{\fnm{Zahra Gh.} \sur{Esfahani}}\email{ghasemi.zahra@gmail.com}

\author[2]{\fnm{Bao} \sur{Le}}\email{uac6qw@virginia.edu}

\author[2]{\fnm{Per B.} \sur{Sederberg}}\email{pbs5u@virginia.edu}

\affil*[1]{\orgdiv{Department of Psychological and Brain Sciences},
\orgname{Boston University}, \orgaddress{\street{610 Commonwealth Ave},
\city{Boston}, \postcode{02215}, \state{MA}, \country{USA}}}
\affil[2]{\orgdiv{Department of Psychology},
\orgname{University of Virginia}, \orgaddress{\street{409 McCormick Road},
\city{Charlottesville}, \postcode{22904}, \state{VA}, \country{USA}}}

\abstract{
Firing across populations of neurons in many regions of the mammalian brain
maintains a temporal memory, a neural timeline of the recent past.
Behavioral results demonstrate that people can 
both remember the past and anticipate the future
over an analogous internal timeline. 
This paper presents a mathematical framework for building this timeline of the
future.  We assume that the input to the system is a time series of
symbols---sparse tokenized representations of the present---in
continuous time.   The goal is to record pairwise temporal relationships
between symbols over a wide range of time scales.  We assume that the brain
has access to a temporal memory in the form of the real Laplace transform.
Hebbian associations with a diversity of synaptic time scales are
formed between the past timeline and the present symbol.  The associative
memory stores the convolution between the past and the present.   Knowing the
temporal relationship between the past and the present allows one to infer
relationships between the present and the future.   
With appropriate normalization, this Hebbian associative matrix can store a
Laplace successor representation and a Laplace predecessor representation
from which measures of temporal contingency can be evaluated.  The diversity
of synaptic time constants allows for learning of non-stationary statistics as
well as joint statistics between triplets of symbols.
This framework synthesizes a number of recent neuroscientific
findings including results from dopamine
neurons in the mesolimbic forebrain.  
} 

\keywords{Temporal memory, prediction, Laplace transform, convolution}

%

\maketitle

\section{Introduction}

Consider the experience of listening to a familiar melody.  
As the song unfolds, notes feel as if they recede away from the present, an
almost spatial experience.  According to  \citet{Huss66}  
``points of temporal duration recede, as points of a
stationary object in space recede when I {`go away from the object'}.''  
For a familiar melody, \citet{Huss66} argues that events predicted in the
future also have an analogous spatial extent, a phenomenon he referred to as
\emph{protention}.
This experience is consistent with the hypothesis that the brain maintains an
inner timeline extending from the distant past towards the present and from
the present forwards into the  future.  In addition to introspection and
phenomenological analysis, one can reach similar conclusions from
examination of data in carefully controlled cognitive psychology experiments
\citep{TigaEtal22}.

The evolutionary utility of an extended timeline for future events is obvious.
Knowing what will happen when in the future allows for selection of an
appropriate action in the present.
Indeed, much of computational neuroscience presumes that the fundamental goal of the
cortex is to predict the future
\citep{Clar13,FrisKieb09,Fris10,RaoBall99,PalmEtal15}.   

In AI, a great deal of
research focuses on reinforcement learning (RL) algorithms that attempt to
optimize future  outcomes within a particular planning horizon
\citep{KeEtal18,DabnEtal20} without a temporal memory.  From the
perspective of psychology, RL is a natural extension of the Rescorla-Wagner
model \citep{RescWagn72} an associative model for classical conditioning
\citep{SuttBart81,SchuEtal97,WaelEtal01}.   
Associative models describe connections between a pair of stimuli (or stimulus
and an outcome etc) as a simple scalar value.  
In simple associative models, variables that affect the strength of an
association, such as the number of pairings between stimuli, or attention,
etc, must all combine to affect a single scalar value.  Thus, although the strength
of an association can fall off with the time between stimuli, the association
itself does not actually convey information about time \emph{per se}
\citep{Gall21a}.  


Cognitive psychologists have argued  that classical conditioning does not reflect
atomic associations between stimuli, but rather explicit storage and retrieval
of temporal relationships
\citep{CoheEich93,ArceEtal05,BalsGall09,GallEtal19,Namb21}.  
In this view, behavioral associations in classical conditioning reflect
learning of temporal \emph{contingencies} between stimuli, such that knowing
that a particular stimulus was experienced in the present changes our
expectations for the time at which an outcome will be experienced
\citep{JeonEtal22,FloeEtal24}.
Such a theory clearly requires a temporal memory in order to learn
temporal relationships between stimuli.

This paper presents a formal hypothesis for how populations of neurons could
learn and express temporal relationships between symbols, ignoring
similarity structure within stimuli.  We assume the existence of a temporal memory
expressed in the firing of neurons with an effectively continuous spectrum of
time constants, forming the Laplace transform of the recent past
\citep{AtanEtal23,BrigEtal20,KantEtal24,TsaoEtal18,ZuoEtal23}.
Neurophysiological results suggest that the temporal memory expressed in
neural firing extends at least several minutes \citep{TsaoEtal18}.
We additionally hypothesize a neural timeline of the future expressed as Laplace
transform \citep{CaoEtal24}.  The present is part of both the past
and the future, so that the current symbol is simultaneously the
most recent part of the past and the most imminent part of the future.
Hebbian associations between
the Laplace transform for the past and the present symbol store temporal
relationships between symbols.   In addition, a continuous spectrum of
\emph{synaptic} time scales enable learning of temporal relationships
over time scales  much longer than a few minutes.  This spectrum of synaptic
time constants also enables learning of higher-order relationships among
symbols expressed as their joint statistics.

\section{Constructing neural timelines of the past and future}
We take as input a finite set of discrete symbols,
\stimone{}, \stimtwo{}, etc., that are occasionally presented for an instant
in continuous time.  
There are consistent  temporal relationships between some of the symbols, such
that knowing one symbol was presented at time $t$ may provide information
about the occurrence of another symbol at time $t+\tau$.
For convenience we assume that the time between repetitions of any given symbol is
much longer than the temporal relationships that are to be discovered and much
longer than the longest time constant $1/\smin$.  Much like the assumptions
necessary to write out the Rescorla-Wagner model \citep{Namb21,Gall21}, this
set of assumptions allows us to imagine that experience is segmented into a series of
discrete trials and that each symbol can be presented at most once per
trial. This assumption allows easy interpretation of quantities that we will
derive. 

\subsection{The present}
Let us take as input a vector valued function of time $\mathbf{f}(t)$.
The notation $\mathbf{v}$ refers to a vector with each element a real number,
$\mathbf{v}'$ is a transposed vector, so that $\mathbf{u}'\mathbf{v}$ is the
inner product, a scalar, and
$\mathbf{u}\mathbf{v}'$ is the outer product, a matrix.
We assume a tokenized representation between symbols, so that
$\mathbf{\basetwo}'\mathbf{\baseone} = \delta_{\basetwo,\baseone}$ where 
$\delta_{ij}$ is the Kronecker delta function.
We write
$\mathbf{f}_t$ for the symbol available at time $t$. 
At instants $t$ when no stimulus is
presented, $\mathbf{f}_t = \mathbf{0}$, the zero vector.   
If we present a specific symbol at a specific time $t_o$, this adds to $\mathbf{f}(t)$ the
basis vector for that symbol multiplied by a delta function over time centered
at $t_o$.  At most times, the input is zero. 
We will occasionally refer to the moment on a particular trial when
\stimone{} is presented, $\mathbf{f}_{t} = \mathbf{\baseone}$ as $t_x$. 

%

We write $\mathbf{f}_t(\tau)$ to describe the
true past that led up to time $t$. The continuous variable $\tau$ runs from
$0^-$, corresponding to the moment of the past closest to the present
backwards to $-\infty$, corresponding
to the distant past.
Whereas $\mathbf{f}_t$ is the symbol available in the present at a particular
instant $t$,
$\mathbf{f}_t(\tau), \tau \in (-\infty, 0)$ is the timeline that led up to time $t$. 
Under the assumption that every symbol is presented at most once per trial,
each component of $\mathbf{f}_t(\tau)$ over the interval $\tau<0$ is either a delta function at some
particular $\tau$ or zero everywhere.  The goal of the associative memory is
to provide a guess about the future that will follow time $t$,
$\mathbf{f}_t(\tau), \tau \in (0, \infty)$ (Figure~\ref{fig:notation}) given
the symbol available in the present.  

The symbol provided in the present $\mathbf{f}_t$ is available to both the
past and the future.  The present enters the past timeline at its most recent
point.  In this formulation, the present is also available as the most
rearward portion of the future timeline.  By associating the past to the
rearward portion of the future, we can learn temporal relationships between symbols
separated in time.  By probing these associations with the present---as the
most recent part of the past timeline---we can construct an extended estimate
of the future.

\subsection{Laplace neural manifolds for the past and the future}
We estimate both the past and the future as functions over neural manifolds.  
Each manifold is a population of processing elements---neurons---each of which
is indexed by a position in a coordinate space. We treat the coordinates as
effectively continuous and locally Euclidean.  At each moment, each neuron
is mapped onto  a scalar value  corresponding to its firing rate over a macroscopic
period of time on the order of say 100~ms.  We propose that the past
and the future are represented by separate manifolds that interact with one
another.

The representations for both the past and the future each utilize two connected
manifolds.  We refer to one kind of manifold, indexed by an effectively
continuous variable $s$, as a Laplace space.   The other kind of manifold,
indexed by an effectively continuous variable $\taustar$, is referred to as an
inverse space.  Taken together, we refer to these paired representations as a
Laplace Neural Manifold.  The representations of the  past follow previous
work in theoretical neuroscience \citep{ShanHowa13,HowaEtal14}, cognitive psychology
\citep{HowaEtal15,SaleEtal22}, and neuroscience \citep{BrigEtal20,CaoEtal22}.   

\subsubsection{Laplace spaces for remembered past and predicted future}
\begin{figure}
	\centering
	\begin{tabular}{lr}
		\textbf{A}\\
		&\includegraphics[width=0.4\textwidth]{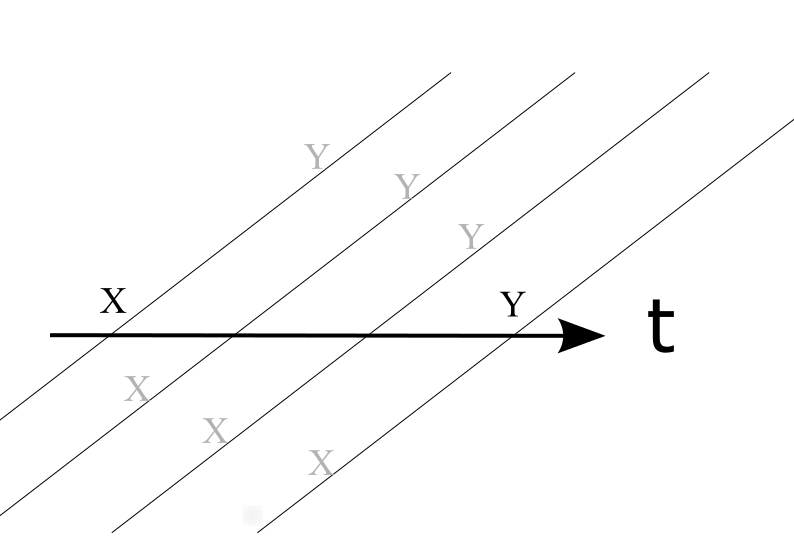}
	\end{tabular}

	~\\
	\begin{tabular}{lclc}
		\textbf{B} && \textbf{C} \\
		
		&\includegraphics[width=0.3\columnwidth]{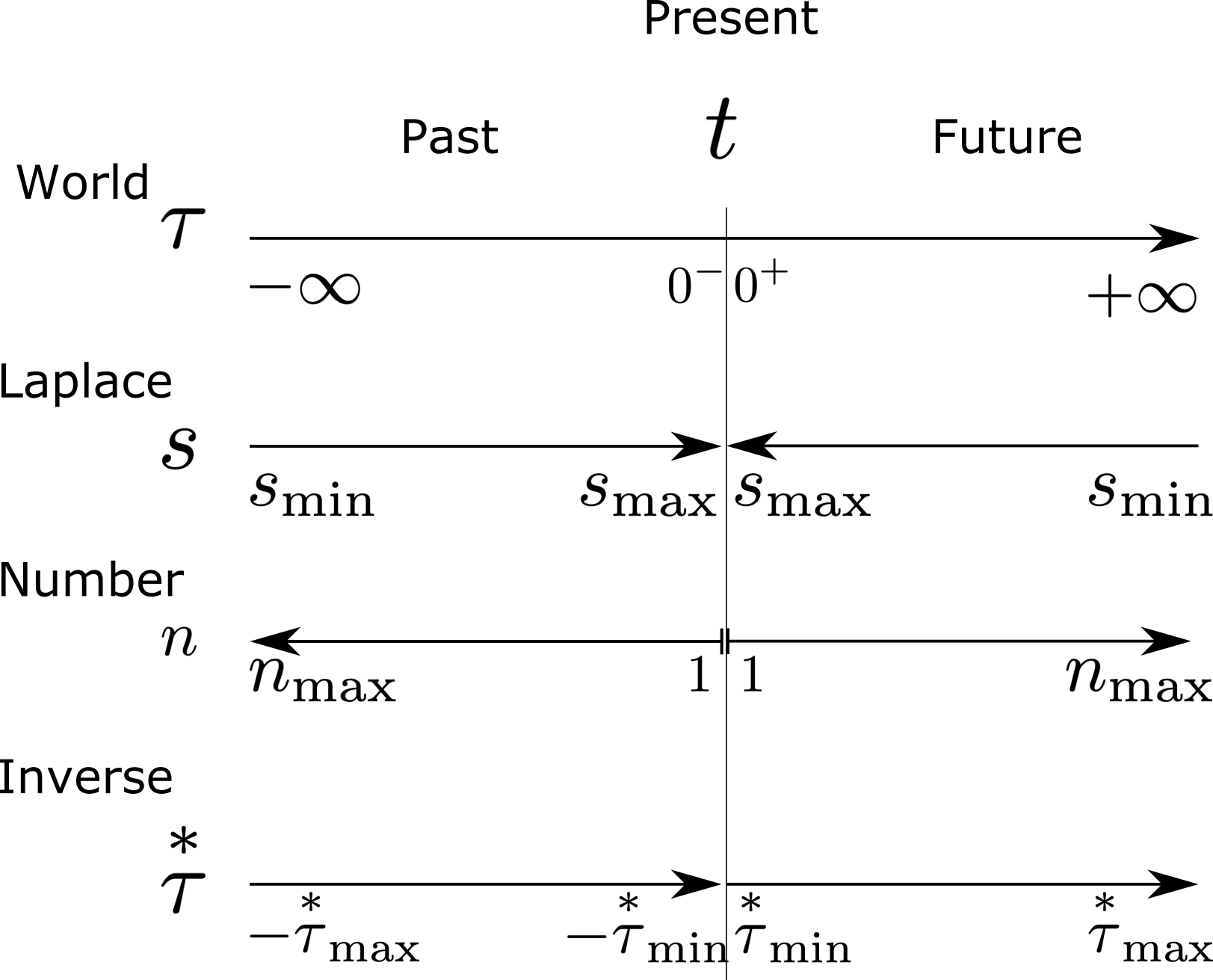}
		&&
		\includegraphics[width=0.4\columnwidth]{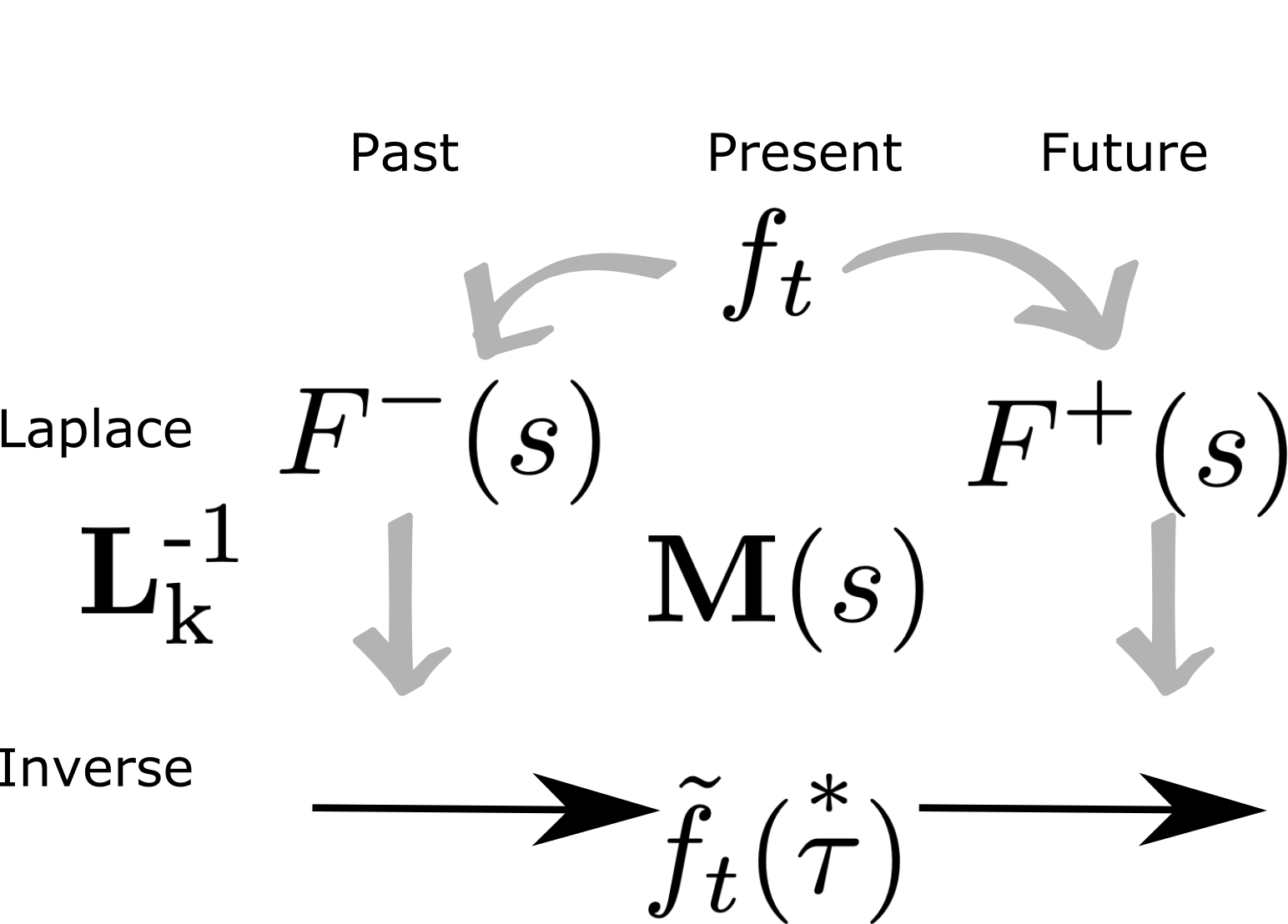}
	\end{tabular}
	\caption{
		Guide to notation.
		\textbf{A.} 
		Time measured externally is drawn as a 
		horizontal line; the ``internal timeline'' available to the
		agent at each moment is drawn as a diagonal line.  
		The remembered past at time $t$ is drawn below the horizontal;
		the predicted future is drawn above the horizontal.  
		Locations on the internal timeline are spaced to suggest  
		logarithmic compression.
		Consider a case in which \stimone{} and \stimtwo{} are
		presented many times with a consistent temporal relationship.
		If \stimone{} is
		presented at $t_\baseone$ and then 
		\stimtwo{} is presented at some later time $t_\basetwo$. 
		After \stimone{} is presented at $t=t_\baseone$, it recedes
		into the past, so for $t > t_\baseone$ we find that
		$\mathbf{f}_t(t-t_\baseone) = \baseone{}$.
		At the moment \stimtwo{} is presented, 
		$f_{t_\basetwo}(t_\basetwo - t_\baseone) = \baseone$.  
		After the relationship between \stimone{} and \stimtwo{} is
		learned, then after \stimone{} is presented \stimtwo{} is
		predicted a time $t_\basetwo - t_\baseone$ in the future.
		As time proceeds after presentation of \stimone{}, the
		predicted occurrence of \stimtwo{} should approach closer and
		closer to the present.
		\textbf{B.} Sign conventions.  At the present moment 
		$t$, objective time $\tau$ runs from $-\infty$ to $\infty$.
		$\tau=0$ corresponds to time $t$.
		The real Laplace domain variable $s$ runs from $0^+$ to
		$+\infty$ for both past and future, approximated as 
		$\smin$ and 
		$\smax$.  The
		units of $s$ are $t^{-1}$; the values corresponding to
		different points of the timeline are
		shown in the same vertical alignment.
		Cell number for Laplace and inverse spaces $n$ are aligned
		with one another.
		The variable $\taustar$ describes position along the inverse
		spaces.  It is in register with $\tau$ and derived from $s$.
		\textbf{C.} The stimulus available in the present,
		$\mathbf{f}_t$ provides input to two sets of neural manifolds.  One
		set of neural manifolds represents the past; the other
		estimates the future.  $\mat{M}(s)$ stores temporal
		relationships between events.
		\label{fig:notation}
	}
\end{figure}

The Laplace space
corresponding to the past, which we write as $\PastLaplbold{t}(s)$ encodes
the Laplace transform of $\mathbf{f}_t(-\tau)$, the past leading from the
present at time $t$ back towards the infinite past:
\begin{equation}
	\PastLaplbold{t}(s) = 
  \int_0^{\infty} e^{-s\tau} \mathbf{f}_t(-\tau) d\tau
=	\L{\mathbf{f}_t(-\tau)}(s),\  \tau \leq 0
	\label{eq:FLf}
\end{equation}
We restrict $s$ to real values on the positive line \cite[but
see][]{AghaEtal23}.\footnote{
The sign convention here is
distinct from prior papers that did not require $\tau$
(and $\taustar$ which will be introduced shortly) to be defined on both sides of zero
(Fig.~\ref{fig:notation}).
}
Many neurons tile the $s$ axis continuously for each symbol.  
To the extent that we can ensure a set of exponential receptive fields with a
continuous spectrum of $s$ values, we have established that
$\PastLaplbold{t}(s)$ is the Laplace transform of the past.
Exponential receptive fields over the past with a continuous spectrum of time
constants have been observed in many brain regions and species
\citep{AtanEtal23,BrigEtal20,TsaoEtal18,ZuoEtal23,DansEtal23,CaoEtal24}.

The index $s$ assigned to a neuron corresponds to the inverse
of its functional time constant.  
The Laplace space
corresponding to the future, which we write as $\mathbf{\FutLapl}(s)$ is an
attempt to estimate the Laplace transform of the future,
$\L{\mathbf{f}_t(\tau)}(s)$ over the interval $\tau \geq 0$.
Thus, there is a natural mapping between
$1/s$ and $|\tau|$ within both the past and the future.  By convention, $s$ is
positive for both the past and the future so that $\PastLaplbold{t}(s)$ is the
Laplace transform of $\mathbf{f}_t(-\tau)$ for  $\tau \leq 0$  whereas
$\FutLaplbold{t}(s)$ is the Laplace transform of
$\mathbf{f}_t(\tau)$ for  $\tau \geq 0$.  

Although $s$ is effectively continuous, this does not require that neurons sample
$s$ evenly.  Following previous work in psychology
\cite[e.g.,][]{ChatBrow08,Pian16,HowaShan18}, neuroscience
\citep{GuoEtal21,CaoEtal22}, and theoretical neuroscience
\citep{LindFage96,ShanHowa13}, we assume
that $s$ is sampled on a logarithmic scale.
Let $n$ be the neuron number, starting from the largest value of
$\smax$ nearest $\tau=0$ and extending out from the present.
We obtain a logarithmic scale by choosing $ds/dn = -s$.

%

\subsubsection{Updating Laplace spaces in real time}
Suppose that we have arranged for one particular component
of $\PastLaplbold{t}(s)$ or $\FutLaplbold{t}(s)$ to hold the Laplace transform of
one particular symbol, which we write as $f_t(\tau)$.   Suppose further that
$f_t(\tau)$ is zero in the neighborhood of $\tau=0$.   Consider how
this component, which we write as $\PastLapl{}(s)$ or $\FutLapl{}(s)$, should
update as time passes.  Let us pick some minimal increment of time $\Delta t$
on the order of, say, 100~ms.
At time $t+\Delta t$, information in $f_t(\tau)$ for $\tau \leq
0$  recedes further away from the present, so that $\PastLapl{t+\Delta t}(s) =
\L{f_{t}(\tau+\Delta t)}$.
In contrast, at time $t+\Delta t$, information in the future $f_t(\tau)$ for
$\tau> 0$ comes closer to
the present, so that $\FutLapl{t+\Delta t}(s) = \L{f_{t}(\tau-\Delta t)}$.
More generally, suppose that $F_t(s)$ is the Laplace transform of a function
over some variable $x$, $F_t(s) = \L{f_t(x)}(s)$.  
Defining $\alpha \equiv \frac{\Delta x}{\Delta t}$, we can update  
$F_t(s)$ as 
\begin{equation}
	F_{t+\Delta t}(s)  = \L{\mathcal{T}_{\alpha\Delta t } f_t(x)}(s)
				  =  e^{-s\alpha\Delta t} F_t(s)
	\label{eq:alphaODE}
\end{equation}
where $\mathcal{T}$ is the translation operator, $\mathcal{T}_a f(x) = f(x+a)$
and we have used the expression for the Laplace transform of translated
functions.  Equation~\ref{eq:alphaODE} describes a recipe for updating both
$F_t^{\pm}(s)$ with $\alpha_\pm$ in the absence of new input.  
Using the sign convention developed here, we fix   $\alpha_- =
1$ for $\PastLapl{}(s)$ and fix $\alpha_+ = -1$  for $\FutLapl{}(s)$.
It is possible to incorporate changes into the rate of flow of subjective time
by letting $\alpha_\pm$ change in register, such that $\alpha_+(t) =
-\alpha_-(t)$ for all $t$. 
The expression in Eq.~\ref{eq:alphaODE} holds more generally
and can be used to update Laplace transforms over many continuous variables of
interest for cognitive neuroscience \citep{HowaEtal14,HowaEtal18,HowaHass20}.

We are in a position to explain how $\PastLaplbold{t}(s)$ comes to represent
the Laplace transform of $\mathbf{f}_t(-\tau)$ over the interval $\tau \in
(-\infty,0)$; a discussion of how
$\FutLaplbold{t}(s)$ comes to estimate the future requires more development
and will be postponed.
When a symbol is presented at  time $t$, it enters timeline of the past at
$\tau=0$.  So, incorporating the Laplace transform 
of the most recent part of the past  with the past that is already
available  and then evolving to time $t + \Delta t$ we have
\begin{eqnarray}
	\PastLaplbold{t+\Delta t}(s) &=& e^{-s \Delta t}\left[ \PastLaplbold{t}(s) + \L{
		\delta\left(0\right)\mathbf{f}_t}\right] \nonumber\\
	& =& e^{-s \Delta t} \PastLaplbold{t}(s) + e^{-s \Delta t}
	\mathbf{f}_t.
	\label{eq:EvolvePast}
\end{eqnarray}
At time $t+\Delta t$, the input from time $t$ is encoded as the Laplace
transform of that symbol a time $\Delta t$ in the past.  At each subsequent
time step, an additional factor of $e^{-s \Delta t}$ accumulates.  As
time passes, the input from time $t$ is always stored as Laplace transform of
a delta function at the appropriate place on the timeline.  Because this is
true for all stimuli that enter $\PastLaplbold{}(s)$, we conclude that
$\PastLaplbold{t}(s)$ encodes the Laplace transform of the past
$\mathbf{f}_t(-\tau)$ over the interval $\tau \leq 0$.

\begin{figure}
	\centering
	\includegraphics[width=0.9\columnwidth]{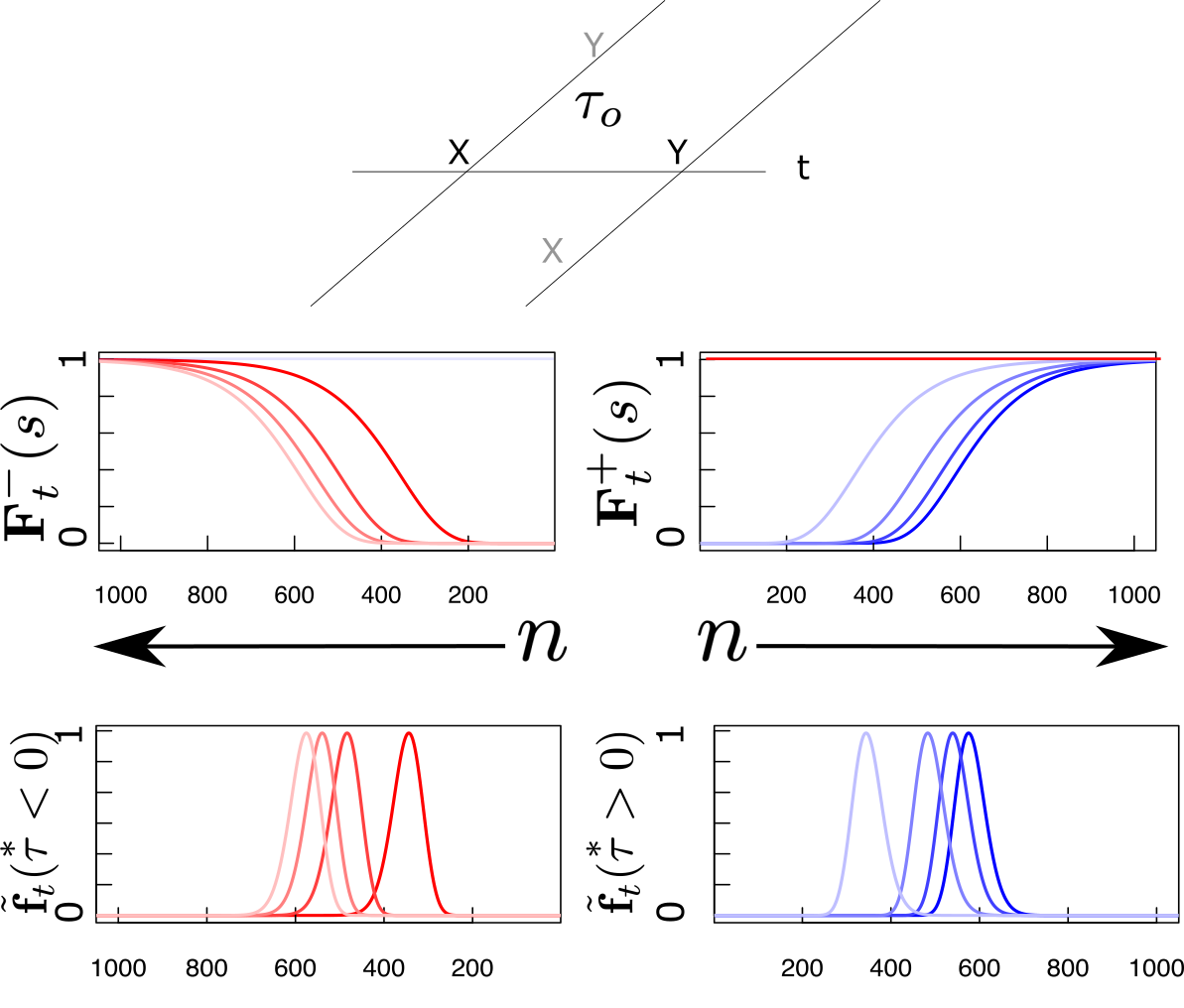}
	\caption{Neural manifolds that construct a logarithmically-compressed
	internal timeline of the past and the future.  Top: A temporal
	relationship exists
	between  \stimone{} and \stimtwo{} such that \stimtwo{} always follows
	\stimone{} after a delay of $\tau_o$ seconds.   Consider how the
	internal timeline ought to behave after \stimone{} is presented at
	$t=0$.  At time $t$, the past should include \stimone{} $t$ seconds in
	the past and \stimtwo{} $\tau_o-t$ seconds in the future.
	Middle and bottom:
	Samples of the timeline at evenly-spaced moments between zero and
	$\tau_o$.  At each moment, there is a pattern of activity over neurons
	indexed by $n$.  The state of the timeline at earlier moments, closer
	to $t=0$, are darker and later
	moments closer to $t=\tau_o$  are lighter.  Red lines are
	neurons coding for \stimone{} (primarily in the past except precisely
	at $t$), blue lines are neurons coding for
	\stimtwo{} (primarily in the future except precisely at $t=\tau_o$). 
	Middle: Laplace spaces for the past (left) and future (right) shown as
	a function of cell number $n$; Bottom: inverse spaces, constructed
	using the Post approximation, for the past
	(left) and future (right) shown as a function of log time.
	Exactly at time $t=0$, \stimone{} is available a time $0^+$
	in the future (dark horizontal red line, middle right).  Similarly,
	exactly at $t=\tau_o$, \stimtwo{} is available a time $0^-$ in the
	past (light horizontal blue line, middle left).
	\label{fig:timeline}
	}
\end{figure}

The middle panel of Figure~\ref{fig:timeline} illustrates the profile of
activity over  $\PastLaplbold{t}$  and $\FutLaplbold{t}$, shown as a function
of cell number $n$, resulting from the
Laplace transform of a delta function at various moments in time.
In the
middle panel, the $s$  axis 
for the past is reversed to allow appreciation of the relationship between
past time $\tau \leq 0$ and $\PastLapl{}$.
Note that the Laplace transform of a delta function has a characteristic shape
as a function of cell number that merely translates  as time
passes.
Note that the magnitude of the translation of $F^{\pm}[n]$ depends on the
value of $\tau_o$. 
It can be shown that for a delta function $ F^{\pm}_{t + \Delta t}[n] =
F_t^{\pm}[n+\Delta n]$ with $\Delta n = \alpha_\pm \frac{\Delta t}{\tau_o}$.
This can be appreciated by noting that the distances between successive lines
in the middle panel of Fig.~\ref{fig:timeline} are not constant despite
the fact that they correspond to the same time displacement.  Whereas
$\Delta n$ goes down as time passes for $\PastLapl{}[n]$ as the past becomes
more remote from the present, $\Delta n$ increases with the passage of time for $\FutLapl{}[n]$
as the future grows closer to the present.   

There are implementational challenges to building a neural circuit that
obeys Eq.~\ref{eq:alphaODE}; these challenges are especially serious when 
$\alpha <0$, which requires activation to grow exponentially.
If one is willing to
restrict the representation of each symbol to the Laplace transform of a delta
function at a single point in time, it is straightforward to implement
a continuous attractor network \citep{KhonFiet21} to allow the ``edge'' in the
Laplace transform as a function of $n$ to translate appropriately.
\citet{DaniHowa24} constructed a simple continuous attractor network to
demonstrate the feasibility of this approach.


\subsubsection{Inverse spaces for remembered past and predicted future}
The mammalian brain also contains ``time cells''  
with circumscribed receptive fields 
\citep{PastEtal08,MacDEtal11,TigaEtal18a,SchoEtal23}.
Time cells resemble a ``direct'' estimate of the
past and are reasonably well approximated as: 
\begin{equation}
\ftilde_{t}(\taustar) = 
	\int_{0}^\infty
	\Phi\left(\frac{\tau}{\taustar}\right)f_t(-\tau) d\tau
	\label{eq:ftilde}
\end{equation}
where $\Phi(x)$ is a unimodal function with its maximum at 1 and $\taustar$ is
here defined to be negative (Fig.~\ref{fig:notation}).   
$\ftilde_{t}(\taustar)$ estimates the true past in the neighborhood of $\taustar$.
As $\Phi()$ becomes more and more sharp,  
approaching a
delta function, $\ftilde_t(\taustar)$ goes to $f_t(\taustar)$.  
In
this sense $\ftilde(\taustar)$ is like the inverse Laplace transform of
$F_t(s)$.  However, because receptive fields depend only on the ratio of
$\tau/\taustar$, and because neurons sample the $\taustar$ axis
logarithmically, $\ftilde(n)$ is a convolution of $f(\log
\tau)$
and another function of $\log \tau$ that controls the blur. 

The bottom panel of Figure~\ref{fig:timeline}  shows a graphical depiction of
the inverse space for the past and the future during the interval between
presentation of \stimone{} and \stimtwo{}.  
The inverse spaces approximate the past, $\ftilde_t(\taustar)$ for $\taustar < 0$ 
and the future,
$\ftilde_t(\taustar)$ for $\taustar > 0$  on a log scale.  
As the delta function corresponding
to the time of \stimone{} recedes into the
past, the corresponding bump of activity in $\mathbf{\baseone}'\ftildebold_t(n)$
also moves, keeping its shape but moving more and more slowly as \stimone{}
recedes further and further into the past.  In the future, the delta
function corresponding to the predicted time of \stimtwo{} should start a time
$\tau_o$ in the future and come closer to the present as time passes.  
As the prediction for \stimtwo{} approaches the
present, the corresponding bump of activity in
$\mathbf{\basetwo}'\ftildebold_t(n)$ keeps its shape but the speed of the bump
accelerates rather than slowing with the passage of time.


It is in principle possible to construct the inverse space from the Laplace
space \emph{via} a linear feedforward operator.
Previous papers \cite[e.g.,][]{ShanHowa13} have made use of the Post
approximation to the inverse Laplace transform to construct the inverse space
from the Laplace space.  This is not neurally reasonable \citep{Gosm18}; the
Post approximation is difficult to
implement even in artificial neural networks
\cite[e.g.,][]{TanoEtal20,JacqEtal21}.  A more
robust approach would be a continuous attractor network
\cite[for a review see][]{KhonFiet21} that takes
input as the derivative of $F$ with respect to $n$.  The width of the bump
in $\ftilde$ would depend on internal connections between neurons in $\ftilde$
and global inhibition would stabilize the activity over $\ftilde$.  In this
case, moving the bump in different directions, corresponding to $\alpha >0$
and $\alpha < 0$ is analogous to moving a bump of activity in a ring
attractor  in different directions. A companion
paper fleshes out these ideas \citep{DaniHowa24}.

\subsection{Predicting the future from the past}
The previous subsection describes how to evolve the Laplace manifold for the
past.  
If we could somehow initialize the
representation of the future appropriately
then we could use the same approach to evolve the Laplace manifold for the
future during periods when no symbol is experienced. Initializing the future
will be accomplished \emph{via} learned temporal relationships between the
past and the future. 

The model has access to the Laplace transform of the past, as described above.
We define the present so that it overlaps with both the most recent part of
the past and the most imminent, or ``rearward,'' part of the future.
We form Hebbian associations between the Laplace transform of the past and the
Laplace transform of the rearward portion of the future.
Recall that products of 
Laplace transforms are the Laplace transform of the convolution of these
functions.
Because there is a reflection between the definition of $\PastLaplbold{t}(s)$ and
$\FutLaplbold{t}(s)$, the convolution of these two functions measures distances 
between time points in the past and the
present.
Later the present stimulus, taken as the Laplace transform of the most recent
part of the past, can be used to recover the Laplace transform of an extended
future timeline.

There are two sets of weights storing these associations, $\mathbf{M}(s)$ and
$\mathbf{\bar{M}}(s)$.  Each of these weights learn associations between
the Laplace transform of the past, $\PastLaplbold{t}(s)$, and the present
stimulus $\mathbf{f}_t$.  The two sets of weights are normalized differently.
Roughly speaking, $\mathbf{M}(s)$ stores the  Laplace transform of the future
conditionalized on the present symbol.  In contrast $\mathbf{\bar{M}}(s)$
stores the Laplace transform of the past conditionalized on the present
symbol.  With the assumptions that let us consider discrete trials, these
transforms are understandable as pairwise statistics of events corresponding
to a presentation of each symbol on a trial.   We will see that taken together
$\mathbf{M}(s)$ and $\mathbf{\bar{M}}(s)$ enable us to estimate the
associative and temporal contingency between each pair of symbols
conditionalized on each other symbol. 

The learning rate and forgetting rate for the sets of weights fixes a time
horizon for learning over trials.
By choosing a continuous spectrum of forgetting rates $\learn$ and learning
rates $1-\learn$, both 
$\mat{M}(\learn,s)$ and $\mat{\bar{M}}(\learn,s)$ retain a memory 
for the history as a function of trials.
Continuous forgetting allows the weights to implement a
discrete approximation to the Laplace transform.  This property of
$\mathbf{M}(\learn,s)$ and $\mathbf{\bar{M}}(\learn,s)$ means that it is in
principle possible to aggregate joint statistics between stimuli.

\subsubsection{Encoding $\mat{M}(s)$}
The moment a nonzero stimulus $\mathbf{f}_t$ is experienced,
we assume it is available to both $\PastLaplbold{}$ and $\FutLaplbold{}$, triggering a
number of operations which presumably occur sequentially within a small window of
time on the order of 100~ms.  
First, the present stimulus updates a prediction for the future  \emph{via} a
set of connections $\mat{M}$ organized by $s$.  Then these connections are
updated by associating the past to the present.  Finally the present stimulus
is added to the representation of the past.   For ease of exposition we will
first focus on describing the connections between the past and the future.  

\begin{figure}
	\centering
	\includegraphics[width=0.6\columnwidth]{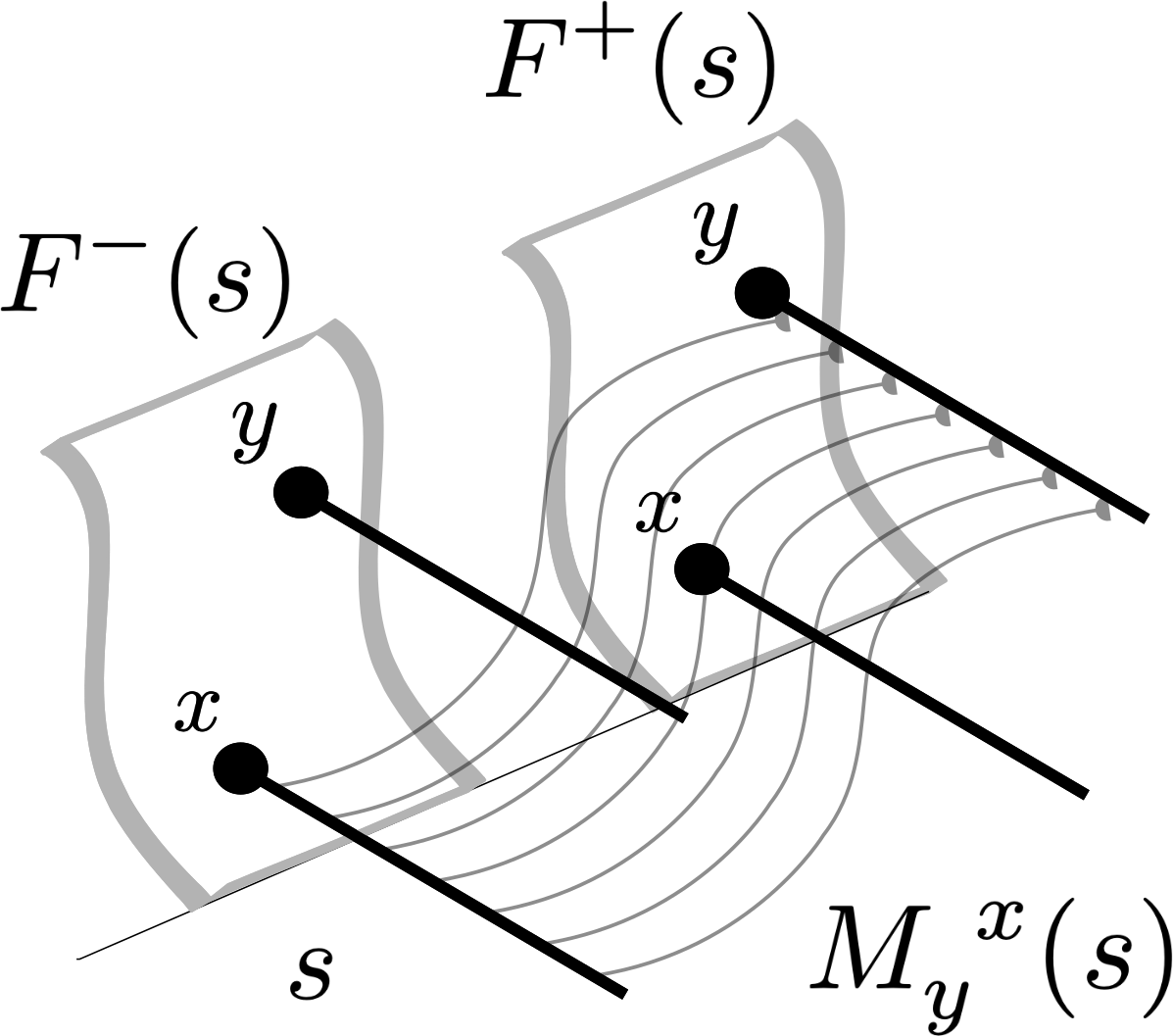}
	\caption{
		Schematic figure illustrating $M_y^{\ x} (s)$.
		$\PastLapl{}(s)$ and $\FutLapl{}(s)$ components for all
		the possible symbols, here shown schematically as sheets.
		Two symbols \stimone{} and \stimtwo{} are shown in both
		$\PastLapl{}(s)$ and $\FutLapl{}(s)$.   Each
		symbol is associated with a population of neurons spanning a
		continuous set of $s$ values, shown as the heavy lines in this
		cartoon.  $\mat{M}(s)$ describes the
		connections between each symbol in $\PastLaplbold{}(s)$
		to each symbol in $\FutLaplbold{}(s)$
		for each value of $s$.  
		The curved lines $M_y^{\ x}(s)$ illustrate the set of weights
		connecting units corresponding to \stimone{} in $\PastLapl{}$ to
		units corresponding \stimtwo{} in
		$\FutLapl{}$.  Connections exist only between units with the
		same values of $s$.  The strength of the connections in
		$\MXY$   vary as a function of $s$ in a way
		that reflects the pairwise history between \stimone{} and
		\stimtwo{}.
		\label{fig:Mofsfig}
	}
\end{figure}
We write $\mat{M}(s)$ for a set of connections that associates the Laplace
transform of the past to the
Laplace transform of the future (Fig.~\ref{fig:Mofsfig}).  
We postpone discussion of the other set of weights $\mat{\bar{M}}(s)$.
For any particular
value $s_o$, $\mat{M}(s_o)$ is a matrix describing connections from each symbol in
$\PastLaplbold{}(s_o)$ to each symbol 
in $\FutLaplbold{}(s_o)$.  For each
pair of symbols, say \stimone{} and \stimtwo{}, we write $M_{\basetwo}^{\
\baseone}(s_o)$ for the strength of the connection \emph{from} the cell corresponding
to $\mathbf{\baseone}$ with $s=s_o$ in
$\PastLaplbold{}$ \emph{to} the cell corresponding to  $\mathbf{\basetwo}$ in 
$\FutLaplbold{}$ with
$s=s_o$.
$\mat{M}(s)$ does not include connections between neurons with different
values of $s$.  
On occasion it will be useful to think of the set of
connections between
a pair of symbols over all values of $s$, which we write as $M_{\basetwo}^{\
\baseone}(s)$.  Similarly, we write $\mathbf{M}^{\basetwo}(s)$ for the set of
connections \emph{from} \stimtwo{} in $\PastLapl{}$ to all stimuli in
$\FutLapl{}$ over all values of $s$.  We write $\mathbf{M}_{\basetwo}(s)$  for
the set of connections \emph{to} \stimtwo{} in $\FutLapl{}$ from all symbols 
and all values of $s$.  In this paper, the superscripting and subscripting of
$\MXY(s)$ has no significance beyond a visual aid to help keep the indices
straight.

\begin{figure}
	\centering
		\includegraphics[width=0.9\columnwidth]{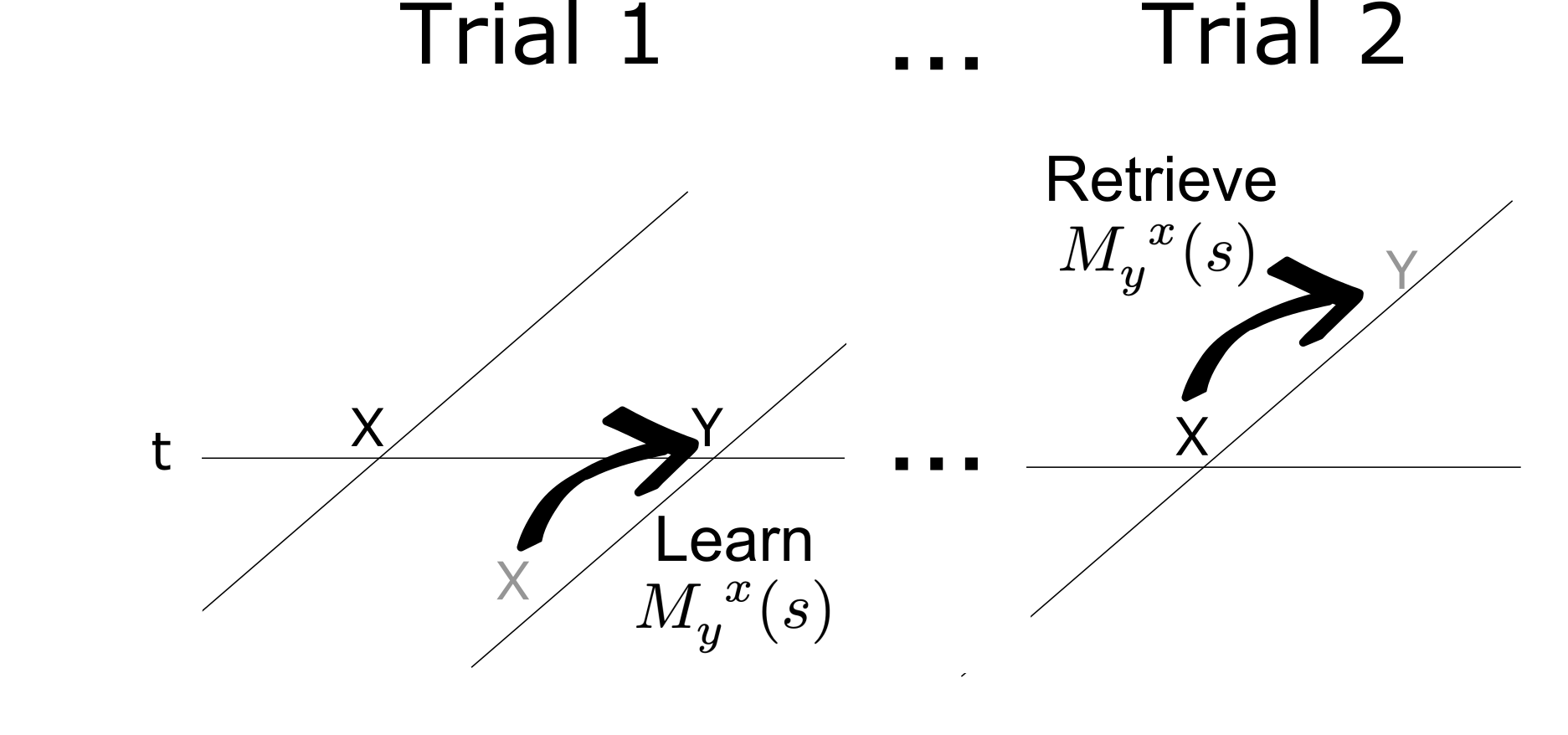}
	\caption{
		Learning and expressing pairwise associations with
		$\mat{M}(s)$.  The horizontal line is time; the diagonal lines
		indicate the internal timeline at the moments they intersect.
		Memory for the past is below the horizontal line; prediction
		of the future is above. When \stimone{} is 
		presented for the first time, it predicts nothing.  When
		\stimtwo{} is presented, the past contains a memory for
		\stimone{} in the past. When \stimtwo{} is presented, $\MXY$
		stores the temporal relationship between \stimone{} in the
		past and \stimtwo{} in the present---the rearward part of the
		future. In addition to storing learned relationships,
		connections from each item decay each time it was presented
		(not shown).  When \stimone{} is repeated much later in time,
		the stored connections in $\MXY$ retrieve a prediction of
		\stimtwo{} in the future.  \label{fig:Mschem}
	}
\end{figure}

When a particular stimulus \stimtwo{} is presented the
connections to and from that stimulus in $\mat{M}(s)$ are updated.
When \stimtwo{} is presented, the connections from \stimtwo{} in the past
towards all stimuli in the present are updated as
\begin{equation}
	\mathbf{M}_{}^{\basetwo}(s) \rightarrow \learn  \mathbf{M}_{}^{\basetwo}(s)
	\label{eq:Mfromforget}
\end{equation}
That is, the connections from \basetwo{}  to every other
stimulus for each value of $s$ are all scaled down by a value $\learn$.  
Later we will
consider the implications of a continuous spectrum of $\learn$ values;
for now let us  just treat $\learn$ as a fixed parameter restricted to be
between zero and one.  When \stimtwo{} is presented, it  momentarily becomes
available at the ``rearward part'' of the future.  In much the same way that
the present enters the past (Eq.~\ref{eq:EvolvePast}) at $\tau=0^-$, we also
assume that the present is also available momentarily in the future at
$\tau=0$.
When $\stimtwo{}$ is presented, the connections from each symbol in the past
to \stimtwo{} in the future are updated as 
\begin{equation}
	\mathbf{M}_{\basetwo}(s)  \rightarrow   \mathbf{M}_{\basetwo}(s)  +
	(1-\learn) \PastLaplbold{t}(s)
	\label{eq:MtoY}
\end{equation}
Connections involving symbols that are not present in the history retained by $\PastLaplbold{t}(s)$
are not updated.  We can understand Eq.~\ref{eq:MtoY} as a Hebbian association
between the units in $\PastLaplbold{}(s)$, whose current activation is given
by $\PastLaplbold{t}(s)$ and the units in the future $\FutLaplbold{}(s)$
corresponding to the present stimulus
\stimtwo{} (see Fig.~\ref{fig:Mschem}).  
More generally, we can understand this learning rule as
strengthening connections from the past $\PastLaplbold{t}(s)$
to the rearward part of the future, $\L{\delta(0)}(s) \mathbf{f}_t =
e^{-s 0} \mathbf{f}_t = \mathbf{f}_t$. 
Because the second term is the product of two Laplace transforms,
it can also be understood as the  Laplace transform of a convolution, here,
the convolution of the present with the past.\footnote{Because of the sign
conventions adopted here, $\PastLapl{t}(s)$ is the Laplace transform of
$f_t(-\tau)$ whereas $\FutLapl{t}(s)$ is the transform of $f_t(\tau)$. Viewed
in this light it is more precise to think of Eq.~\ref{eq:MtoY}
as learning the Laplace transform of the cross-correlation between the present
and the past.}  Convolution has long been used as an associative operation in
mathematical psychology \citep{Murd82,JoneMewh07,KatoCapl17}, neural networks
\citep{Plat95,Elia13,BlouEtal16}, and
computational neuroscience \citep{SteiSomp22}. 

\subsubsection{$\mat{M}(s)$ is a Laplace successor representation}

From examination of Eqs.~\ref{eq:Mfromforget}~and~\ref{eq:MtoY}, we see that 
after each trial $\MXY$ is multiplied by $\learn$ when \stimone{} was
presented.  For trials on which  \stimtwo{}
was also presented,  $(1-\learn) e^{-s \tau_o}$ is added to $\MXY$.
Writing $h[i]$ as an indicator variable for the history of presentations of
\stimtwo{} on the trial $i$ steps in the past we find at the conclusion of a
trial that
\begin{equation}
	\MXY = (1-\learn) e^{-s \tau_o} \sum_i \learn^{i} h[i] .
	\label{eq:MXYseries}
\end{equation}
Note that if $P(\basetwo|\baseone)= 1$, then after an infinitely long series of trials $\sum_i
h[i] \learn^i = \frac{1}{1-\learn}$  and $\MXY = e^{-s\tau_o}$ for all choices
of $\learn$.  Following similar logic, if we relax the assumption that
$P(\basetwo|\baseone)=1$ and take the limit as $\learn$ goes to 1, we find
that $\MXY= P(\basetwo|\baseone) e^{-s\tau_o}$.

Now let us relax the assumption that the time lag between \stimone{} and
\stimtwo{} always takes the same value.
Let the lag be a random
variable $\tau_{\baseone\basetwo}$ subject to the constraint that
$\tau_{\baseone\basetwo}$ is always $>0$.  This is not a fundamental
restriction; if $\tau_{\baseone\basetwo}$ changed sign, those observations
would contribute to $\MYX$ instead of $\MXY$.  Now, again taking the limit as
$\learn \rightarrow 1$,  we find 
\begin{equation}
	\MXY = P\left(\basetwo | \baseone\right) E\left[e^{-s \tau_{\baseone
	\basetwo}} \right] 
	 =  P\left(\basetwo | \baseone\right)  \L{\tau_{\baseone
	\basetwo}}(s)
\label{eq:MXYContingent}
\end{equation}
where we have used the definition for the Laplace transform of a random
variable, again with the understanding that we restrict $s$ to be real and
positive.

Equation~\ref{eq:MXYContingent} illustrates several important properties of
$\mat{M}(s)$.
First, we can see that $\MXY$ provides complete information about the
distribution of temporal lags between \stimone{}  preceding \stimtwo{}.  This
can be further appreciated by noting that the Laplace transform of the random
variable on the
right hand side is the moment generating function of $-\tau_{\baseone
\basetwo}$. Keeping the computation in the Laplace domain means that there is
no blur introduced by going into the inverse space as in previous attempts to
build a model for predicting the future
\citep{ShanEtal16,TigaEtal19a,GohEtal22}. 
Second, because $\L{\tau_{\baseone \basetwo}}(s=0) = 1$ as long as the
expectation of $\tau_{\baseone \basetwo}$ is finite, 
$\MXYbare(s=0) = P(\basetwo|\baseone)$ and $\mathbf{M}(s=0)$ captures the
pairwise probabilities between all symbols.    

In the limit as $\learn \rightarrow 1$,
$\mat{M}(s)$ is closely related to the successor representation
\citep{Daya93,GersEtal12,MomeEtal17,StacEtal17,CarvEtal24} with a
continuous distribution of discount rates
\citep{KurtRedi09,MomeHowa18,TanoEtal20,MassEtal23,SousEtal23}.  More precisely, if one
assumes a complete compound serial representation of the past
and a fixed action policy, then computes the successor representation from RL
\citep{Daya93,GersEtal12}, but with a continuous spectrum of discount rates
$\gamma$, one would obtain $\mat{M}(s)$ with the identification $s= -\log
\gamma$.  However, computing $\mat{M}(s)$ does not require temporal difference
learning.  In RL language,  $\PastLaplbold{}(s)$ is an ensemble of eligibility
traces with a continuous spectrum of forgetting rates.  Associating this
multiscale eligibility trace to outcomes is sufficient to compute
$\mat{M}(s)$, which we might refer to as a Laplace successor representation.

\subsubsection{$\bar{\mathbf{M}}(s)$ is a Laplace {predecessor} representation}

It is straightforward to construct a Laplace predecessor representation
\citep{NambStub21}  using $\PastLaplbold{}(s)$, the Laplace transform of the
past, and Hebbian learning. 
We write out a new set of connections $\bar{\mat{M}}(s)$.
Adapting Eqs.~\ref{eq:Mfromforget}~and~\ref{eq:MtoY},
when each item \stimtwo{} is presented 
\begin{eqnarray}
	\mathbf{\bar{M}}^{\basetwo}(s)  & \rightarrow&  \learn \mathbf{\bar{M}}^{\basetwo}(s)  +
	(1-\learn) \PastLaplbold{t}(s) \label{eq:MtoYalt}
\end{eqnarray}
That is, when \stimtwo{} is presented at time $t$ and \stimone{} is available in
$\PastLaplbold{t}(s)$,  $\barMYX$ is incremented.  
Following similar steps as for $\mat{M}(s)$, in the limit as  $\learn
\rightarrow 1$, we get  
\begin{equation}
	\barMYX 
	= P\left(\baseone | \basetwo\right)\L{\tau_{\baseone
	\basetwo}}(s),
\label{eq:MXYContingentalt}
\end{equation}
which can be compared to Equation~\ref{eq:MXYContingent}.
Thus, with learning as in Eq.~\ref{eq:MtoYalt}, we can refer to
$\mat{\bar{M}}(s)$ as a Laplace predecessor representation.   

Note that the convention of $\bar{\mathbf{M}}(s)$ is different than
$\mat{M}(s)$.  Whereas $\MXY$ describes relationships between  \stimone{}
preceding \stimtwo{}, $\barMXY$ describes relationships between \stimtwo{}
preceding \stimone{}.  In this sense $\bar{\mathbf{M}}(s)$ is like $\mathbf{M}^T(s)$.
In addition one must also account for the reflection operator involved in the
definition of $\PastLaplbold{t}(s)$  as compared to $\FutLaplbold{t}(s)$ and
the different marginalization. 

The foregoing makes clear that if the brain has access to
$\PastLaplbold{}(s)$---an eligibility trace with a continuum of time
horizons---it is straightforward to compute either a successor representation
or a predecessor representation in a way that maintains complete information
about the temporal relationships between stimuli.   This approach does not
require selecting a single time horizon or time constant for either
representation \citep{FloeEtal24}.  


\subsubsection{Measures of contingency using $\mat{M}(s)$ and
$\bar{\mat{M}}(s)$}
Information contained in $\mat{M}(s)$ and $\bar{\mat{M}}(s)$ can be used to
not only describe pairwise relationships between stimuli but also to assess
contingency between symbols, allowing solutions to the temporal credit
assignment problem.  The goal here is not to propose a specific
measure of contingency---there are undoubtedly a multiplicity of such rules
that could be used for cognitive and neural modeling---but simply to sketch
out the properties of $\mat{M}(s)$ and $\bar{\mat{M}}(s)$.  We continue
attending to the limit as $\learn \rightarrow 1$.

\begin{figure}
	\centering
	\begin{tabular}{lcclc}
		\textbf{a} &&& \textbf{b}\\
		\includegraphics[width=0.45\textwidth]{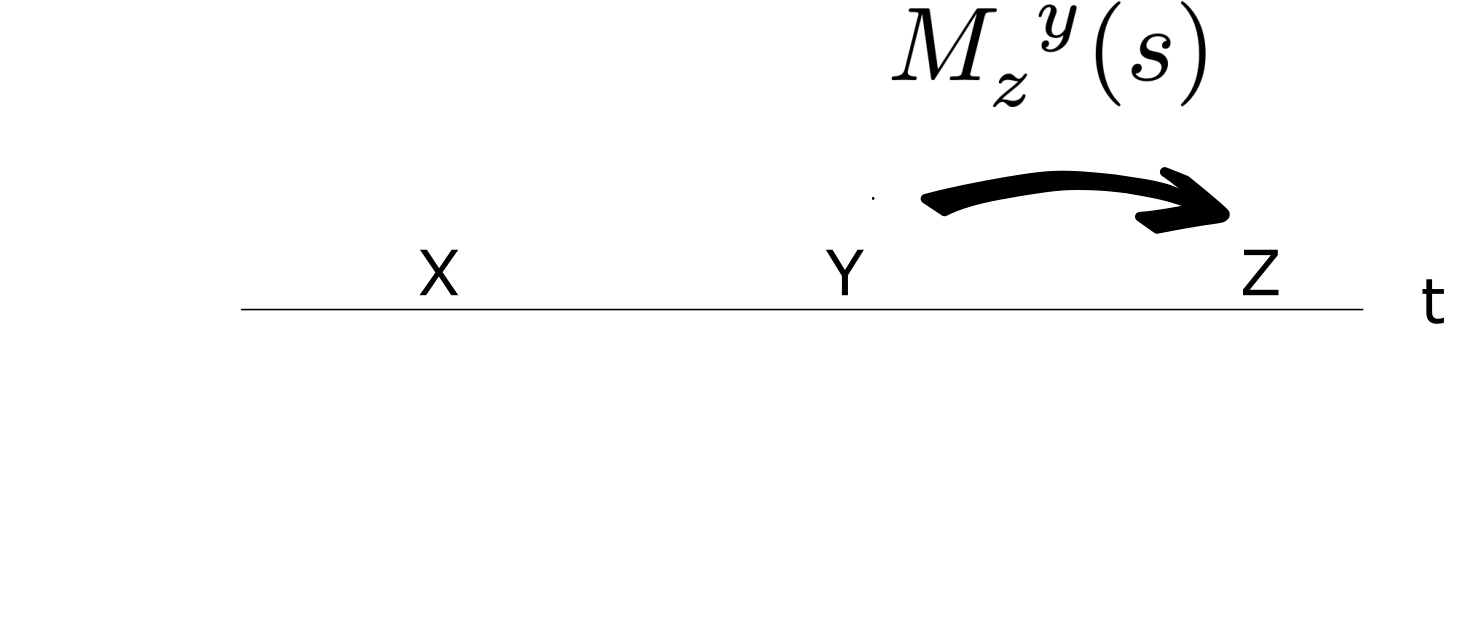}
		&&&
		\includegraphics[width=0.45\textwidth]{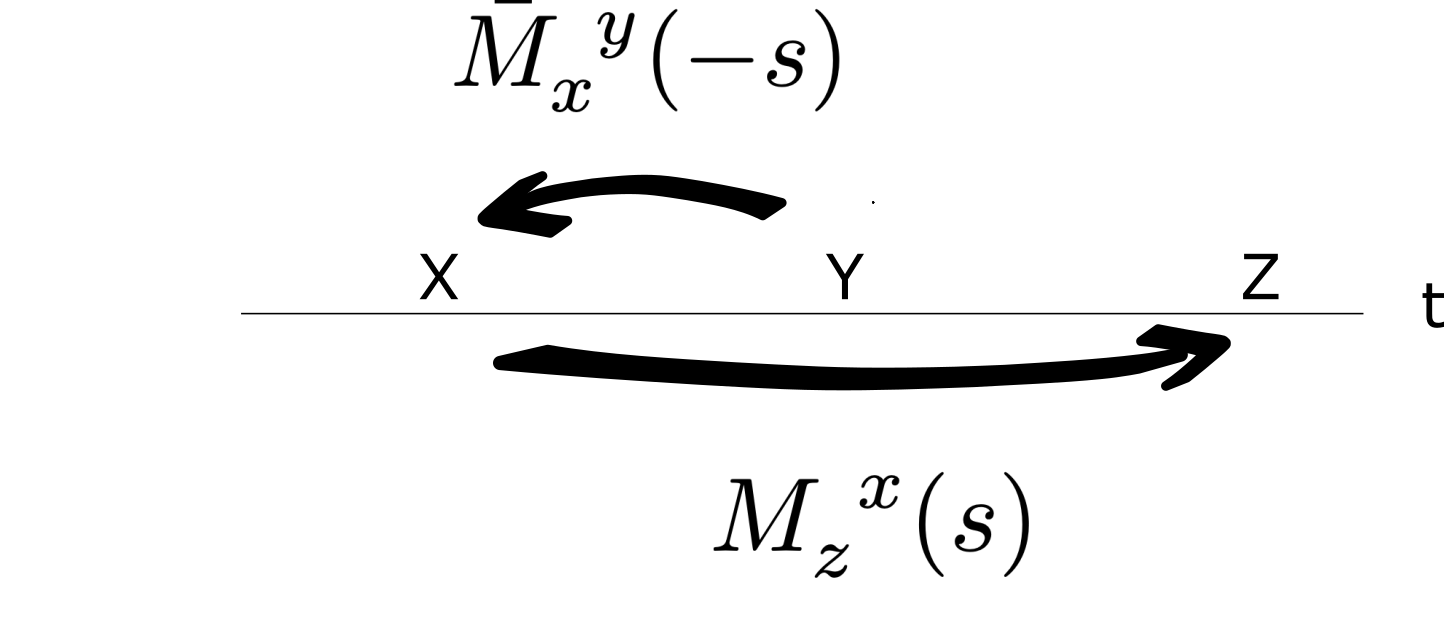}
	\end{tabular}
	\caption{
		Measuring contingency by comparing pairwise relationships
		between \stimtwo{} and \stimthree{} to
		pairwise relationships conditionalized on \stimone{}. 
		\textbf{a.}  Equation~\ref{eq:LaplaceDiff} captures the
		Laplace transform of the random
		variable $\tau_{\basetwo\basethree}$.  By assumption, on each trial 
		$\tau_{\basetwo\basethree} =   \tau_{\baseone \basethree} -
		\tau_{\baseone \basetwo}$.  
		\textbf{b.}  Equation~\ref{eq:LaplaceCrossCorr} captures the
		convolution of $\tau_{\baseone \basethree}$ and
		$-\tau_{\baseone \basetwo}$.  If these intervals are
		independent across trials then $\MYZbare(s) = \MXZbare(s)
		\barMYXbare(-s)$.
		\label{fig:MbarM}
	}
\end{figure}

For this illustration, let us restrict our attention to relationships between  
three symbols \stimone{}, \stimtwo{} and \stimthree{}.  We assume for
simplicity that, if they are presented on a trial, the three stimuli are presented in order
on each trial.  Let us
refer to the time lags between symbols as random variables
$\tau_{\baseone \basetwo}$, $\tau_{\basetwo \basethree}$; on trials where all
three symbols are observed  $\tau_{\baseone \basethree} = \tau_{\baseone
\basetwo} + \tau_{\basetwo \basethree}$.  
For convenience let's assume that the distributions are chosen such that the
relative times of presentation do not overlap. 
We denote the probabilities of each symbol occuring on a trial such that
$P(\basethree|\basetwo)$ gives the conditional probability that \stimthree{}
is observed on a trial given that \stimtwo{} is also observed on that trial.

We are interested in how much ``credit'' to allocate \stimtwo{} for the occurrence
and timing of \stimthree{}, taking into account \stimone{}.  
We will  compare $ \MYZbare(s)$, which describes the future occurrences of
\stimthree{} conditionalized on \stimtwo{} in the present to 
$\MXZbare(s)   \barMYXbare(-s)$ (Fig.~\ref{fig:MbarM}).  This quantity is $\MXZbare(s)$---the future
of \stimthree{} predicted by
\stimone{}---multiplied by $\barMYXbare(-s)$---the past occurrence of
\stimone{} predicted by knowing that \stimtwo{} is in the present.  That is,
$\MXZbare(s)   \barMYXbare(-s)$  describes the future of \stimthree{} predicted by the past
occurrence of \stimone{}  that is observed when 
\stimtwo{} is in the present.
The reflection operator allows the integration of these two timelines in a way
that can be compared to the future of \stimthree{} given that \stimtwo{} was
observed in the present \citep{ColeEtal95}. 

We will work through the implications of this high level description under very simple
circumstances.
Recall that  under the circumstances described in this subsection,
\begin{eqnarray}
	\MYZbare(s) &=& P(\basethree|\basetwo) \ \L{\tau_{\basetwo\basethree}}(s)\\
	 &=& P(\basethree|\basetwo) \  \L{\tau_{\baseone\basethree} - \tau_{\baseone\basetwo}}(s)
	 \label{eq:LaplaceDiff}
\end{eqnarray}
Using properties of the Laplace transform we can rewrite  $\MXZbare(s)
\barMXYbare(-s) $
as
\begin{eqnarray}
	\MXZbare(s)   \barMXYbare(-s) &=& P(\basethree|\baseone) \L{ \tau_{\baseone\basethree}}(s) 
		\ P(\baseone|\basetwo) \L{-\tau_{\baseone\basetwo}}(s)\\
		&=& P(\basethree|\baseone)P(\baseone|\basetwo)\  \L{\tau_{\baseone\basethree} \ast
		\left(-\tau_{\baseone\basetwo}\right)}(s).
	\label{eq:LaplaceCrossCorr}
\end{eqnarray}  
The second term describes the Laplace transform of the
convolution  of $\tau_{\baseone\basethree}$ and $-\tau_{\baseone \basetwo}$.  
Because the sum of two independent random variables is equal to their
convolution, the Laplace transforms in
Eqs~\ref{eq:LaplaceDiff}~and~\ref{eq:LaplaceCrossCorr} will enable us to
assess the dependence between the times of presentations of \stimone{}, \stimtwo{}, and
\stimthree{}.

\paragraph{Associative contingency at $s=0$}
$\mat{M}(s=0)$ gives information about the pairwise
probabilities between each pair of symbols. 
Suppose that \stimone{}, \stimtwo{}  and \stimthree{} occur on different
trials.  Is the occurrence of \stimthree{} predicted by \stimtwo{} or
\stimone{}?  Or some more complex situation? 

From Eqs.~\ref{eq:MXYContingent}~and~\ref{eq:MXYContingentalt} and basic
properties of random variables, we could compare
\begin{equation}
	\MYZbare(s=0) = P(\basethree|\basetwo)
	\label{eq:AssocContYZ}
\end{equation}
to 
\begin{equation}
\MXZbare(s=0) \barMYXbare(s=0) =  P(\basethree|\baseone)P(\baseone|\basetwo) 
	\label{eq:AssocContXZXY}
\end{equation}
If Eqs~\ref{eq:AssocContYZ}~and~\ref{eq:AssocContXZXY} are
equal to one another, then credit for \stimthree{} should go to \stimone{} rather than
\stimtwo{}.  To the extent they differ, then \stimtwo{} should get credit for
the occurrence of \stimthree{}.

Of course there are limits to how well the future can be predicted with
pairwise statistics.  More generally, we would like to consider joint
statistics.  This requires estimating higher order probabilities, e.g., 
$P(\baseone,\basethree|\basetwo)$.  We establish later that joint statistics
can be estimated  from $\mat{M}(\learn,s)$.   In an environment where  joint
statistics are important, predicting the future using simple pairwise
relationships is untenable.  However, it should be  possible to recode the symbols
into a new set of symbols that can be used to predict the future using
pairwise relationships.

\paragraph{Temporal contingency}
So that we can focus on temporal contingency, 
let us assome that all three stimuli are presented on each trial so that
$P(\basetwo|\baseone) = P(\basethree|\basetwo) = P(\basethree|\baseone)=1$.
Because $\MXY$ contains information about every moment of the
distribution $\tau_{\baseone\basetwo}$
it is
straightforward to ask whether the distribution of times for \stimthree{}
conditionalized on \stimtwo{} is higher or lower entropy than the distribution
conditionalized on \stimone{}.  It is also possible to use $\mat{M}(s)$ and
$\mat{\bar{M}}(s)$  to capture more subtle temporal relationships.


Recall that the distribution of
the sum of two random variables equals the convolution of those random
variables if they are independent of one another.
Thus comparing the distribution of
$\tau_{\basetwo\basethree}=\tau_{\baseone\basethree} -
\tau_{\baseone\basetwo}$ to the distribution of the convolution
$\tau_{\baseone\basethree} \ast (-\tau_{\baseone\basetwo})$ allows us to 
assess the dependency across trials of the timing of the three stimuli.
Equation~\ref{eq:LaplaceDiff} shows that the Laplace transform of 
$\tau_{\baseone\basethree} - \tau_{\baseone\basetwo}$
is stored in $\MYZ$,
whereas 
Equation~\ref{eq:LaplaceCrossCorr} shows that the Laplace transform of 
$\tau_{\baseone\basethree} - \tau_{\baseone\basetwo}$
is stored in $\MXZ\barMYXbare(-s)$.  Comparing these two quantities allows us
to assess the dependence between the times of occurrence of \stimone{} and
\stimthree{} conditionalized on \stimtwo{} in the present.  


\subsection{Continuum of $\learn$ allows a temporal memory \emph{across} trials}
For the past several subsections we have considered the limit where $\learn
\rightarrow 1$. That limit is not physically realizable.  How should we choose
the value of $\learn$?  The answer is that we should not choose a single value
of  $\learn$.
In much the same way we treat $s$ as a continuous variable rather
than treating it as a parameter to be estimated from the data, we can also
treat $\learn$ as a continuous variable. Continuous 
$s$ means that $\PastLaplbold{}(s)$ maintains a temporal
memory of the entire past.  Similarly, continuous $\learn$  enables
$\mat{M}(\learn,s)$ to retain complete information about pairwise
relationships as a function of trial history.  Similar relationships can be
worked out for $\mat{\bar{M}}(\learn,s)$ but we focus  
on $\mat{M}(\learn,s)$ here for simplicity.

Equation~\ref{eq:MXYseries},
which describes the situation where $\tau_{\baseone \basetwo}$ is equal to
$\tau_o$ on each trial, can be rewritten as
\[
	\MXYbare(\learn,s) = (1-\learn) e^{-s \tau_o} \Z{h[i]}\left(\learn^{-1}\right) 
\]
where $\Z{}(z)$ is the Z-transform, the discrete analog of the Laplace
transform \citep{Ogat70}.  An analogous relationship can be written for
$\bar{\mathbf{M}}(s)$.

Although the notation is a bit more unwieldy, allowing $\tau_{\baseone
\basetwo}$ to vary across trials we see that the trial history of timing is
also retained by $\mat{M}(\learn, s)$.
Writing the delay between \stimone{} and \stimtwo{}  on the trial $i$ steps in
the past as $\tau[i]$, and $H[i](s) \equiv h[i] e^{-s \tau[i]}$ we can write
\begin{eqnarray}
	\MXYbare(\learn,s) &=& (1-\learn)
	\Z{H[i](s)}\left(\learn^{-1}\right).
	\label{eq:alphaz}
\end{eqnarray}
We understand the Z-transform to be taken over the discrete variable $i$ and
not the continuous variable $s$.

\begin{figure}
	\centering
	\includegraphics[width=0.95\columnwidth]{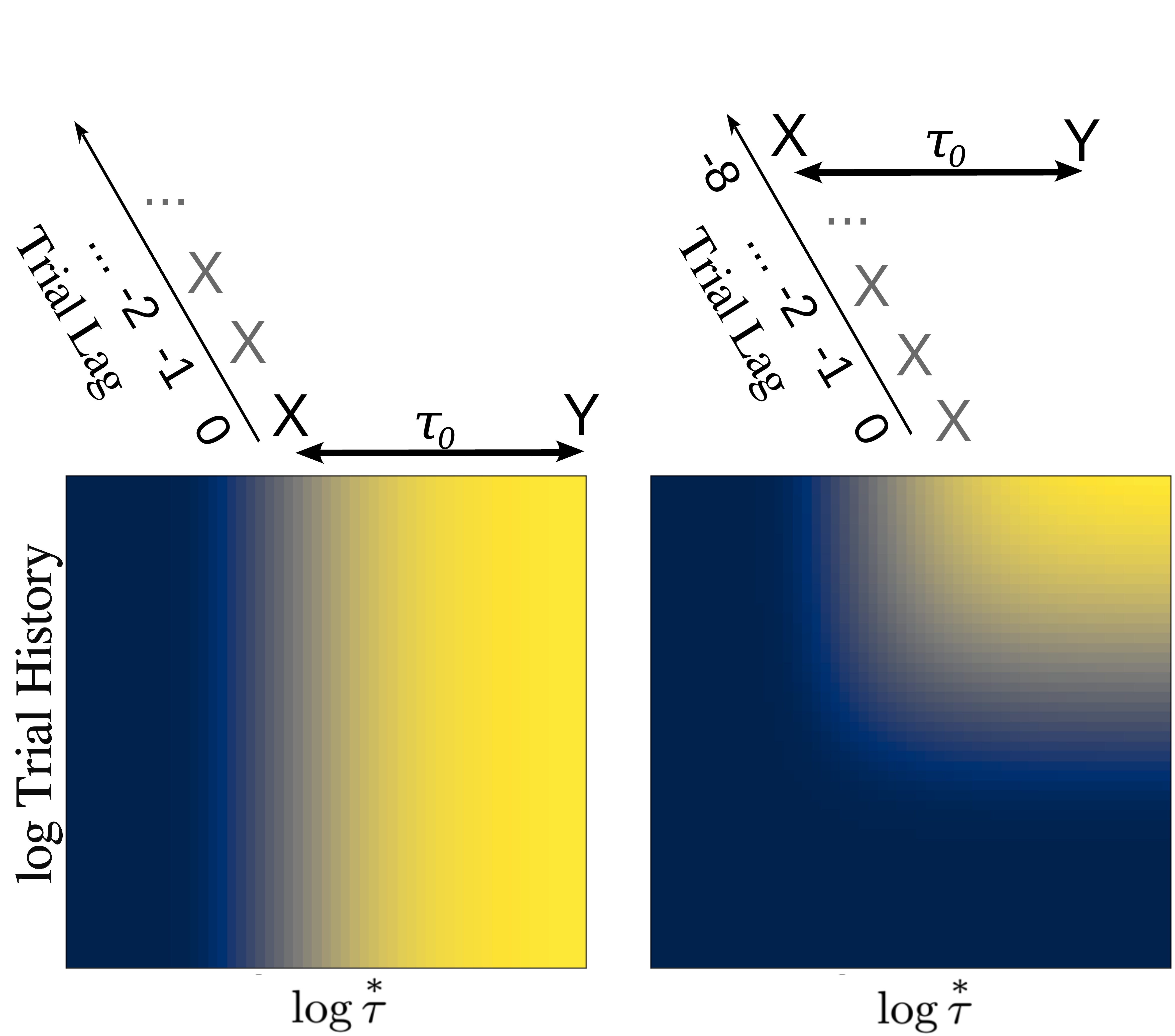}
	\caption{
		$\mat{M}(\learn,s)$ contains information about both time
		within a trial and trial history.
		Left: Consider a single pairing of \stimone{} and \stimtwo{}
		on the most recent trial. The heatmap shows the degree to which \stimtwo{} is
		cued by \stimone{} by  $\frac{\mathbf{\basetwo}'\mat{M}(\learn,s)
		\mathbf{\baseone}}{1-\learn}$  projected onto log time.  
		The profile as a function of $\log \tau$ is identical to the
		profile for future time in
		Figure~\ref{fig:timeline}.  If the pairing between \stimone{}
		and \stimtwo{} had a longer delay, the edge would be further
		to the right.
		Right:  The single pairing of \stimone{} and \stimtwo{}
		is followed by an additional series of trials on which
		\stimone{} was presented by itself.  Now there is an edge in
		both trial history and time within trial.  Additional trials
		with only \stimone{} would push this edge further towards the
		top of the graph.  Additional trials with \stimone{} and
		\stimtwo{} paired would be added to this plot with a time
		delay that reflects the timing of the pairing. 
	\label{fig:Mrhosfig}
	}
\end{figure}

Because the Z-transform is in principle invertible, information about the
entire trial history has been retained by virtue of having a continuum of
forgetting rates $\learn$.
Figure~\ref{fig:Mrhosfig} illustrates the ability to extract the trial history
including timing information of events that follow \stimone{} from
$\mat{M}(\learn, s) \mathbf{\baseone}$.  

This illustrates a remarkable property of Laplace-based temporal memory.
Although each synaptic matrix with a specific value of $\learn_o$ forgets
exponentially with a fixed time horizon (the time constant is given by $(-\log
\learn_o)^{-1}$), the set of matrices with a continuum of $\learn$ retains
information about the \emph{entire} trial
history.  Although each matrix has a specific time horizon, the set of all
matrices with continuous values of $\learn$ has a continuity of time horizons,
tiling the entire trial history.  In practice there must be some numerical
imprecision in the biological instantiation of $\mat{M}(\learn,s)$.  In
principle however, a continuum of forgetting rates $\learn$ means that the
past is not forgotten. Instead the past, as a function of trial history,
has been written across the continuum of $\learn$. 

\subsection{Estimating three point correlation functions from
Z-transform}
A great deal of information can be extracted from the trial histories encoded in
$\mat{M}(\learn,s)$  and $\mat{\bar{M}}(\learn,s)$.   $\MYZ$ contains the
two-point probability distribution of \stimtwo{} and
\stimthree{}.  It would be preferable to predict the occurrence and timing of
\stimthree{}  using the three-point probability distribution.
 
Because $\mat{M}(\learn,s)$ contains information about the paired trial
history, in principle we can extract information about the three-point
correlation function.  The problem of estimating the three-point correlation
function between stimuli is straightforward if one has access to the trial
history of both the past conditionalized on the present and the future
conditionalized on the present.   This information is contained in
$\barMYXbare(\learn,s)$ and $\MYZbare(\learn,s)$ respectively.

For instance,
if  \stimthree{} only occurs on trials on which both
\stimone{} and \stimtwo{} are presented, but not on trials when only one of
them are presented, then we should observe a positive
correlation between the trial history encoded in $\MYZbare(\learn,s=0)$ and
$\barMYXbare(\learn,s=0)$.
Similarly, one can imagine that the joint timing of the presentations of
\stimone{} and \stimtwo{} predicts the timing of \stimthree{}, as if all three 
symbols are being generated by a process that can unfold at different rates.

Access to the joint statistics between symbols
can in principle be leveraged to provide a much more complete prediction of
the future, especially when integrated into deep networks that recode the
symbols into new sets of symbols.  Moreover, continuous values of $\learn$
may allow networks built using $\mat{M}(\learn,s)$ and  $\bar{\mat{M}}(\learn,s)$
to respond to non-stationary statistics.

\subsection{Updating the future}
Let us return to the problem of generating a prediction of the immediate
future.  We again restrict our attention to the limit as $\learn$ goes to 1
and assume the system has experienced a very long sequence of trials with the
same underlying statistics.  Moreover, we assume for the present that only
pairwise relationships are important, so we can neglect the  temporal credit
assignment problem, and construct the Laplace transform of the future that
predicted solely on the basis of the present stimulus. 

There are two problems that need to be resolved to write an analog of
Eq.~\ref{eq:EvolvePast} for $\FutLaplbold{t+\Delta t}(s)$.  First, we can only
use Eq.~\ref{eq:alphaODE} to update $\FutLaplbold{t}(s)$ if 
$\FutLaplbold{t}(s)$  is already the Laplace transform of a predicted future;
we must create a circumstance that makes that true.  
Second, we need to address the situation where a prediction reaches the
present.  
Because of the discontinuity at $\tau=0$ special considerations are necessary
to allow the time of a stimulus to pass from the future to the past.

\subsubsection{Predicting the  future with the present}
Equation~\ref{eq:MXYContingent} indicates that the weights in $\MXY$ record the
future occurrences of \stimtwo{} given that \stimone{} occurs in
the present.  $\MXY$ captures both the probability that \stimtwo{}
will follow \stimone{} as well as the distribution of temporal delays at which
\stimtwo{} is expected to occur.  This information is encoded as a probability
times the Laplace transform of a random variable.
If we only need to consider \stimone{} in predicting the future, then $\MXY$
is precisely how we would like to initialize the future prediction for
\stimtwo{} in $\FutLaplbold{t}(s)$ after
\stimone{} is presented (Fig.~\ref{fig:Mschem}).  

We probe $\mat{M}(s)$ with the ``immediate past.'' 
When \stimone{} is presented it enters $\PastLaplbold{t}(s)$ as  
$ \L{\delta\left(0\right)  \mathbf{\baseone}} (s) $.  Multiplying $\mat{M}(s)$
from the right with the immediate past, yields a prediction for the future.
\begin{equation}
	\mat{M}(s)\ e^{-s 0} \ \mathbf{\baseone} = \mat{M}(s) \mathbf{\baseone} = 
	P\left(\basetwo|\baseone\right) \L{\tau_{\baseone \basetwo}}(s)\ \mathbf{\basetwo}
	\label{eq:MXYexample}
\end{equation}
More generally, the input
to the future at time $t$ should be given by $\mat{M}(s) \L{\delta(0)
\mathbf{f}_t}$.  For concision we write this as $\mat{M}(s) \mathbf{f}_t$.
Because the past stored in $\mat{M}(s)$ was a probability times the Laplace
transform the distribution of a random variable, the future recovered in
this way is also understandable as a probability times the Laplace transform
of a random variable. If only Laplace transforms of delta functions can be
represented in $\FutLaplbold{t}(s)$, then we can imagine sampling from this
distribution of future times, perhaps with a preference for times more near to
the future.

 
\subsubsection{Continuity of the predicted future through $\tau=0$}
The neural representation described here approximates a continuous timeline 
by stitching together separate Laplace neural manifolds for the past
and the future.  With the passage of time, information in the future moves
ever closer to the present.  As time passes and a prediction reaches the
present,  this discontinuity must be addressed.  Otherwise, the firing rates
will grow exponentially without bound.

We can detect predictions that have reached the present by examining 
$\FutLaplbold{t}(s=\infty)$, which only rises from zero when $\tau \rightarrow
0$.  In practice, we would use  $\smax$ which should be on the
order of $(\Delta t)^{-1}$.  If the future that is being represented is the
Laplace transform of a delta function, then we can simply take components for
which $\FutLaplbold{t}(\smax) > 0$ to zero for all $s$ at
the next time step.  More generally, if the future that is represented is not
simply a delta function, the linearity of the Laplace transform allows us to
subtract  $\FutLaplbold{t}(s=\infty)$ from all $s$ values without affecting
the evolution at subsequent time points.

If a prediction reaches the present and is observed, then no further action is
needed.  If a prediction reaches the present, but is not observed, we can
trigger an observation of  a ``not symbol'', written e.g., \not{\stimone} to
describe the observation of a failed prediction for a stimulus \stimone{}.
Although we won't pursue it here, one could allow ``not symbols'' to be
predicted by stimuli and to predict other stimuli, allowing for the model to
provide appropriate predictions for a relatively complex set of contingencies
using only pairwise relationships.

\subsubsection{Evolving $\FutLaplbold{t+\Delta t}(s)$}
Integrating these two additional factors allows us to write a general
expression for evolving $\FutLaplbold{t}(s)$ to $\FutLaplbold{t+\Delta t}(s)$.
\begin{eqnarray}
	\FutLaplbold{t+\Delta t}(s) &=& e^{s \Delta t} \FutLaplbold{t}(s)  -
	\FutLaplbold{t}(s=\infty)
	+
	\mat{M}(s) \mathbf{f}_t.
	\label{eq:EvolveFuture}
\end{eqnarray}
If the future is expressed as a delta function,  continuous
attractor networks with an edge are sufficient to support this evolution
\citep{DaniHowa24}.  Because the future is in general more complex than a
delta function, and predictions for distant parts of the future can change as
events happen in the present, additional considerations are necessary.

\section{Neural predictions}


Regions as widely separated as the cerebellum
\citep{WagnLuo20,DeZeEtal21}, striatum \cite[e.g.,][]{MeerEtal11}, PFC
\cite[e.g.,][]{RainEtal99,NingEtal22}, OFC \cite[e.g.,][]{NambEtal19,SchoEtal98,YounShap11}, hippocampus
\citep{FerbShap03,DuveEtal23} and thalamus
\citep{KomuEtal01} contain active representations that code for the future.
One can find evidence of predictive signals extending over long periods of
time that modulate firing in primary visual cortex
\citep{GavoBear14,KimEtal19,HomaEtal22,YuEtal22}.  
Prediction apparently involves a substantial proportion of the brain.
Coordinating activity and plasticity over
such a wide region would require careful synchronization
\citep{HassEtal02,HamiEtal21}.  The timescale of this synchronization,
presumably on the order of 100~ms, fixes $\Delta t$, places a bound on the
fastest timescales $1/s$ that can be observed, and operationalizes the
duration of the ``present.''

Given the widespread nature of predictive signal, we will not attempt to
map specific equations onto specific brain circuits.  Rather we will
illustrate the observable properties implied by these equations with an eye
towards facilitating future empirical work.  The predictions fall into two
categories.  One set of predictions describes properties of ongoing firing of
neurons participating in Laplace Neural Manifolds for past and future time.
Another set of predictions are a direct consequence of the properties of
learned weights.  We also briefly discuss the model in this paper in the
context of recent empirical work on the computational basis of the dopamine
signal \citep{JeonEtal22}.

\subsection{Active firing neurons}
This paper proposes the existence of Laplace Neural Manifolds to code for the identity
and time of \emph{future} events.  This implies there should be two
related manifolds, one implementing the Laplace space and one implementing
the inverse space.  Previous neuroscientific work has shown evidence for
Laplace and inverse spaces for a timeline for the past.   The properties of
the proposed neural manifolds for future time can be understood by analogy to
the neural manifolds for the past.

\subsubsection{Single-cell properties of neurons coding for the past}
So-called temporal context cells observed in the entorhinal cortex
\citep{TsaoEtal18,BrigEtal20} are triggered by a particular event and then
relax exponentially back to baseline firing with a variety of time constants.
The firing of temporal context cells is as one would expect for a population
coding $\PastLapl{}(s)$.  So-called time cells observed in the hippocampus
\citep{PastEtal08,MacDEtal11,TaxiEtal20,ShahEtal22,ShikEtal21,SchoEtal22} and many other brain regions
\cite[e.g.,][]{TigaEtal18a,TigaEtal17b,MellEtal15,BakhEtal17,AkhlEtal16,JinEtal09,SubrSmit24}
fire sequentially as events recede into the past, as one would expect from
neurons participating in $\ftilde(\taustar)$ for $\taustar < 0$.  
Time cells are consistent with qualitative and quantitative predictions,
including the conjecture that time constants are distributed along a
logarithmic scale \citep{CaoEtal22}.

\subsubsection{Single-cell and population-level properties of neurons coding
for the past and the future} 
\begin{figure}
	\centering
	\begin{tabular}{ll}
		\textbf{A} & \textbf{B}\\
		\includegraphics[width=0.45\columnwidth]{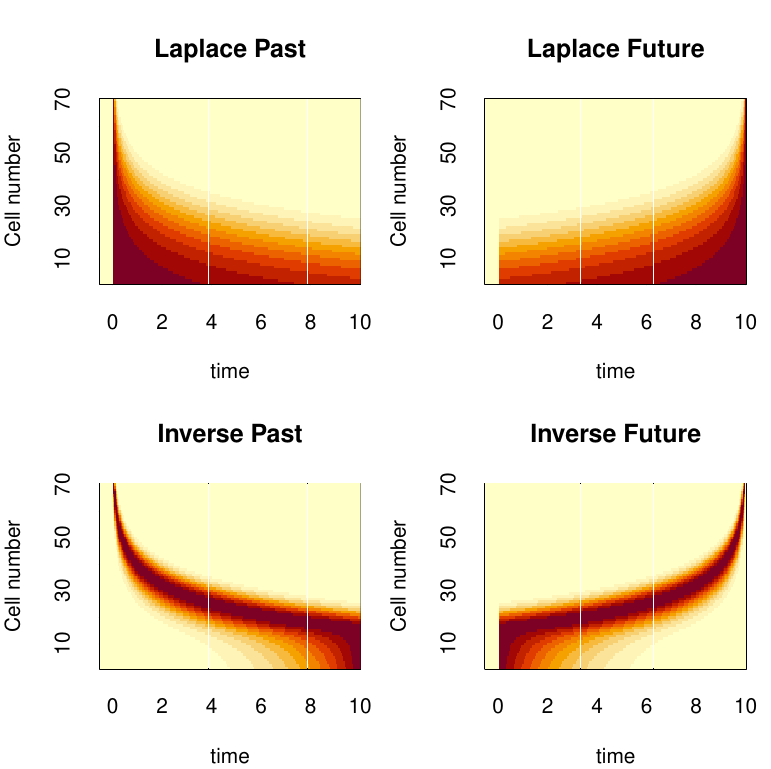}
		&\includegraphics[width=0.45\columnwidth]{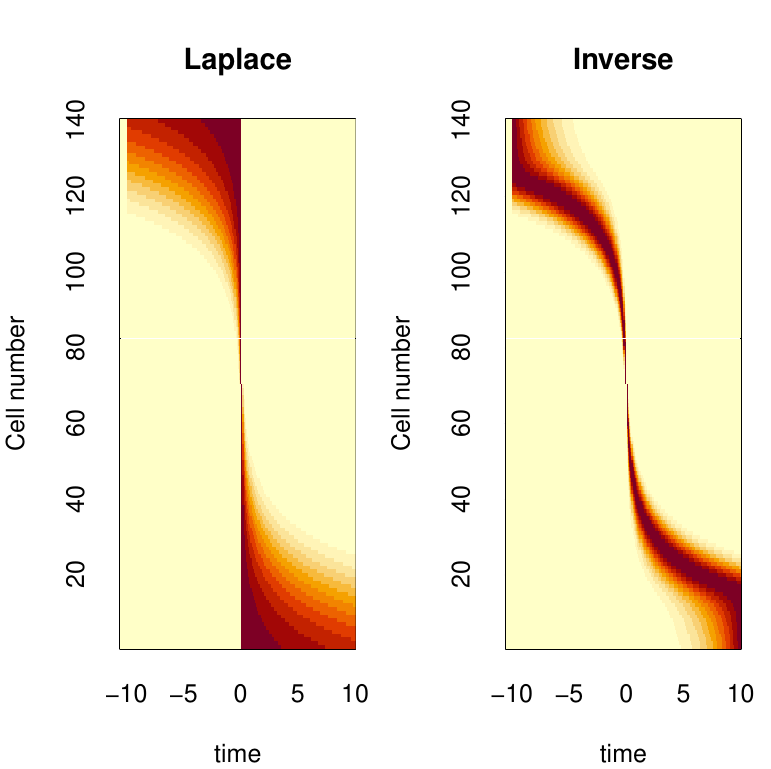}
	\end{tabular}
	\caption{Predicted firing for Laplace and inverse spaces plotted as
	heatmaps. 	
	\textbf{A.}  Consider an experiment in which \stimone{} precedes
	\stimtwo{} separated by 10~s.  The top row shows firing as a function
	of time for cells in the Laplace space for the past (left) and the
	future (right).  Note that the cells in
	$\PastLapl{t}(s)$ peak at time
	zero and then decay exponentially.  In contrast cells in
	$\FutLapl{t}(s)$ peak at 10~s and ramp up exponentially.  The bottom
	row shows firing as a function of time for cells in the Inverse space. 
	\textbf{B.}
	Consider an experiment in which \stimtwo{} is predicted to occur at
	time zero and then recedes into the past.  Cells
	coding for both past and future are recorded together and sorted
	on the average time at which they fire.
	 Left:  For Laplace spaces, neurons in $\FutLapl{t}(s)$ are sorted to
	 the top of the figure and neurons $\PastLapl{t}(s)$ are sorted to the
	 bottom of the figure. Right: Inverse spaces show similar properties
	 but give rise to a characteristic ``pinwheel'' shape.
	 \label{fig:FutureTimeHeatmaps}
	}
\end{figure}

In situations where the future can be predicted,
$\FutLapl{}(s)$ and
$\ftilde(\taustar > 0)$ should behave as mirror images of the corresponding
representations of the past.  Figure~\ref{fig:FutureTimeHeatmaps}A illustrates
the firing of cells coding for a stimulus remembered in the past (left) and
predicted in the future (right).  Neurons participating in the Laplace space,
sorted on their values of $s$, are shown in the top; neurons participating in
the inverse space, sorted on their values of $\taustar$ are shown on the
bottom.

The firing of neurons constituting the Laplace space shows a characteristic
shape when plotted as a function of time in this simple experiment.  Neurons
coding for the past are triggered shortly after presentation of the stimulus
and then relax exponentially with a variety of rates.  Neurons coding for the
future ramp up, peaking as the predicted time of occurrence
grows closer. The ramps have different characteristic time constants.
Different populations are triggered by the presentation of
different symbols (not shown)  so that the identity of the remembered and
predicted symbols as well as their timing can be decoded from 
populations coding $\PastLaplbold{}(s)$ and $\FutLaplbold{}(s)$.
The largest value of $1/s$ in the figure is chosen to be a bit longer than the
delay in the experiment, resulting in a subset of neurons that appear to fire
more or less constantly throughout the delay \citep{EnelEtal20}.  

The firing of neurons constituting the  inverse space also shows a
characteristic shape when plotted as a function of time in this simple
experiment.  Neurons tile the
delay, with more cells firing early in the interval with more narrow receptive
fields.  The logarithmic compression of
$n$ results in a characteristic ``backwards J'' shape for the past and a
mirror image ``J'' shape for the future.   Again, different populations would
code for different stimuli in the past and in the future (not shown) so that
the identity of the remembered and predicted stimuli and their time from the
present could be decoded from a population coding $\ftildebold(\taustar)$.
Figure~\ref{fig:FutureTimeHeatmaps}B shows firing that would be expected for a
population that includes cells coding for the same stimulus, say \stimtwo{}, both in
the past and the future around the time of a predicted occurrence of that
symbol. 

\subsubsection{Plausible anatomical locations for an internal future timeline}
This computational hypothesis should evaluated with carefully
planned analyses.  However, the published literature shows evidence that is at
least roughly
in line with the hypothesis of neural manifolds for future time.
Firing that ramps systematically upward in anticipation of important outcomes
including planned movements has been observed in (at least) mPFC
\citep{HenkEtal21}, ALM \citep{InagEtal18}, cerebellum \citep{GarcEtal24}, and thalamus
\citep{KomuEtal01}.   \citet{KomuEtal01} showed evidence for ramping
firing in the thalamus that codes for outcomes in a Pavlovian
conditioning experiment.   
Two recent papers show evidence that ramping neurons during motor preparation
in ALM \citep{InagEtal18,AffaEtal24} and interval timing in mPFC
\citep{HenkEtal21,CaoEtal24} do so with a
continuous spectrum of time constants.  

For instance, \citet{CaoEtal24},
reanalyzing data originally published by \citet{HenkEtal21} observed the
firing of neurons during the reproduction phase of an interval reproduction
task.  On each trial, the animal is exposed to a delay of $T$ seconds, which
must then be reproduced.
Let us refer to the moment the reproduction phase begins  as $t=0$. 
Now at time $t<T$, the beginning of the interval is $\tau=t$ seconds in the past
and the planned movement is a time $\tau = T-t$ seconds in the future.
Figure~\ref{fig:ExpResults}A shows that some neurons in mPFC ramped down as
$e^{-st}$ with a continuum of rate constants $s$ and other neurons ramped up as
$e^{-s(T-t)}$, again with a continuum of rate constants $s$.  There are
potentially important differences between the empirical results in the paper
and the theoretical model presented here---for instance many of the neurons
coding for the time of the future planned movements rescaled the timecourse of
their firing depending on the value of  $T$ on that trial---but the overall
correspondence to the predictions described here is striking.  In at least
some regions, in some tasks, the ongoing firing of cortical neurons codes the time of
planned future events \emph{via} real Laplace transform of the future. 

There is also circumstantial neurophysiological evidence for sequential firing
leading to predicted future events as predicted by $\ftilde(\taustar)$ for
$\taustar> 0$ coding for future events.
Granule cells in cerebellum appear to fire in sequence in the time leading up
to an anticipated reward \citep{WagnEtal17,WagnLuo20}.  
During performance of a task in which monkeys must perform a sequence of
movements, neurons firing in sequence that decoded the time of \emph{future}
movements were observed in PFC but not in posterior parietal cortex  \citep{WataEtal23}.
OFC may be another good
candidate region to look for ``future time cells.'' OFC has  long been argued
by many authors to code for the identity of predicted outcomes
\citep{HikoWata00,SchoRoes05,MainKepe09}.  More recently \citet{EnelEtal20}
showed sequential activation in OFC during a task in which it was possible to
predict the value of a reward that was delayed for several seconds.  Finally,
it should be noted that the properties of $\ftilde(\taustar)$ over the future  are a
temporal analog of spatial ``distance-to-goal'' cells that have been observed in spatial
navigation studies \citep{SareEtal17,GautTank18}.

\subsection{Predictions from weight matrix $\mat{M}(\learn,s)$}
\begin{figure}
	\centering
		 \includegraphics[width=0.8\columnwidth]{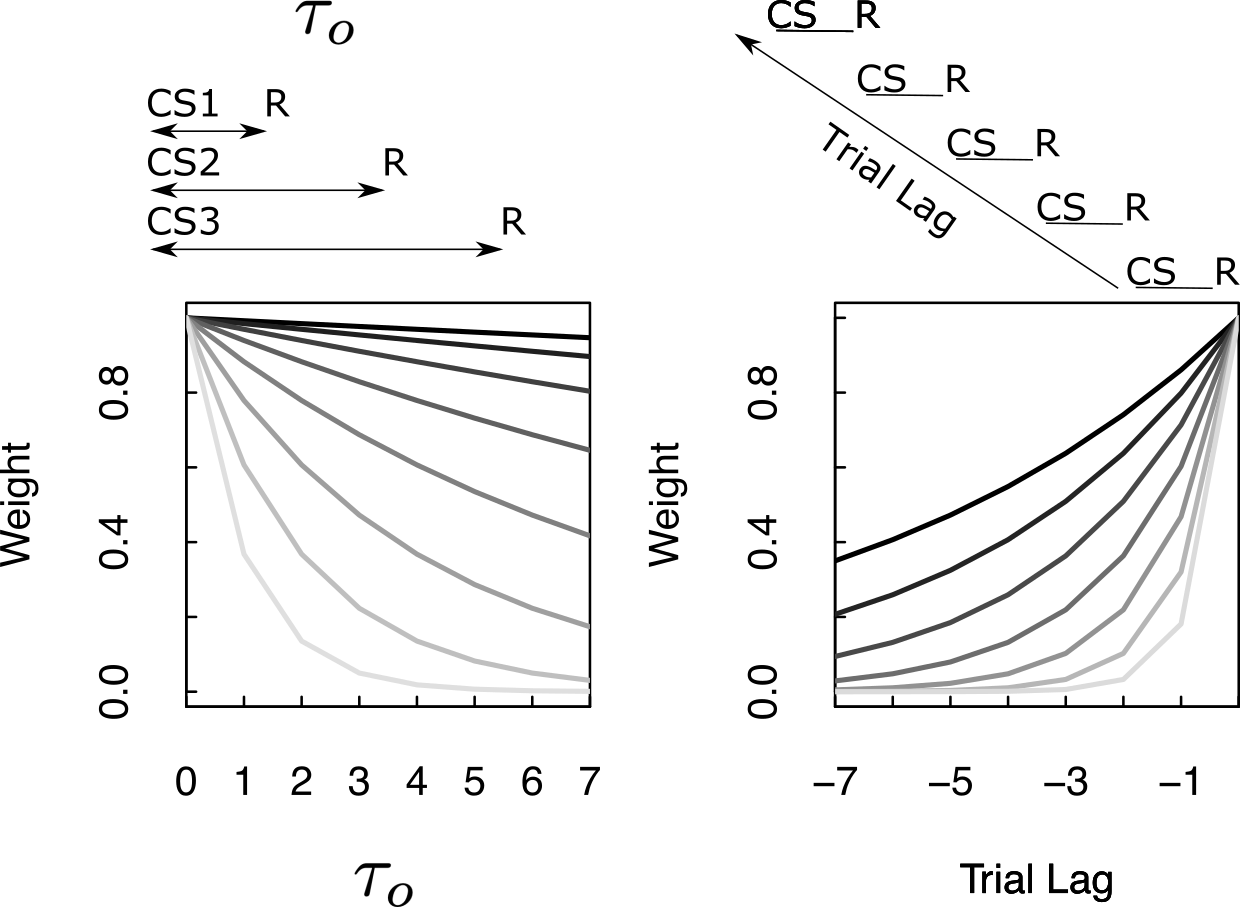}
	\caption{ Neural predictions derived from properties of $\mat{M}(s)$.
	\textbf{Left.} 
	Plot of the magnitude of the entry in $\mat{M}_r(\learn=1,s)$
	connecting each of the
	conditioned stimuli \textsc{cs} to the outcome \textsc{r}  as a
	function of the $\tau_o$ corresponding to that \textsc{cs}.  Different
	lines correspond to entries with different values of $s$.  Weights
	corresponding to different values of $s$ show exponential discounting
	as $\tau_o$ is manipulated, with a variety of discount rates.
	\textbf{Right.} 
	Plot of the magnitude of $\mat{M}(\learn,s=0)$ associated with a
	single pairing of \textsc{cs} and \textsc{r} a certain number of
	trials in the past.  Different lines show the results for different
	values of $\learn$.  For clarity, these curves have been normalized
	such that they have the same value at trial lag zero.
	\label{fig:Weights}
	}
\end{figure}


\subsubsection{Properties of weights due to $s$}
Consider an experiment in which different symbols, denoted \textsc{cs1},
\textsc{cs2}, etc, precede an outcome \textsc{r} by a delay
$\tau_o$.  The value of $\tau_o$ changes across the different symbols
(Figure~\ref{fig:Weights}A).  Ignoring $\learn$ for the moment, the strength
of the connections from each
\textsc{cs} to
\textsc{r}  depend on the value of $\tau_o$ for that stimulus and the
value of $s$ for each synapse: $e^{-s \tau_o}$.  When a particular \textsc{cs}
is presented at time $t$, the amount of information that flows along each
synapse is $e^{-s\tau_o}$ and the pulse of input to $\FutLaplbold{t+\Delta
t}(s) - \FutLaplbold{t}(s)$ corresponding to the outcome is $e^{-s\tau_o}$.

\begin{figure}
	\centering
	\includegraphics[width=.9\textwidth]{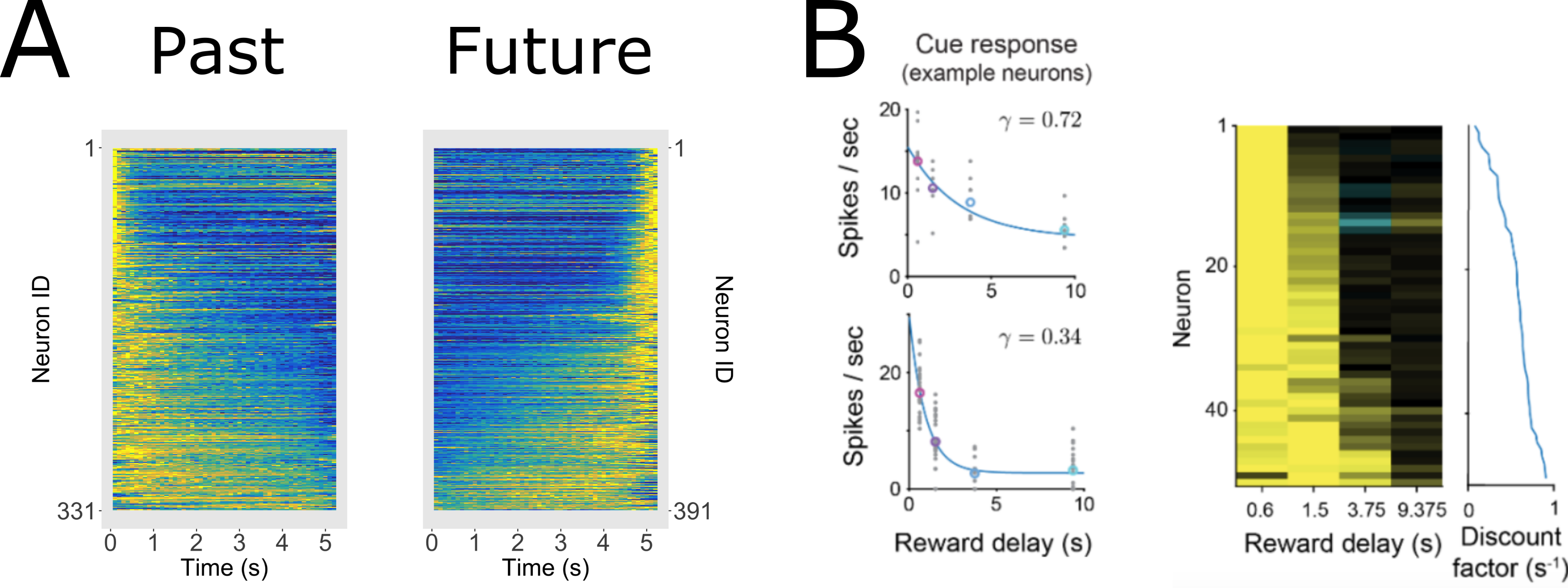}
	\caption{
		Recent observation of key neural predictions of this approach.
		\textbf{A.} 
		In this interval reproduction experiment, rodents had to
		reproduce a delay of some duration $T$.  Calling the start of
		the reproduction period $t=0$, at time $t$,  the beginning of the
		interval is $t$ seconds in past while the planned end of the
		reproduction period is $T-t$ seconds in the future. 
		Firing of neurons in rodent mPFC have properties
		resembling those predicted for $\PastLapl{t}(s)$ and 
		$\FutLapl{t}(s)$, showing exponential decay/ramping with a
		continuous spectrum of time constants.
		Compare to Figure~\ref{fig:FutureTimeHeatmaps}A.
		Adapted from Cao, et al., (2024).
		\textbf{B.} 	In a classical conditioning experiment,
		different stimuli predicted a rewarding outcome at different
		delays.  Firing of dopamine neurons was recorded following
		each of the stimuli.  The change in firing as function of the
		time until the reward was fitted with an exponential curve
		(left), indexed by the discount rate $\gamma$.    Across
		dopamine neurons, a wide range of time constants was observed
		(right).  The results are as one would expect  if the dopamine
		system projects information about the time of future events to
		the rest of the brain \emph{via} $\mat{M}(s)$.  Adapted from
		Masset, et al., (2023).  \label{fig:ExpResults}
	}
\end{figure}

Thus, considering each connection as a function of $\tau_o$, firing should go
down exponentially as a function of $\tau_o$ with a rate constant that depends
on the value of $s$.  This pattern of results aligns well with 
experimental results observed in mid-brain dopamine neurons
\citep{MassEtal23,SousEtal23}.  
It has long been known that firing of dopamine neurons, averaged over neurons,
around the time of the conditioned stimulus goes down with delay
\citep{FiorEtal08}. 
\citet{MassEtal23} measured the firing of dopamine
neurons to different stimuli that predicted reward delivery at different delays.
This study showed that there was a heterogeneity of
exponential decay rates in the firing of dopamine neurons in this paradigm
(Fig.~\ref{fig:ExpResults}B),  
much as illustrated in Fig.~\ref{fig:Weights}A.  
In the context of TDRL, this
finding is consistent with a continuous spectrum of exponential discount
rates \citep{MomeHowa18,TanoEtal20}.   In any event, these findings
\citep{SousEtal23,MassEtal23} are clear evidence that the phasic firing of
midbrain dopamine neurons at the time of a predictive stimulus codes for the
Laplace transform of the time until future reward.

\nocite{HenkEtal21,CaoEtal24}

\subsubsection{Properties of weights due to $\learn$}
A continuum of  forgetting rates $\learn$ predicts a range of trial history
effects.
Figure~\ref{fig:Weights}B shows the weights in $\mat{M}(\learn)$ over past
trials that result from different values of $\learn$.   This is simply
$\learn^i$  where $i$ is the trial recency with values normalized such that
the weight at the most recent trial is 1.
The weights $\mat{M}(\learn)$ record the Z-transform of the trial history of reinforcement.  
Many papers show dependence on previous trial outcomes in response to a cue
stimulus in learning and decision-making experiments
\citep{BernEtal11,MorcHarv16,ScotEtal17,AkraEtal18,HattEtal19,HattKomi22}.
These studies show history-dependent effects in a wide range of brain regions
and often show a continuous spectrum of decay rates within a brain region
\cite[see especially][]{BernEtal11,DansEtal23}.   Notably, distributions of time constants for trial history
effects cooccur with distributions of ongoing activity in multiple brain
regions \citep{SpitEtal20}.


\subsection{Dopamine and learning}
The connection between TDRL and neuroscience related to dopamine has been one
of the great triumphs of computational neuroscience \citep{SchuEtal97}. 
The standard account is that the firing of dopamine neurons signals reward
prediction error (RPE) which drives plasticity.  Despite its remarkable
success at predicting the findings of many behavioral and neurophysiological
experiments, the RPE account has been under increasing strain over recent
years.  The standard account did not predict the existence of a number of
striking effects, including increasing dopamine firing during delay under
uncertainty \citep{FiorEtal03}, dopamine ramps in spatial experiments
\citep{HoweEtal13}, dopamine waves \citep{HamiEtal21}, and heterogeneity
of dopamine responses across neurons and brain regions
\citep{DabnEtal20,MassEtal22,WeiEtal21}, although many of these phenomena can be
accommodated within the RPE framework with elaboration.
\citep{Gers17a,KimEtal20,GardEtal18,LeeEtal22} 
\citet{JeonEtal22} reported the results of several experiments that
flatly contradict the standard model.  These experiments were proposed to
evaluate an alternative hypothesis for dopamine firing in the brain.

\citet{JeonEtal22} propose that dopamine signals whether the current stimulus
is a cause of reward.  The model developed there, referred to as ANCCR,
assesses the contingency between a stimulus and outcomes.  
$\mat{M}(\learn,s)$ and $\bar{\mat{M}}(s)$ contain information about
the contingencies---temporal and otherwise---between a symbol and possible
outcomes.  Both ANCCR and the framework developed in this paper are inspired
by a similar critique of Rescorla-Wagner theory and TDRL \citep{Gall21}.
In order to make a complete account of
the experiments in the \citep{JeonEtal22} paper, the current framework would
have to be elaborated in several ways.  
However, the current framework does not require one to 
specify an intrinsic timescale of association \emph{a priori}. Perhaps it
is possible to develop a generalization of the current framework that does not
rely on the simplifying assumption of discrete trials in order to yield
readily interpretable measures of contingency.  


\section{Discussion}

This paper takes a phenomenological approach to computational neuroscience.
The strategy is to write down equations that, if the brain could somehow obey
them, would be consistent with a wide range of observed cognitive and neural
phenomena.  
The phenomenological equations make concrete predictions that  can be
evaluated with cognitive and neurophysiological experiments.
To the extent the predictions hold, the question of how the brain manages
to obey these phenomenological equations could then  become a separate subject of
inquiry.  The
phenomenological equations require a number of capabilities of neural
circuits, both at the level of synapses and in terms of ongoing neural
activity. We make those explicit here.

\subsection{Circuit assumptions for synaptic weights}
$\mat{M}(\learn, s)$ and $\mat{\bar{M}}(\learn,s)$ require that the
brain uses  continuous variables, $\learn$ and $s$, to organize connections
between many neurons, most likely spanning multiple brain regions.
For the phenomenological equations to be viable, these continuous variables
should be deeply embedded in the functional architecture of the brain.  For
instance, in order to invert the integral transforms, it is necessary to 
compute a derivative over these continuous variables.  This suggests 
a gradient in these continuous variables  should be anatomically identifiable.
Conceivably anatomical gradients in gene expression and functional
architecture \cite[e.g.,][]{PhilEtal19,GuoEtal21,RoyEtal22} could generate 
anatomical gradients in $s$ and/or $\learn$. Perhaps part of the function of
traveling waves of activity such as theta oscillations
\citep{LubeSiap09,PateEtal12,ZhanJaco15} or dopamine waves \citep{HamiEtal19} is
to make anatomical gradients salient.



\subsection{Circuit assumptions for ongoing activity}

At the neural level, this framework assumes the existence of circuits that can
maintain activity of a  Laplace Neural Manifold over time.  There is 
evidence that the brain has found some solution to this problem
\citep{TsaoEtal18,BrigEtal20,AtanEtal23,ZuoEtal23,CaoEtal24}.
Exponential growth of firing, as proposed by Eq.~\ref{eq:EvolveFuture} 
seems on its face to be a computationally risky proposition \citep[but
see][]{DaniHowa24}.
However, this proposal does create testable predictions.
Moreover, firing rates that increase monotonically as a function of one or
another
environmental variable are widely
observed.  For instance border cells as an
animal approaches a barrier \citep{SolsEtal08} and evidence accumulation cells
\citep{RoitShad02} both increase monotonically.  If this monotonic increase in
firing reflects movement of an edge along a Laplace Neural Manifold, 
the characteristic time scale of the increase should be heterogeneous across
neurons.  If the brain has access to a circuit with paired $\alpha$'s, it
could reuse this circuit to construct cognitive models for spatial navigation
\citep{HowaEtal14}, evidence accumulation \citep{HowaEtal18}, and perhaps
cognitive computation more broadly \citep{HowaHass20}.
Consistent with this hypothesis, monotonic cells in spatial navigation and
evidence accumulation---border cells and evidence accumulation cells---have
sequential analogues \citep{WilsMcNa93,MorcHarv16,KoayEtal22} as one would
expect if they reflect a Laplace space that is coupled with an inverse space.

Perhaps part of the solution to implementing these equations in the brain is
to restrict the kinds of functions that can be represented over the Laplace
Neural Manifold.  A continuous attractor network that can maintain and
evolve the Laplace transform of a single delta function per basis vector can
readily be constructed \citep{DaniHowa24}.  In this case, each component of 
$\PastLaplbold{t}(s)$ and $\FutLaplbold{t}(s)$ would be at any moment the
Laplace transform of a delta function; $\mat{M}(s)$ and $\mat{\bar{M}}(s)$
would still be able to store distributions over multiple presentations.
In this case when an item is presented perhaps $\FutLaplbold{t+\Delta t}(s)$
could update by sampling from the distribution expressed by $\mat{M}(s)
\mathbf{f}_t$.  

\subsection{Generalizing beyond time}
%
It should be possible to extend the current framework to multiple dimensions
beyond time, including real space and abstract spaces
\citep{HowaEtal14,HowaEtal18}.  Properties of the Laplace domain enable
data-independent operators that enable efficient computation
\citep{HowaHass20}.  For instance, given that a state of a Laplace neural
manifold is the Laplace transform of a function, we can construct the Laplace
transform of the translated function \citep[Eq.~\ref{eq:alphaODE}, see also][]{ShanEtal16}.  
Critically, the translation operator is independent of
the function to be translated.
Restricting our attention to Laplace transforms of delta functions, 
we can construct the sum or difference using convolution and cross
correlation respectively \citep{HowaEtal15,HowaHass20}.  
The binary operators for addition and subtraction also do not need to be learned.
Perhaps the control theory that governs behavior is analogous to generic
spatial navigation in a continuous space.  


\subsection{Scale-covariance as a design goal}
Because the $s$ values are sampled along a logarithmic scale, all of the
quantities in this paper are 
scale-covariant.  Rescaling time, taking
$\tau_{\baseone\basethree} \rightarrow a  \tau_{\baseone\basethree}$,
$\tau_{\baseone\basetwo} \rightarrow a \tau_{\baseone \basetwo}$, etc,
simply takes $s \rightarrow s/a$.
Because the $s$ values are chosen in a geometric series, rescaling time simply
translates along the $n$ axis.
All the components of the model,  $\PastLaplbold{}$,
$\FutLaplbold{}$, $\mat{M}$, and $\mat{\bar{M}}$,
all use the same kind of logarithmic scale for time.
All of the components of the model are
time scale-covariant, responding to rescaling time with a translation
over cell number. Thus any measure that integrates over $n$ (and is not
subject to edge effects) is scale-invariant. 

Empirically, there is not a characteristic time scale to
associative learning  \citep{BalsGall09,Gers22,BurkEtal23,GallShah24}; any model that requires
choice of a time scale for learning to proceed is thus incorrect.  Logarithmic
time scales are observed neurally \citep{CaoEtal22,GuoEtal21}.  Logarithmic
time scales can be understood as a commitment to a world with power law
statistics \citep{WeiStoc12a,Pian16} or as an attempt to function in many
different environments without a strong prior on the time scales it will
encounter \citep{HowaShan18}.  

Recent work has shown that the use of logarithmic time scales also enables
scale-invariant CNNs for vision \citep{JansLind21}
and audition \citep{JacqEtal22}.  For instance,  \citep{JacqEtal22} trained 
deep CNNs to categorize spoken digits.    When tested on digits presented at
very different speeds than the training examples (imagine someone saying the
word ``seven'' stretched over four seconds), the deep CNN with a
logarithmic time axis generalized perfectly.  Rescaling time
translates the neural representation at each layer; convolution is translation
equivariant; including a maxpool operation over the convolutional layer
renders the entire CNN translation-invariant.  Time is not only
important in speech perception \cite[e.g.,][]{LernEtal14} but vision as well
\citep{RussEtal22} suggesting that these ideas can be incorporated into a wide
range of sensory systems.  

\subsection{Convolution, relational memory and cognitive graphs}
There is a long-standing tension in psychology between accounts of learning
based on simple associations and cognitive representations. 
For instance \citet{Tolm48} contrasted behaviorist accounts of
stimulus-response associations with  a ``cognitive map'' studying the
behavior of rats in spatial mazes.  This paper has already touched on this
tension between association and temporal contingency---which requires metric temporal
relationships between stimuli---in the study of Pavlovian learning and reward
systems in the brain.   \citep{FodoPyly88} used analogous arguments in an
early critique of connectionism that echoes to the present day in contemporary
debates about
whether large language models ``understand'' language or not.  Continuing interest in
``neurosymbolic'' artificial intelligence can be seen as an extension of this
longstanding  debate \citep{Marc18a}.

For researchers studying episodic memory and neural representations in the
hippocampus, 
cognitive maps rather than simple atomic associations have
long been the dominant view \citep{OKeeNade78}.
\citet{CoheEich93}
emphasized that cognitive maps  are more general than
spatial maps of the physical environment and can be used to describe other
forms of relationships.  In their view a \emph{relational} representation
``maintains the `compositionality' of the items, that is, the encoding of items
both as perceptually distinct `objects' and as parts of larger scale `scenes'
and `events' that capture the relevant relations between them.''  
In the view of \citet{CoheEich93}, relational memory is critical for flexible,
context-dependent expression of stored knowledge, in much the same way that
rats can take a novel shortcut to a reward in a pre-learned maze
\citep{Tolm48}.  These ideas about relational memory have led to
``neurosymbolic'' computational models developed with specific attention to
hippocampal function \citep{WhitEtal20}.

The convolutions stored in  the Laplace domain in $\mat{M}(s)$ and
$\bar{\mat{M}}(s)$ are precisely 
relational representations.  The convolution of
two functions $f\ast g$ is neither $f$, nor $g$, but is composed from them. 
The convolution between the function
\textsc{\stimone{} was $\tau$ seconds in the past} and the function
\textsc{\stimtwo{} is in the rearward portion of the future}
describes an ``event'' including \stimone{} and \stimtwo{} in a particular
relationship.  If we substitute functions of physical space rather than
functions of time, it would be straightforward to understand this convolution
as a ``scene'' as
  proposed by \citet{CoheEich93}.
Because simple Hebbian association of
Laplace representations is is sufficient to perform convolution, it is
straightforward to at least write down neural models for relatively complex
data structures in the Laplace domain.  

Not only do convolutions provide a way to implement relational representations
as envisioned by \citet{CoheEich93}, they also lend themselves to flexible
\emph{expression} of memory.
Convolution has an inverse operation, cross-correlation.   So,
if $h = f \ast g$, then $h \# f \simeq g$, where $\#$ is the
cross-correlation.  
This property 
enables symbolic
computation \citep{Gayl04,SchlEtal22}.  For instance, consider convolutions of delta
functions.  If $f$ is a delta function at $x_f$ and $g$ is a delta function at
$x_g$, then $(f \ast g)(x)$ is a delta function at $x_f + x_g$.  
Convolution of delta functions is thus mapped onto addition and cross-correlation---which is just
convolution with reflection of one of the functions along the $x$ axis---maps
onto subtraction. 
With a bit of creativity to deal with positive and negative numbers, not
unlike the treatment of a timeline that continues from $-\infty$ to $\infty$
used here, one can build a computational system that implements the group
describing the reals under addition, clearly meeting the requirements for a
symbolic computer. 
Coming back to the hippocampus, navigation in a physical space requires vector subtraction.
For instance to know how to get from physical location $x_f$ to
physical location $x_g$, we must be able to compute $x_g - x_f$.  One can also
perform spatial navigation in abstract spaces using the same data-independent
operators \citep{GallKing11,HowaHass20}.

Laplace Neural Manifolds are thus well-suited to not only learn, represent,
and store relationships between stimuli but also to flexibly re-express
relational information in a context-appropriate manner using data-independent
operators.  These two properties make Laplace Neural Manifolds ideal for
cognitive maps of both real and abstract spaces.

\paragraph*{Conflict of interest statement}

MWH and PBS are co-founders of Cognitive Scientific AI, Inc., which could
benefit indirectly from publication of this manuscript.

\backmatter

\bmhead{Acknowledgements}
Address correspondence to Marc Howard.  The authors gratefully acknowledge
discussions with Karthik Shankar, Vijay Namboodiri, Joe McGuire, Nao Uchida
and the Uchida lab, and Kia Nobre  as well as careful review of previous
versions of the manuscript by  Nicole Howard, Hallie Botnovcan, Aakash Sarkar,
and an anonymous reviewer.   This paper is dedicated to the memory of Karthik
Shankar.\\
arxiv://2302.10163

\bibliography{bibdesk}

\begin{thebibliography}{}
\renewcommand{\doi}[1]{\url{https://doi.org/#1}}
\bibcommenthead

\bibitem [\protect \citeauthoryear {%
Affan%
\ \protect \BOthers {.}}{%
Affan%
\ \protect \BOthers {.}}{%
{\protect \APACyear {2024}}%
}]{%
AffaEtal24}
\APACinsertmetastar {%
AffaEtal24}%
\begin{APACrefauthors}%
Affan, R.O.%
, Bright, I.M.%
, Pemberton, L.%
, Cruzado, N.A.%
, Scott, B.B.%
\BCBL {} Howard, M.%
\end{APACrefauthors}%
\unskip\
\newblock
\APACrefYearMonthDay{2024}{}{}.
\newblock
{\BBOQ}\APACrefatitle {Ramping dynamics in the frontal cortex unfold over
  multiple timescales during motor planning} {Ramping dynamics in the frontal
  cortex unfold over multiple timescales during motor planning}.{\BBCQ}
\newblock
\APACjournalVolNumPages{bioRxiv}{}{}{2024--02,}
\newblock

\newblock

\PrintBackRefs{\CurrentBib}

\bibitem [\protect \citeauthoryear {%
Aghajan%
, Kreiman%
\BCBL {}\ \BBA {} Fried%
}{%
Aghajan%
\ \protect \BOthers {.}}{%
{\protect \APACyear {2023}}%
}]{%
AghaEtal23}
\APACinsertmetastar {%
AghaEtal23}%
\begin{APACrefauthors}%
Aghajan, Z.M.%
, Kreiman, G.%
\BCBL {} Fried, I.%
\end{APACrefauthors}%
\unskip\
\newblock
\APACrefYearMonthDay{2023}{}{}.
\newblock
{\BBOQ}\APACrefatitle {Minute-scale periodicity of neuronal firing in the human
  entorhinal cortex} {Minute-scale periodicity of neuronal firing in the human
  entorhinal cortex}.{\BBCQ}
\newblock
\APACjournalVolNumPages{Cell Reports}{42}{11}{,}
\newblock

\newblock

\PrintBackRefs{\CurrentBib}

\bibitem [\protect \citeauthoryear {%
Akhlaghpour%
\ \protect \BOthers {.}}{%
Akhlaghpour%
\ \protect \BOthers {.}}{%
{\protect \APACyear {2016}}%
}]{%
AkhlEtal16}
\APACinsertmetastar {%
AkhlEtal16}%
\begin{APACrefauthors}%
Akhlaghpour, H.%
, Wiskerke, J.%
, Choi, J.Y.%
, Taliaferro, J.P.%
, Au, J.%
\BCBL {} Witten, I.%
\end{APACrefauthors}%
\unskip\
\newblock
\APACrefYearMonthDay{2016}{}{}.
\newblock
{\BBOQ}\APACrefatitle {Dissociated sequential activity and stimulus encoding in
  the dorsomedial striatum during spatial working memory} {Dissociated
  sequential activity and stimulus encoding in the dorsomedial striatum during
  spatial working memory}.{\BBCQ}
\newblock
\APACjournalVolNumPages{eLife}{5}{}{e19507,}
\newblock

\newblock

\PrintBackRefs{\CurrentBib}

\bibitem [\protect \citeauthoryear {%
Akrami%
, Kopec%
, Diamond%
\BCBL {}\ \BBA {} Brody%
}{%
Akrami%
\ \protect \BOthers {.}}{%
{\protect \APACyear {2018}}%
}]{%
AkraEtal18}
\APACinsertmetastar {%
AkraEtal18}%
\begin{APACrefauthors}%
Akrami, A.%
, Kopec, C.D.%
, Diamond, M.E.%
\BCBL {} Brody, C.D.%
\end{APACrefauthors}%
\unskip\
\newblock
\APACrefYearMonthDay{2018}{}{}.
\newblock
{\BBOQ}\APACrefatitle {Posterior parietal cortex represents sensory history and
  mediates its effects on behaviour} {Posterior parietal cortex represents
  sensory history and mediates its effects on behaviour}.{\BBCQ}
\newblock
\APACjournalVolNumPages{Nature}{554}{7692}{368--372,}
\newblock

\newblock

\PrintBackRefs{\CurrentBib}

\bibitem [\protect \citeauthoryear {%
Arcediano%
, Escobar%
\BCBL {}\ \BBA {} Miller%
}{%
Arcediano%
\ \protect \BOthers {.}}{%
{\protect \APACyear {2005}}%
}]{%
ArceEtal05}
\APACinsertmetastar {%
ArceEtal05}%
\begin{APACrefauthors}%
Arcediano, F.%
, Escobar, M.%
\BCBL {} Miller, R.R.%
\end{APACrefauthors}%
\unskip\
\newblock
\APACrefYearMonthDay{2005}{}{}.
\newblock
{\BBOQ}\APACrefatitle {Bidirectional associations in humans and rats.}
  {Bidirectional associations in humans and rats.}{\BBCQ}
\newblock
\APACjournalVolNumPages{Journal of Experimental Psychology: Animal Behavior
  Processes}{31}{3}{301-18,}
\newblock

\newblock

\PrintBackRefs{\CurrentBib}

\bibitem [\protect \citeauthoryear {%
Atanas%
\ \protect \BOthers {.}}{%
Atanas%
\ \protect \BOthers {.}}{%
{\protect \APACyear {2023}}%
}]{%
AtanEtal23}
\APACinsertmetastar {%
AtanEtal23}%
\begin{APACrefauthors}%
Atanas, A.A.%
, Kim, J.%
, Wang, Z.%
, Bueno, E.%
, Becker, M.%
, Kang, D.%
\BDBL {}others%
\end{APACrefauthors}%
\unskip\
\newblock
\APACrefYearMonthDay{2023}{}{}.
\newblock
{\BBOQ}\APACrefatitle {Brain-wide representations of behavior spanning multiple
  timescales and states in C. elegans} {Brain-wide representations of behavior
  spanning multiple timescales and states in c. elegans}.{\BBCQ}
\newblock
\APACjournalVolNumPages{Cell}{186}{19}{4134--4151,}
\newblock

\newblock

\PrintBackRefs{\CurrentBib}

\bibitem [\protect \citeauthoryear {%
Bakhurin%
\ \protect \BOthers {.}}{%
Bakhurin%
\ \protect \BOthers {.}}{%
{\protect \APACyear {2017}}%
}]{%
BakhEtal17}
\APACinsertmetastar {%
BakhEtal17}%
\begin{APACrefauthors}%
Bakhurin, K.I.%
, Goudar, V.%
, Shobe, J.L.%
, Claar, L.D.%
, Buonomano, D.V.%
\BCBL {} Masmanidis, S.C.%
\end{APACrefauthors}%
\unskip\
\newblock
\APACrefYearMonthDay{2017}{}{}.
\newblock
{\BBOQ}\APACrefatitle {Differential encoding of time by prefrontal and striatal
  network dynamics} {Differential encoding of time by prefrontal and striatal
  network dynamics}.{\BBCQ}
\newblock
\APACjournalVolNumPages{Journal of Neuroscience}{37}{4}{854--870,}
\newblock

\newblock

\PrintBackRefs{\CurrentBib}

\bibitem [\protect \citeauthoryear {%
Balsam%
\ \BBA {} Gallistel%
}{%
Balsam%
\ \BBA {} Gallistel%
}{%
{\protect \APACyear {2009}}%
}]{%
BalsGall09}
\APACinsertmetastar {%
BalsGall09}%
\begin{APACrefauthors}%
Balsam, P.D.%
\BCBT {}\ \BBA {} Gallistel, C.R.%
\end{APACrefauthors}%
\unskip\
\newblock
\APACrefYearMonthDay{2009}{}{}.
\newblock
{\BBOQ}\APACrefatitle {Temporal maps and informativeness in associative
  learning.} {Temporal maps and informativeness in associative
  learning.}{\BBCQ}
\newblock
\APACjournalVolNumPages{Trends in Neuroscience}{32}{2}{73--78,}
\newblock

\newblock

\PrintBackRefs{\CurrentBib}

\bibitem [\protect \citeauthoryear {%
Bernacchia%
, Seo%
, Lee%
\BCBL {}\ \BBA {} Wang%
}{%
Bernacchia%
\ \protect \BOthers {.}}{%
{\protect \APACyear {2011}}%
}]{%
BernEtal11}
\APACinsertmetastar {%
BernEtal11}%
\begin{APACrefauthors}%
Bernacchia, A.%
, Seo, H.%
, Lee, D.%
\BCBL {} Wang, X.J.%
\end{APACrefauthors}%
\unskip\
\newblock
\APACrefYearMonthDay{2011}{}{}.
\newblock
{\BBOQ}\APACrefatitle {A reservoir of time constants for memory traces in
  cortical neurons.} {A reservoir of time constants for memory traces in
  cortical neurons.}{\BBCQ}
\newblock
\APACjournalVolNumPages{Nature Neuroscience}{14}{3}{366-72,}
\newblock

\newblock

\PrintBackRefs{\CurrentBib}

\bibitem [\protect \citeauthoryear {%
Blouw%
, Solodkin%
, Thagard%
\BCBL {}\ \BBA {} Eliasmith%
}{%
Blouw%
\ \protect \BOthers {.}}{%
{\protect \APACyear {2016}}%
}]{%
BlouEtal16}
\APACinsertmetastar {%
BlouEtal16}%
\begin{APACrefauthors}%
Blouw, P.%
, Solodkin, E.%
, Thagard, P.%
\BCBL {} Eliasmith, C.%
\end{APACrefauthors}%
\unskip\
\newblock
\APACrefYearMonthDay{2016}{}{}.
\newblock
{\BBOQ}\APACrefatitle {Concepts as semantic pointers: A framework and
  computational model} {Concepts as semantic pointers: A framework and
  computational model}.{\BBCQ}
\newblock
\APACjournalVolNumPages{Cognitive science}{40}{5}{1128--1162,}
\newblock

\newblock

\PrintBackRefs{\CurrentBib}

\bibitem [\protect \citeauthoryear {%
Bright%
\ \protect \BOthers {.}}{%
Bright%
\ \protect \BOthers {.}}{%
{\protect \APACyear {2020}}%
}]{%
BrigEtal20}
\APACinsertmetastar {%
BrigEtal20}%
\begin{APACrefauthors}%
Bright, I.M.%
, Meister, M.L.R.%
, Cruzado, N.A.%
, Tiganj, Z.%
, Buffalo, E.A.%
\BCBL {} Howard, M.W.%
\end{APACrefauthors}%
\unskip\
\newblock
\APACrefYearMonthDay{2020}{}{}.
\newblock
{\BBOQ}\APACrefatitle {A temporal record of the past with a spectrum of time
  constants in the monkey entorhinal cortex} {A temporal record of the past
  with a spectrum of time constants in the monkey entorhinal cortex}.{\BBCQ}
\newblock
\APACjournalVolNumPages{Proceedings of the National Academy of
  Sciences}{117}{}{20274-20283,}
\newblock

\newblock

\PrintBackRefs{\CurrentBib}

\bibitem [\protect \citeauthoryear {%
Burke%
\ \protect \BOthers {.}}{%
Burke%
\ \protect \BOthers {.}}{%
{\protect \APACyear {2023}}%
}]{%
BurkEtal23}
\APACinsertmetastar {%
BurkEtal23}%
\begin{APACrefauthors}%
Burke, D.A.%
, Jeong, H.%
, Wu, B.%
, Lee, S.A.%
, Floeder, J.R.%
\BCBL {} K~Namboodiri, V.M.%
\end{APACrefauthors}%
\unskip\
\newblock
\APACrefYearMonthDay{2023}{}{}.
\newblock
{\BBOQ}\APACrefatitle {Few-shot learning: temporal scaling in behavioral and
  dopaminergic learning} {Few-shot learning: temporal scaling in behavioral and
  dopaminergic learning}.{\BBCQ}
\newblock
\APACjournalVolNumPages{BioRxiv}{}{}{2023--03,}
\newblock

\newblock

\PrintBackRefs{\CurrentBib}

\bibitem [\protect \citeauthoryear {%
Cao%
, Bladon%
, Charczynski%
, Hasselmo%
\BCBL {}\ \BBA {} Howard%
}{%
Cao%
\ \protect \BOthers {.}}{%
{\protect \APACyear {2022}}%
}]{%
CaoEtal22}
\APACinsertmetastar {%
CaoEtal22}%
\begin{APACrefauthors}%
Cao, R.%
, Bladon, J.H.%
, Charczynski, S.J.%
, Hasselmo, M.%
\BCBL {} Howard, M.%
\end{APACrefauthors}%
\unskip\
\newblock
\APACrefYearMonthDay{2022}{}{}.
\newblock
{\BBOQ}\APACrefatitle {Internally generated time in the rodent hippocampus is
  logarithmically compressed} {Internally generated time in the rodent
  hippocampus is logarithmically compressed}.{\BBCQ}
\newblock
\APACjournalVolNumPages{eLife}{https://doi.org/10.7554/eLife.75353}{}{,}
\newblock

\newblock

\PrintBackRefs{\CurrentBib}

\bibitem [\protect \citeauthoryear {%
Cao%
, Bright%
\BCBL {}\ \BBA {} Howard%
}{%
Cao%
\ \protect \BOthers {.}}{%
{\protect \APACyear {{\protect \BIP {}}}}%
}]{%
CaoEtal24}
\APACinsertmetastar {%
CaoEtal24}%
\begin{APACrefauthors}%
Cao, R.%
, Bright, I.M.%
\BCBL {} Howard, M.W.%
\end{APACrefauthors}%
\unskip\
\newblock
\APACrefYearMonthDay{{\protect \BIP {}}}{}{}.
\newblock
{\BBOQ}\APACrefatitle {Ramping cells in rodent mPFC encode time to past and
  future events via real Laplace transform} {Ramping cells in rodent mpfc
  encode time to past and future events via real laplace transform}.{\BBCQ}
\newblock
\APACjournalVolNumPages{Proceedings of the National Academy of Sciences}{}{}{,}
\newblock

\newblock

\PrintBackRefs{\CurrentBib}

\bibitem [\protect \citeauthoryear {%
Carvalho%
, Tomov%
, de Cothi%
, Barry%
\BCBL {}\ \BBA {} Gershman%
}{%
Carvalho%
\ \protect \BOthers {.}}{%
{\protect \APACyear {2024}}%
}]{%
CarvEtal24}
\APACinsertmetastar {%
CarvEtal24}%
\begin{APACrefauthors}%
Carvalho, W.%
, Tomov, M.S.%
, de Cothi, W.%
, Barry, C.%
\BCBL {} Gershman, S.J.%
\end{APACrefauthors}%
\unskip\
\newblock
\APACrefYearMonthDay{2024}{}{}.
\newblock
{\BBOQ}\APACrefatitle {Predictive representations: building blocks of
  intelligence} {Predictive representations: building blocks of
  intelligence}.{\BBCQ}
\newblock
\APACjournalVolNumPages{Neural Computation}{}{}{1--74,}
\newblock

\newblock

\PrintBackRefs{\CurrentBib}

\bibitem [\protect \citeauthoryear {%
Chater%
\ \BBA {} Brown%
}{%
Chater%
\ \BBA {} Brown%
}{%
{\protect \APACyear {2008}}%
}]{%
ChatBrow08}
\APACinsertmetastar {%
ChatBrow08}%
\begin{APACrefauthors}%
Chater, N.%
\BCBT {}\ \BBA {} Brown, G.D.A.%
\end{APACrefauthors}%
\unskip\
\newblock
\APACrefYearMonthDay{2008}{}{}.
\newblock
{\BBOQ}\APACrefatitle {From universal laws of cognition to specific cognitive
  models} {From universal laws of cognition to specific cognitive
  models}.{\BBCQ}
\newblock
\APACjournalVolNumPages{Cognitive Science}{32}{1}{36-67,}
\newblock
\begin{APACrefDOI} \doi{10.1080/03640210701801941} \end{APACrefDOI}
\newblock

\newblock

\PrintBackRefs{\CurrentBib}

\bibitem [\protect \citeauthoryear {%
Clark%
}{%
Clark%
}{%
{\protect \APACyear {2013}}%
}]{%
Clar13}
\APACinsertmetastar {%
Clar13}%
\begin{APACrefauthors}%
Clark, A.%
\end{APACrefauthors}%
\unskip\
\newblock
\APACrefYearMonthDay{2013}{}{}.
\newblock
{\BBOQ}\APACrefatitle {Whatever next? Predictive brains, situated agents, and
  the future of cognitive science} {Whatever next? predictive brains, situated
  agents, and the future of cognitive science}.{\BBCQ}
\newblock
\APACjournalVolNumPages{Behavioral and Brain Sciences}{36}{03}{181--204,}
\newblock

\newblock

\PrintBackRefs{\CurrentBib}

\bibitem [\protect \citeauthoryear {%
Cohen%
\ \BBA {} Eichenbaum%
}{%
Cohen%
\ \BBA {} Eichenbaum%
}{%
{\protect \APACyear {1993}}%
}]{%
CoheEich93}
\APACinsertmetastar {%
CoheEich93}%
\begin{APACrefauthors}%
Cohen, N.J.%
\BCBT {}\ \BBA {} Eichenbaum, H.%
\end{APACrefauthors}%
\unskip\
\newblock
\APACrefYear{1993}.
\newblock
\APACrefbtitle {Memory, amnesia, and the hippocampal system} {Memory, amnesia,
  and the hippocampal system}.
\newblock
\APACaddressPublisher{Cambridge, MA}{The MIT Press}.
\PrintBackRefs{\CurrentBib}

\bibitem [\protect \citeauthoryear {%
Cole%
, Barnet%
\BCBL {}\ \BBA {} Miller%
}{%
Cole%
\ \protect \BOthers {.}}{%
{\protect \APACyear {1995}}%
}]{%
ColeEtal95}
\APACinsertmetastar {%
ColeEtal95}%
\begin{APACrefauthors}%
Cole, R.P.%
, Barnet, R.C.%
\BCBL {} Miller, R.R.%
\end{APACrefauthors}%
\unskip\
\newblock
\APACrefYearMonthDay{1995}{}{}.
\newblock
{\BBOQ}\APACrefatitle {Temporal Encoding in Trace Conditioning} {Temporal
  encoding in trace conditioning}.{\BBCQ}
\newblock
\APACjournalVolNumPages{Animal Learning \& Behavior}{23}{2}{144-153,}
\newblock

\newblock

\PrintBackRefs{\CurrentBib}

\bibitem [\protect \citeauthoryear {%
Dabney%
\ \protect \BOthers {.}}{%
Dabney%
\ \protect \BOthers {.}}{%
{\protect \APACyear {2020}}%
}]{%
DabnEtal20}
\APACinsertmetastar {%
DabnEtal20}%
\begin{APACrefauthors}%
Dabney, W.%
, Kurth-Nelson, Z.%
, Uchida, N.%
, Starkweather, C.K.%
, Hassabis, D.%
, Munos, R.%
\BCBL {} Botvinick, M.%
\end{APACrefauthors}%
\unskip\
\newblock
\APACrefYearMonthDay{2020}{}{}.
\newblock
{\BBOQ}\APACrefatitle {A distributional code for value in dopamine-based
  reinforcement learning} {A distributional code for value in dopamine-based
  reinforcement learning}.{\BBCQ}
\newblock
\APACjournalVolNumPages{Nature}{}{}{1--5,}
\newblock

\newblock

\PrintBackRefs{\CurrentBib}

\bibitem [\protect \citeauthoryear {%
Daniels%
\ \BBA {} Howard%
}{%
Daniels%
\ \BBA {} Howard%
}{%
{\protect \APACyear {submitted}}%
}]{%
DaniHowa24}
\APACinsertmetastar {%
DaniHowa24}%
\begin{APACrefauthors}%
Daniels, B.C.%
\BCBT {}\ \BBA {} Howard, M.W.%
\end{APACrefauthors}%
\unskip\
\newblock
\APACrefYearMonthDay{submitted}{}{}.
\newblock
{\BBOQ}\APACrefatitle {Continuous Attractor Networks for Laplace Neural
  Manifold Representations of Sparse Functions} {Continuous attractor networks
  for laplace neural manifold representations of sparse functions}.{\BBCQ}
\newblock
\APACjournalVolNumPages{Computational Brain and
  Behavior}{https://arxiv.org/abs/2406.04545}{}{,}
\newblock

\newblock

\PrintBackRefs{\CurrentBib}

\bibitem [\protect \citeauthoryear {%
Danskin%
\ \protect \BOthers {.}}{%
Danskin%
\ \protect \BOthers {.}}{%
{\protect \APACyear {2023}}%
}]{%
DansEtal23}
\APACinsertmetastar {%
DansEtal23}%
\begin{APACrefauthors}%
Danskin, B.P.%
, Hattori, R.%
, Zhang, Y.E.%
, Babic, Z.%
, Aoi, M.%
\BCBL {} Komiyama, T.%
\end{APACrefauthors}%
\unskip\
\newblock
\APACrefYearMonthDay{2023}{}{}.
\newblock
{\BBOQ}\APACrefatitle {Exponential history integration with diverse temporal
  scales in retrosplenial cortex supports hyperbolic behavior} {Exponential
  history integration with diverse temporal scales in retrosplenial cortex
  supports hyperbolic behavior}.{\BBCQ}
\newblock
\APACjournalVolNumPages{Science Advances}{9}{48}{eadj4897,}
\newblock
\begin{APACrefDOI} \doi{10.1126/sciadv.adj4897} \end{APACrefDOI}
\newblock
\begin{APACrefURL} {https://www.science.org/doi/abs/10.1126/sciadv.adj4897}
  \end{APACrefURL}
\newblock
{\href{https://arxiv.org/abs/https://www.science.org/doi/pdf/10.1126/sciadv.adj4897}{{https://www.science.org/doi/pdf/10.1126/sciadv.adj4897}}}
\newblock

\PrintBackRefs{\CurrentBib}

\bibitem [\protect \citeauthoryear {%
Dayan%
}{%
Dayan%
}{%
{\protect \APACyear {1993}}%
}]{%
Daya93}
\APACinsertmetastar {%
Daya93}%
\begin{APACrefauthors}%
Dayan, P.%
\end{APACrefauthors}%
\unskip\
\newblock
\APACrefYearMonthDay{1993}{}{}.
\newblock
{\BBOQ}\APACrefatitle {Improving generalization for temporal difference
  learning: The successor representation} {Improving generalization for
  temporal difference learning: The successor representation}.{\BBCQ}
\newblock
\APACjournalVolNumPages{Neural Computation}{5}{4}{613--624,}
\newblock

\newblock

\PrintBackRefs{\CurrentBib}

\bibitem [\protect \citeauthoryear {%
De~Zeeuw%
, Lisberger%
\BCBL {}\ \BBA {} Raymond%
}{%
De~Zeeuw%
\ \protect \BOthers {.}}{%
{\protect \APACyear {2021}}%
}]{%
DeZeEtal21}
\APACinsertmetastar {%
DeZeEtal21}%
\begin{APACrefauthors}%
De~Zeeuw, C.I.%
, Lisberger, S.G.%
\BCBL {} Raymond, J.L.%
\end{APACrefauthors}%
\unskip\
\newblock
\APACrefYearMonthDay{2021}{}{}.
\newblock
{\BBOQ}\APACrefatitle {Diversity and dynamism in the cerebellum} {Diversity and
  dynamism in the cerebellum}.{\BBCQ}
\newblock
\APACjournalVolNumPages{Nature neuroscience}{24}{2}{160--167,}
\newblock

\newblock

\PrintBackRefs{\CurrentBib}

\bibitem [\protect \citeauthoryear {%
Duvelle%
, Grieves%
\BCBL {}\ \BBA {} van~der Meer%
}{%
Duvelle%
\ \protect \BOthers {.}}{%
{\protect \APACyear {2023}}%
}]{%
DuveEtal23}
\APACinsertmetastar {%
DuveEtal23}%
\begin{APACrefauthors}%
Duvelle, {\'E}.%
, Grieves, R.M.%
\BCBL {} van~der Meer, M.A.%
\end{APACrefauthors}%
\unskip\
\newblock
\APACrefYearMonthDay{2023}{}{}.
\newblock
{\BBOQ}\APACrefatitle {Temporal context and latent state inference in the
  hippocampal splitter signal} {Temporal context and latent state inference in
  the hippocampal splitter signal}.{\BBCQ}
\newblock
\APACjournalVolNumPages{Elife}{12}{}{e82357,}
\newblock

\newblock

\PrintBackRefs{\CurrentBib}

\bibitem [\protect \citeauthoryear {%
Eliasmith%
}{%
Eliasmith%
}{%
{\protect \APACyear {2013}}%
}]{%
Elia13}
\APACinsertmetastar {%
Elia13}%
\begin{APACrefauthors}%
Eliasmith, C.%
\end{APACrefauthors}%
\unskip\
\newblock
\APACrefYear{2013}.
\newblock
\APACrefbtitle {How to build a brain: A neural architecture for biological
  cognition} {How to build a brain: A neural architecture for biological
  cognition}.
\newblock
\APACaddressPublisher{}{Oxford University Press}.
\PrintBackRefs{\CurrentBib}

\bibitem [\protect \citeauthoryear {%
Enel%
, Wallis%
\BCBL {}\ \BBA {} Rich%
}{%
Enel%
\ \protect \BOthers {.}}{%
{\protect \APACyear {2020}}%
}]{%
EnelEtal20}
\APACinsertmetastar {%
EnelEtal20}%
\begin{APACrefauthors}%
Enel, P.%
, Wallis, J.D.%
\BCBL {} Rich, E.L.%
\end{APACrefauthors}%
\unskip\
\newblock
\APACrefYearMonthDay{2020}{}{}.
\newblock
{\BBOQ}\APACrefatitle {Stable and dynamic representations of value in the
  prefrontal cortex} {Stable and dynamic representations of value in the
  prefrontal cortex}.{\BBCQ}
\newblock
\APACjournalVolNumPages{Elife}{9}{}{e54313,}
\newblock

\newblock

\PrintBackRefs{\CurrentBib}

\bibitem [\protect \citeauthoryear {%
Ferbinteanu%
\ \BBA {} Shapiro%
}{%
Ferbinteanu%
\ \BBA {} Shapiro%
}{%
{\protect \APACyear {2003}}%
}]{%
FerbShap03}
\APACinsertmetastar {%
FerbShap03}%
\begin{APACrefauthors}%
Ferbinteanu, J.%
\BCBT {}\ \BBA {} Shapiro, M.L.%
\end{APACrefauthors}%
\unskip\
\newblock
\APACrefYearMonthDay{2003}{}{}.
\newblock
{\BBOQ}\APACrefatitle {Prospective and retrospective memory coding in the
  hippocampus.} {Prospective and retrospective memory coding in the
  hippocampus.}{\BBCQ}
\newblock
\APACjournalVolNumPages{Neuron}{40}{6}{1227-39,}
\newblock

\newblock

\PrintBackRefs{\CurrentBib}

\bibitem [\protect \citeauthoryear {%
Fiorillo%
, Newsome%
\BCBL {}\ \BBA {} Schultz%
}{%
Fiorillo%
\ \protect \BOthers {.}}{%
{\protect \APACyear {2008}}%
}]{%
FiorEtal08}
\APACinsertmetastar {%
FiorEtal08}%
\begin{APACrefauthors}%
Fiorillo, C.D.%
, Newsome, W.T.%
\BCBL {} Schultz, W.%
\end{APACrefauthors}%
\unskip\
\newblock
\APACrefYearMonthDay{2008}{}{}.
\newblock
{\BBOQ}\APACrefatitle {The temporal precision of reward prediction in dopamine
  neurons} {The temporal precision of reward prediction in dopamine
  neurons}.{\BBCQ}
\newblock
\APACjournalVolNumPages{Nature Neuroscience}{}{}{,}
\newblock
\begin{APACrefDOI} \doi{10.1038/nn.2159} \end{APACrefDOI}
\newblock

\newblock

\PrintBackRefs{\CurrentBib}

\bibitem [\protect \citeauthoryear {%
Fiorillo%
, Tobler%
\BCBL {}\ \BBA {} Schultz%
}{%
Fiorillo%
\ \protect \BOthers {.}}{%
{\protect \APACyear {2003}}%
}]{%
FiorEtal03}
\APACinsertmetastar {%
FiorEtal03}%
\begin{APACrefauthors}%
Fiorillo, C.D.%
, Tobler, P.N.%
\BCBL {} Schultz, W.%
\end{APACrefauthors}%
\unskip\
\newblock
\APACrefYearMonthDay{2003}{}{}.
\newblock
{\BBOQ}\APACrefatitle {Discrete coding of reward probability and uncertainty by
  dopamine neurons} {Discrete coding of reward probability and uncertainty by
  dopamine neurons}.{\BBCQ}
\newblock
\APACjournalVolNumPages{Science}{299}{5614}{1898-902,}
\newblock
\begin{APACrefDOI} \doi{10.1126/science.1077349} \end{APACrefDOI}
\newblock

\newblock

\PrintBackRefs{\CurrentBib}

\bibitem [\protect \citeauthoryear {%
Floeder%
, Jeong%
, Mohebi%
\BCBL {}\ \BBA {} Namboodiri%
}{%
Floeder%
\ \protect \BOthers {.}}{%
{\protect \APACyear {2024}}%
}]{%
FloeEtal24}
\APACinsertmetastar {%
FloeEtal24}%
\begin{APACrefauthors}%
Floeder, J.R.%
, Jeong, H.%
, Mohebi, A.%
\BCBL {} Namboodiri, V.M.K.%
\end{APACrefauthors}%
\unskip\
\newblock
\APACrefYearMonthDay{2024}{}{}.
\newblock
{\BBOQ}\APACrefatitle {Mesolimbic dopamine ramps reflect environmental
  timescales} {Mesolimbic dopamine ramps reflect environmental
  timescales}.{\BBCQ}
\newblock
\APACjournalVolNumPages{bioRxiv}{}{}{,}
\newblock

\newblock

\PrintBackRefs{\CurrentBib}

\bibitem [\protect \citeauthoryear {%
Fodor%
\ \BBA {} Pylyshyn%
}{%
Fodor%
\ \BBA {} Pylyshyn%
}{%
{\protect \APACyear {1988}}%
}]{%
FodoPyly88}
\APACinsertmetastar {%
FodoPyly88}%
\begin{APACrefauthors}%
Fodor, J.A.%
\BCBT {}\ \BBA {} Pylyshyn, Z.W.%
\end{APACrefauthors}%
\unskip\
\newblock
\APACrefYearMonthDay{1988}{}{}.
\newblock
{\BBOQ}\APACrefatitle {Connectionism and cognitive architecture: A critical
  analysis} {Connectionism and cognitive architecture: A critical
  analysis}.{\BBCQ}
\newblock
\APACjournalVolNumPages{Cognition}{28}{1}{3--71,}
\newblock

\newblock

\PrintBackRefs{\CurrentBib}

\bibitem [\protect \citeauthoryear {%
Friston%
}{%
Friston%
}{%
{\protect \APACyear {2010}}%
}]{%
Fris10}
\APACinsertmetastar {%
Fris10}%
\begin{APACrefauthors}%
Friston, K.%
\end{APACrefauthors}%
\unskip\
\newblock
\APACrefYearMonthDay{2010}{}{}.
\newblock
{\BBOQ}\APACrefatitle {The free-energy principle: a unified brain theory?} {The
  free-energy principle: a unified brain theory?}{\BBCQ}
\newblock
\APACjournalVolNumPages{Nature Reviews Neuroscience}{11}{}{127-138,}
\newblock

\newblock

\PrintBackRefs{\CurrentBib}

\bibitem [\protect \citeauthoryear {%
Friston%
\ \BBA {} Kiebel%
}{%
Friston%
\ \BBA {} Kiebel%
}{%
{\protect \APACyear {2009}}%
}]{%
FrisKieb09}
\APACinsertmetastar {%
FrisKieb09}%
\begin{APACrefauthors}%
Friston, K.%
\BCBT {}\ \BBA {} Kiebel, S.%
\end{APACrefauthors}%
\unskip\
\newblock
\APACrefYearMonthDay{2009}{}{}.
\newblock
{\BBOQ}\APACrefatitle {Predictive coding under the free-energy principle}
  {Predictive coding under the free-energy principle}.{\BBCQ}
\newblock
\APACjournalVolNumPages{Philosophical Transactions of the Royal Society B:
  Biological Sciences}{364}{1521}{1211-21,}
\newblock
\begin{APACrefDOI} \doi{10.1098/rstb.2008.0300} \end{APACrefDOI}
\newblock

\newblock

\PrintBackRefs{\CurrentBib}

\bibitem [\protect \citeauthoryear {%
C.~Gallistel%
}{%
C.~Gallistel%
}{%
{\protect \APACyear {2021}}%
{\protect \APACexlab {{\protect \BCnt {1}}}}}]{%
Gall21a}
\APACinsertmetastar {%
Gall21a}%
\begin{APACrefauthors}%
Gallistel, C.%
\end{APACrefauthors}%
\unskip\
\newblock
\APACrefYearMonthDay{2021{\protect \BCnt {1}}}{}{}.
\newblock
{\BBOQ}\APACrefatitle {The physical basis of memory} {The physical basis of
  memory}.{\BBCQ}
\newblock
\APACjournalVolNumPages{Cognition}{213}{}{104533,}
\newblock

\newblock

\PrintBackRefs{\CurrentBib}

\bibitem [\protect \citeauthoryear {%
C.~Gallistel%
}{%
C.~Gallistel%
}{%
{\protect \APACyear {2021}}%
{\protect \APACexlab {{\protect \BCnt {2}}}}}]{%
Gall21}
\APACinsertmetastar {%
Gall21}%
\begin{APACrefauthors}%
Gallistel, C.%
\end{APACrefauthors}%
\unskip\
\newblock
\APACrefYearMonthDay{2021{\protect \BCnt {2}}}{}{}.
\newblock
{\BBOQ}\APACrefatitle {Robert {R}escorla: Time, information and contingency}
  {Robert {R}escorla: Time, information and contingency}.{\BBCQ}
\newblock
\APACjournalVolNumPages{Revista de Historia de la
  Psicolog{\'\i}a}{42}{1}{7--21,}
\newblock

\newblock

\PrintBackRefs{\CurrentBib}

\bibitem [\protect \citeauthoryear {%
C.~Gallistel%
, Craig%
\BCBL {}\ \BBA {} Shahan%
}{%
C.~Gallistel%
\ \protect \BOthers {.}}{%
{\protect \APACyear {2019}}%
}]{%
GallEtal19}
\APACinsertmetastar {%
GallEtal19}%
\begin{APACrefauthors}%
Gallistel, C.%
, Craig, A.R.%
\BCBL {} Shahan, T.A.%
\end{APACrefauthors}%
\unskip\
\newblock
\APACrefYearMonthDay{2019}{}{}.
\newblock
{\BBOQ}\APACrefatitle {Contingency, contiguity, and causality in conditioning:
  Applying information theory and Weber's Law to the assignment of credit
  problem.} {Contingency, contiguity, and causality in conditioning: Applying
  information theory and weber's law to the assignment of credit
  problem.}{\BBCQ}
\newblock
\APACjournalVolNumPages{Psychological review}{126}{5}{761,}
\newblock

\newblock

\PrintBackRefs{\CurrentBib}

\bibitem [\protect \citeauthoryear {%
C.R.~Gallistel%
\ \BBA {} King%
}{%
C.R.~Gallistel%
\ \BBA {} King%
}{%
{\protect \APACyear {2011}}%
}]{%
GallKing11}
\APACinsertmetastar {%
GallKing11}%
\begin{APACrefauthors}%
Gallistel, C.R.%
\BCBT {}\ \BBA {} King, A.P.%
\end{APACrefauthors}%
\unskip\
\newblock
\APACrefYear{2011}.
\newblock
\APACrefbtitle {Memory and the computational brain: Why cognitive science will
  transform neuroscience} {Memory and the computational brain: Why cognitive
  science will transform neuroscience}\ (\BVOL~6).
\newblock
\APACaddressPublisher{}{John Wiley \&amp; Sons}.
\PrintBackRefs{\CurrentBib}

\bibitem [\protect \citeauthoryear {%
C.R.~Gallistel%
\ \BBA {} Shahan%
}{%
C.R.~Gallistel%
\ \BBA {} Shahan%
}{%
{\protect \APACyear {2024}}%
}]{%
GallShah24}
\APACinsertmetastar {%
GallShah24}%
\begin{APACrefauthors}%
Gallistel, C.R.%
\BCBT {}\ \BBA {} Shahan, T.A.%
\end{APACrefauthors}%
\unskip\
\newblock
\APACrefYearMonthDay{2024}{}{}.
\newblock
{\BBOQ}\APACrefatitle {Time-scale invariant contingency yields one-shot
  reinforcement learning despite extremely long delays to reinforcement}
  {Time-scale invariant contingency yields one-shot reinforcement learning
  despite extremely long delays to reinforcement}.{\BBCQ}
\newblock
\APACjournalVolNumPages{Proceedings of the National Academy of
  Sciences}{121}{30}{e2405451121,}
\newblock

\newblock

\PrintBackRefs{\CurrentBib}

\bibitem [\protect \citeauthoryear {%
Garcia-Garcia%
\ \protect \BOthers {.}}{%
Garcia-Garcia%
\ \protect \BOthers {.}}{%
{\protect \APACyear {2024}}%
}]{%
GarcEtal24}
\APACinsertmetastar {%
GarcEtal24}%
\begin{APACrefauthors}%
Garcia-Garcia, M.G.%
, Kapoor, A.%
, Akinwale, O.%
, Takemaru, L.%
, Kim, T.H.%
, Paton, C.%
\BDBL {}Wagner, M.J.%
\end{APACrefauthors}%
\unskip\
\newblock
\APACrefYearMonthDay{2024}{}{}.
\newblock
{\BBOQ}\APACrefatitle {A cerebellar granule cell-climbing fiber computation to
  learn to track long time intervals} {A cerebellar granule cell-climbing fiber
  computation to learn to track long time intervals}.{\BBCQ}
\newblock
\APACjournalVolNumPages{Neuron}{}{}{,}
\newblock

\newblock

\PrintBackRefs{\CurrentBib}

\bibitem [\protect \citeauthoryear {%
Gardner%
, Schoenbaum%
\BCBL {}\ \BBA {} Gershman%
}{%
Gardner%
\ \protect \BOthers {.}}{%
{\protect \APACyear {2018}}%
}]{%
GardEtal18}
\APACinsertmetastar {%
GardEtal18}%
\begin{APACrefauthors}%
Gardner, M.P.%
, Schoenbaum, G.%
\BCBL {} Gershman, S.J.%
\end{APACrefauthors}%
\unskip\
\newblock
\APACrefYearMonthDay{2018}{}{}.
\newblock
{\BBOQ}\APACrefatitle {Rethinking dopamine as generalized prediction error}
  {Rethinking dopamine as generalized prediction error}.{\BBCQ}
\newblock
\APACjournalVolNumPages{Proceedings of the Royal Society
  B}{285}{1891}{20181645,}
\newblock

\newblock

\PrintBackRefs{\CurrentBib}

\bibitem [\protect \citeauthoryear {%
Gauthier%
\ \BBA {} Tank%
}{%
Gauthier%
\ \BBA {} Tank%
}{%
{\protect \APACyear {2018}}%
}]{%
GautTank18}
\APACinsertmetastar {%
GautTank18}%
\begin{APACrefauthors}%
Gauthier, J.L.%
\BCBT {}\ \BBA {} Tank, D.W.%
\end{APACrefauthors}%
\unskip\
\newblock
\APACrefYearMonthDay{2018}{}{}.
\newblock
{\BBOQ}\APACrefatitle {A Dedicated Population for Reward Coding in the
  Hippocampus} {A dedicated population for reward coding in the
  hippocampus}.{\BBCQ}
\newblock
\APACjournalVolNumPages{Neuron}{99}{}{179-193.e7,}
\newblock

\newblock

\PrintBackRefs{\CurrentBib}

\bibitem [\protect \citeauthoryear {%
Gavornik%
\ \BBA {} Bear%
}{%
Gavornik%
\ \BBA {} Bear%
}{%
{\protect \APACyear {2014}}%
}]{%
GavoBear14}
\APACinsertmetastar {%
GavoBear14}%
\begin{APACrefauthors}%
Gavornik, J.P.%
\BCBT {}\ \BBA {} Bear, M.F.%
\end{APACrefauthors}%
\unskip\
\newblock
\APACrefYearMonthDay{2014}{}{}.
\newblock
{\BBOQ}\APACrefatitle {Learned spatiotemporal sequence recognition and
  prediction in primary visual cortex} {Learned spatiotemporal sequence
  recognition and prediction in primary visual cortex}.{\BBCQ}
\newblock
\APACjournalVolNumPages{Nature neuroscience}{17}{5}{732--737,}
\newblock

\newblock

\PrintBackRefs{\CurrentBib}

\bibitem [\protect \citeauthoryear {%
Gayler%
}{%
Gayler%
}{%
{\protect \APACyear {2004}}%
}]{%
Gayl04}
\APACinsertmetastar {%
Gayl04}%
\begin{APACrefauthors}%
Gayler, R.W.%
\end{APACrefauthors}%
\unskip\
\newblock
\APACrefYearMonthDay{2004}{}{}.
\newblock
{\BBOQ}\APACrefatitle {Vector symbolic architectures answer {Jackendoff}'s
  challenges for cognitive neuroscience} {Vector symbolic architectures answer
  {Jackendoff}'s challenges for cognitive neuroscience}.{\BBCQ}
\newblock
\APACjournalVolNumPages{arXiv preprint cs/0412059}{}{}{,}
\newblock

\newblock

\PrintBackRefs{\CurrentBib}

\bibitem [\protect \citeauthoryear {%
Gershman%
}{%
Gershman%
}{%
{\protect \APACyear {2017}}%
}]{%
Gers17a}
\APACinsertmetastar {%
Gers17a}%
\begin{APACrefauthors}%
Gershman, S.J.%
\end{APACrefauthors}%
\unskip\
\newblock
\APACrefYearMonthDay{2017}{}{}.
\newblock
{\BBOQ}\APACrefatitle {Dopamine, inference, and uncertainty} {Dopamine,
  inference, and uncertainty}.{\BBCQ}
\newblock
\APACjournalVolNumPages{Neural Computation}{29}{12}{3311--3326,}
\newblock

\newblock

\PrintBackRefs{\CurrentBib}

\bibitem [\protect \citeauthoryear {%
Gershman%
}{%
Gershman%
}{%
{\protect \APACyear {2022}}%
}]{%
Gers22}
\APACinsertmetastar {%
Gers22}%
\begin{APACrefauthors}%
Gershman, S.J.%
\end{APACrefauthors}%
\unskip\
\newblock
\APACrefYearMonthDay{2022}{}{}.
\newblock
{\BBOQ}\APACrefatitle {The molecular memory code and synaptic plasticity: a
  synthesis} {The molecular memory code and synaptic plasticity: a
  synthesis}.{\BBCQ}
\newblock
\APACjournalVolNumPages{arXiv preprint arXiv:2209.04923}{}{}{,}
\newblock

\newblock

\PrintBackRefs{\CurrentBib}

\bibitem [\protect \citeauthoryear {%
Gershman%
, Moore%
, Todd%
, Norman%
\BCBL {}\ \BBA {} Sederberg%
}{%
Gershman%
\ \protect \BOthers {.}}{%
{\protect \APACyear {2012}}%
}]{%
GersEtal12}
\APACinsertmetastar {%
GersEtal12}%
\begin{APACrefauthors}%
Gershman, S.J.%
, Moore, C.D.%
, Todd, M.T.%
, Norman, K.A.%
\BCBL {} Sederberg, P.B.%
\end{APACrefauthors}%
\unskip\
\newblock
\APACrefYearMonthDay{2012}{}{}.
\newblock
{\BBOQ}\APACrefatitle {The successor representation and temporal context} {The
  successor representation and temporal context}.{\BBCQ}
\newblock
\APACjournalVolNumPages{Neural Computation}{24}{6}{1553--1568,}
\newblock

\newblock

\PrintBackRefs{\CurrentBib}

\bibitem [\protect \citeauthoryear {%
Goh%
, Ursekar%
\BCBL {}\ \BBA {} Howard%
}{%
Goh%
\ \protect \BOthers {.}}{%
{\protect \APACyear {2022}}%
}]{%
GohEtal22}
\APACinsertmetastar {%
GohEtal22}%
\begin{APACrefauthors}%
Goh, W.Z.%
, Ursekar, V.%
\BCBL {} Howard, M.W.%
\end{APACrefauthors}%
\unskip\
\newblock
\APACrefYearMonthDay{2022}{}{}.
\newblock
{\BBOQ}\APACrefatitle {Predicting the future with a scale-invariant temporal
  memory for the past} {Predicting the future with a scale-invariant temporal
  memory for the past}.{\BBCQ}
\newblock
\APACjournalVolNumPages{Neural Computation}{34}{642-685}{,}
\newblock

\newblock

\PrintBackRefs{\CurrentBib}

\bibitem [\protect \citeauthoryear {%
Gosmann%
}{%
Gosmann%
}{%
{\protect \APACyear {2018}}%
}]{%
Gosm18}
\APACinsertmetastar {%
Gosm18}%
\begin{APACrefauthors}%
Gosmann, J.%
\end{APACrefauthors}%
\unskip\
\newblock
\APACrefYear{2018}.
\unskip\
\newblock
\APACrefbtitle {An Integrated Model of Contex, Short-Term, and Long-Term
  Memory} {An integrated model of contex, short-term, and long-term memory}\
  \APACtypeAddressSchool {\BUPhD}{}{}.
\unskip\
\newblock
\APACaddressSchool {}{University of Waterloo}.
\PrintBackRefs{\CurrentBib}

\bibitem [\protect \citeauthoryear {%
Guo%
, Huson%
, Macosko%
\BCBL {}\ \BBA {} Regehr%
}{%
Guo%
\ \protect \BOthers {.}}{%
{\protect \APACyear {2021}}%
}]{%
GuoEtal21}
\APACinsertmetastar {%
GuoEtal21}%
\begin{APACrefauthors}%
Guo, C.%
, Huson, V.%
, Macosko, E.Z.%
\BCBL {} Regehr, W.G.%
\end{APACrefauthors}%
\unskip\
\newblock
\APACrefYearMonthDay{2021}{}{}.
\newblock
{\BBOQ}\APACrefatitle {Graded heterogeneity of metabotropic signaling underlies
  a continuum of cell-intrinsic temporal responses in unipolar brush cells}
  {Graded heterogeneity of metabotropic signaling underlies a continuum of
  cell-intrinsic temporal responses in unipolar brush cells}.{\BBCQ}
\newblock
\APACjournalVolNumPages{Nature Communications}{12}{1}{1--12,}
\newblock

\newblock

\PrintBackRefs{\CurrentBib}

\bibitem [\protect \citeauthoryear {%
Hamid%
, Frank%
\BCBL {}\ \BBA {} Moore%
}{%
Hamid%
\ \protect \BOthers {.}}{%
{\protect \APACyear {2019}}%
}]{%
HamiEtal19}
\APACinsertmetastar {%
HamiEtal19}%
\begin{APACrefauthors}%
Hamid, A.A.%
, Frank, M.J.%
\BCBL {} Moore, C.I.%
\end{APACrefauthors}%
\unskip\
\newblock
\APACrefYearMonthDay{2019}{}{}.
\newblock
{\BBOQ}\APACrefatitle {Dopamine waves as a mechanism for spatiotemporal credit
  assignment} {Dopamine waves as a mechanism for spatiotemporal credit
  assignment}.{\BBCQ}
\newblock
\APACjournalVolNumPages{bioRxiv}{}{}{729640,}
\newblock

\newblock

\PrintBackRefs{\CurrentBib}

\bibitem [\protect \citeauthoryear {%
Hamid%
, Frank%
\BCBL {}\ \BBA {} Moore%
}{%
Hamid%
\ \protect \BOthers {.}}{%
{\protect \APACyear {2021}}%
}]{%
HamiEtal21}
\APACinsertmetastar {%
HamiEtal21}%
\begin{APACrefauthors}%
Hamid, A.A.%
, Frank, M.J.%
\BCBL {} Moore, C.I.%
\end{APACrefauthors}%
\unskip\
\newblock
\APACrefYearMonthDay{2021}{}{}.
\newblock
{\BBOQ}\APACrefatitle {Wave-like dopamine dynamics as a mechanism for
  spatiotemporal credit assignment} {Wave-like dopamine dynamics as a mechanism
  for spatiotemporal credit assignment}.{\BBCQ}
\newblock
\APACjournalVolNumPages{Cell}{184}{10}{2733--2749,}
\newblock

\newblock

\PrintBackRefs{\CurrentBib}

\bibitem [\protect \citeauthoryear {%
Hasselmo%
, Bodel\'{o}n%
\BCBL {}\ \BBA {} Wyble%
}{%
Hasselmo%
\ \protect \BOthers {.}}{%
{\protect \APACyear {2002}}%
}]{%
HassEtal02}
\APACinsertmetastar {%
HassEtal02}%
\begin{APACrefauthors}%
Hasselmo, M.E.%
, Bodel\'{o}n, C.%
\BCBL {} Wyble, B.P.%
\end{APACrefauthors}%
\unskip\
\newblock
\APACrefYearMonthDay{2002}{}{}.
\newblock
{\BBOQ}\APACrefatitle {A proposed function for hippocampal theta rhythm:
  {S}eparate phases of encoding and retrieval enhance reversal of prior
  learning} {A proposed function for hippocampal theta rhythm: {S}eparate
  phases of encoding and retrieval enhance reversal of prior learning}.{\BBCQ}
\newblock
\APACjournalVolNumPages{Neural Computation}{14}{}{793-817,}
\newblock

\newblock

\PrintBackRefs{\CurrentBib}

\bibitem [\protect \citeauthoryear {%
Hattori%
, Danskin%
, Babic%
, Mlynaryk%
\BCBL {}\ \BBA {} Komiyama%
}{%
Hattori%
\ \protect \BOthers {.}}{%
{\protect \APACyear {2019}}%
}]{%
HattEtal19}
\APACinsertmetastar {%
HattEtal19}%
\begin{APACrefauthors}%
Hattori, R.%
, Danskin, B.%
, Babic, Z.%
, Mlynaryk, N.%
\BCBL {} Komiyama, T.%
\end{APACrefauthors}%
\unskip\
\newblock
\APACrefYearMonthDay{2019}{}{}.
\newblock
{\BBOQ}\APACrefatitle {Area-specificity and plasticity of history-dependent
  value coding during learning} {Area-specificity and plasticity of
  history-dependent value coding during learning}.{\BBCQ}
\newblock
\APACjournalVolNumPages{Cell}{177}{7}{1858--1872,}
\newblock

\newblock

\PrintBackRefs{\CurrentBib}

\bibitem [\protect \citeauthoryear {%
Hattori%
\ \BBA {} Komiyama%
}{%
Hattori%
\ \BBA {} Komiyama%
}{%
{\protect \APACyear {2022}}%
}]{%
HattKomi22}
\APACinsertmetastar {%
HattKomi22}%
\begin{APACrefauthors}%
Hattori, R.%
\BCBT {}\ \BBA {} Komiyama, T.%
\end{APACrefauthors}%
\unskip\
\newblock
\APACrefYearMonthDay{2022}{}{}.
\newblock
{\BBOQ}\APACrefatitle {Context-dependent persistency as a coding mechanism for
  robust and widely distributed value coding} {Context-dependent persistency as
  a coding mechanism for robust and widely distributed value coding}.{\BBCQ}
\newblock
\APACjournalVolNumPages{Neuron}{110}{3}{502--515,}
\newblock

\newblock

\PrintBackRefs{\CurrentBib}

\bibitem [\protect \citeauthoryear {%
Henke%
\ \protect \BOthers {.}}{%
Henke%
\ \protect \BOthers {.}}{%
{\protect \APACyear {2021}}%
}]{%
HenkEtal21}
\APACinsertmetastar {%
HenkEtal21}%
\begin{APACrefauthors}%
Henke, J.%
, Bunk, D.%
, von Werder, D.%
, H{{\"a}}usler, S.%
, Flanagin, V.L.%
\BCBL {} Thurley, K.%
\end{APACrefauthors}%
\unskip\
\newblock
\APACrefYearMonthDay{2021}{}{}.
\newblock
{\BBOQ}\APACrefatitle {Distributed coding of duration in rodent prefrontal
  cortex during time reproduction} {Distributed coding of duration in rodent
  prefrontal cortex during time reproduction}.{\BBCQ}
\newblock
\APACjournalVolNumPages{Elife}{10}{}{e71612,}
\newblock

\newblock

\PrintBackRefs{\CurrentBib}

\bibitem [\protect \citeauthoryear {%
Hikosaka%
\ \BBA {} Watanabe%
}{%
Hikosaka%
\ \BBA {} Watanabe%
}{%
{\protect \APACyear {2000}}%
}]{%
HikoWata00}
\APACinsertmetastar {%
HikoWata00}%
\begin{APACrefauthors}%
Hikosaka, K.%
\BCBT {}\ \BBA {} Watanabe, M.%
\end{APACrefauthors}%
\unskip\
\newblock
\APACrefYearMonthDay{2000}{}{}.
\newblock
{\BBOQ}\APACrefatitle {Delay activity of orbital and lateral prefrontal neurons
  of the monkey varying with different rewards} {Delay activity of orbital and
  lateral prefrontal neurons of the monkey varying with different
  rewards}.{\BBCQ}
\newblock
\APACjournalVolNumPages{Cerebral cortex}{10}{3}{263--271,}
\newblock

\newblock

\PrintBackRefs{\CurrentBib}

\bibitem [\protect \citeauthoryear {%
Homann%
, Koay%
, Chen%
, Tank%
\BCBL {}\ \BBA {} Berry%
}{%
Homann%
\ \protect \BOthers {.}}{%
{\protect \APACyear {2022}}%
}]{%
HomaEtal22}
\APACinsertmetastar {%
HomaEtal22}%
\begin{APACrefauthors}%
Homann, J.%
, Koay, S.A.%
, Chen, K.S.%
, Tank, D.W.%
\BCBL {} Berry, M.J.%
\end{APACrefauthors}%
\unskip\
\newblock
\APACrefYearMonthDay{2022}{}{}.
\newblock
{\BBOQ}\APACrefatitle {Novel stimuli evoke excess activity in the mouse primary
  visual cortex} {Novel stimuli evoke excess activity in the mouse primary
  visual cortex}.{\BBCQ}
\newblock
\APACjournalVolNumPages{Proceedings of the National Academy of
  Sciences}{119}{5}{e2108882119,}
\newblock

\newblock

\PrintBackRefs{\CurrentBib}

\bibitem [\protect \citeauthoryear {%
Howard%
\ \BBA {} Hasselmo%
}{%
Howard%
\ \BBA {} Hasselmo%
}{%
{\protect \APACyear {2020}}%
}]{%
HowaHass20}
\APACinsertmetastar {%
HowaHass20}%
\begin{APACrefauthors}%
Howard, M.W.%
\BCBT {}\ \BBA {} Hasselmo, M.E.%
\end{APACrefauthors}%
\unskip\
\newblock
\APACrefYearMonthDay{2020}{}{}.
\newblock
{\BBOQ}\APACrefatitle {Cognitive computation using neural representations of
  time and space in the {Laplace} domain} {Cognitive computation using neural
  representations of time and space in the {Laplace} domain}.{\BBCQ}
\newblock
\APACjournalVolNumPages{arXiv preprint arXiv:2003.11668}{}{}{,}
\newblock

\newblock

\PrintBackRefs{\CurrentBib}

\bibitem [\protect \citeauthoryear {%
Howard%
, Luzardo%
\BCBL {}\ \BBA {} Tiganj%
}{%
Howard%
\ \protect \BOthers {.}}{%
{\protect \APACyear {2018}}%
}]{%
HowaEtal18}
\APACinsertmetastar {%
HowaEtal18}%
\begin{APACrefauthors}%
Howard, M.W.%
, Luzardo, A.%
\BCBL {} Tiganj, Z.%
\end{APACrefauthors}%
\unskip\
\newblock
\APACrefYearMonthDay{2018}{}{}.
\newblock
{\BBOQ}\APACrefatitle {Evidence accumulation in a {Laplace} Decision Space}
  {Evidence accumulation in a {Laplace} decision space}.{\BBCQ}
\newblock
\APACjournalVolNumPages{Computational Brain and Behavior}{1}{}{237-251,}
\newblock

\newblock

\PrintBackRefs{\CurrentBib}

\bibitem [\protect \citeauthoryear {%
Howard%
\ \protect \BOthers {.}}{%
Howard%
\ \protect \BOthers {.}}{%
{\protect \APACyear {2014}}%
}]{%
HowaEtal14}
\APACinsertmetastar {%
HowaEtal14}%
\begin{APACrefauthors}%
Howard, M.W.%
, Mac{D}onald, C.J.%
, Tiganj, Z.%
, Shankar, K.H.%
, Du, Q.%
, Hasselmo, M.E.%
\BCBL {} Eichenbaum, H.%
\end{APACrefauthors}%
\unskip\
\newblock
\APACrefYearMonthDay{2014}{}{}.
\newblock
{\BBOQ}\APACrefatitle {A unified mathematical framework for coding time, space,
  and sequences in the hippocampal region} {A unified mathematical framework
  for coding time, space, and sequences in the hippocampal region}.{\BBCQ}
\newblock
\APACjournalVolNumPages{Journal of Neuroscience}{34}{13}{4692-707,}
\newblock
\begin{APACrefDOI} \doi{10.1523/JNEUROSCI.5808-12.2014} \end{APACrefDOI}
\newblock

\newblock

\PrintBackRefs{\CurrentBib}

\bibitem [\protect \citeauthoryear {%
Howard%
\ \BBA {} Shankar%
}{%
Howard%
\ \BBA {} Shankar%
}{%
{\protect \APACyear {2018}}%
}]{%
HowaShan18}
\APACinsertmetastar {%
HowaShan18}%
\begin{APACrefauthors}%
Howard, M.W.%
\BCBT {}\ \BBA {} Shankar, K.H.%
\end{APACrefauthors}%
\unskip\
\newblock
\APACrefYearMonthDay{2018}{}{}.
\newblock
{\BBOQ}\APACrefatitle {Neural Scaling Laws for an Uncertain World} {Neural
  scaling laws for an uncertain world}.{\BBCQ}
\newblock
\APACjournalVolNumPages{Psychologial Review}{125}{}{47-58,}
\newblock
\begin{APACrefDOI} \doi{10.1037/rev0000081} \end{APACrefDOI}
\newblock

\newblock

\PrintBackRefs{\CurrentBib}

\bibitem [\protect \citeauthoryear {%
Howard%
, Shankar%
, Aue%
\BCBL {}\ \BBA {} Criss%
}{%
Howard%
\ \protect \BOthers {.}}{%
{\protect \APACyear {2015}}%
}]{%
HowaEtal15}
\APACinsertmetastar {%
HowaEtal15}%
\begin{APACrefauthors}%
Howard, M.W.%
, Shankar, K.H.%
, Aue, W.%
\BCBL {} Criss, A.H.%
\end{APACrefauthors}%
\unskip\
\newblock
\APACrefYearMonthDay{2015}{}{}.
\newblock
{\BBOQ}\APACrefatitle {A distributed representation of internal time} {A
  distributed representation of internal time}.{\BBCQ}
\newblock
\APACjournalVolNumPages{Psychological Review}{122}{1}{24-53,}
\newblock

\newblock

\PrintBackRefs{\CurrentBib}

\bibitem [\protect \citeauthoryear {%
Howe%
, Tierney%
, Sandberg%
, Phillips%
\BCBL {}\ \BBA {} Graybiel%
}{%
Howe%
\ \protect \BOthers {.}}{%
{\protect \APACyear {2013}}%
}]{%
HoweEtal13}
\APACinsertmetastar {%
HoweEtal13}%
\begin{APACrefauthors}%
Howe, M.W.%
, Tierney, P.L.%
, Sandberg, S.G.%
, Phillips, P.E.%
\BCBL {} Graybiel, A.M.%
\end{APACrefauthors}%
\unskip\
\newblock
\APACrefYearMonthDay{2013}{}{}.
\newblock
{\BBOQ}\APACrefatitle {Prolonged dopamine signalling in striatum signals
  proximity and value of distant rewards} {Prolonged dopamine signalling in
  striatum signals proximity and value of distant rewards}.{\BBCQ}
\newblock
\APACjournalVolNumPages{Nature}{500}{7464}{575--579,}
\newblock

\newblock

\PrintBackRefs{\CurrentBib}

\bibitem [\protect \citeauthoryear {%
Husserl%
}{%
Husserl%
}{%
{\protect \APACyear {1966}}%
}]{%
Huss66}
\APACinsertmetastar {%
Huss66}%
\begin{APACrefauthors}%
Husserl, E.%
\end{APACrefauthors}%
\unskip\
\newblock
\APACrefYear{1966}.
\newblock
\APACrefbtitle {The phenomenology of internal time-consciousness} {The
  phenomenology of internal time-consciousness}.
\newblock
\APACaddressPublisher{Bloomington, IN}{Indiana University Press}.
\PrintBackRefs{\CurrentBib}

\bibitem [\protect \citeauthoryear {%
Inagaki%
, Inagaki%
, Romani%
\BCBL {}\ \BBA {} Svoboda%
}{%
Inagaki%
\ \protect \BOthers {.}}{%
{\protect \APACyear {2018}}%
}]{%
InagEtal18}
\APACinsertmetastar {%
InagEtal18}%
\begin{APACrefauthors}%
Inagaki, H.K.%
, Inagaki, M.%
, Romani, S.%
\BCBL {} Svoboda, K.%
\end{APACrefauthors}%
\unskip\
\newblock
\APACrefYearMonthDay{2018}{}{}.
\newblock
{\BBOQ}\APACrefatitle {Low-dimensional and monotonic preparatory activity in
  mouse anterior lateral motor cortex} {Low-dimensional and monotonic
  preparatory activity in mouse anterior lateral motor cortex}.{\BBCQ}
\newblock
\APACjournalVolNumPages{Journal of Neuroscience}{38}{17}{4163--4185,}
\newblock

\newblock

\PrintBackRefs{\CurrentBib}

\bibitem [\protect \citeauthoryear {%
Jacques%
, Tiganj%
, Howard%
\BCBL {}\ \BBA {} Sederberg%
}{%
Jacques%
\ \protect \BOthers {.}}{%
{\protect \APACyear {2021}}%
}]{%
JacqEtal21}
\APACinsertmetastar {%
JacqEtal21}%
\begin{APACrefauthors}%
Jacques, B.G.%
, Tiganj, Z.%
, Howard, M.W.%
\BCBL {} Sederberg, P.B.%
\end{APACrefauthors}%
\unskip\
\newblock
\APACrefYearMonthDay{2021}{}{}.
\newblock
{\BBOQ}\APACrefatitle {{DeepSITH}: Efficient Learning via Decomposition of What
  and When Across Time Scales} {{DeepSITH}: Efficient learning via
  decomposition of what and when across time scales}.{\BBCQ}
\newblock
 M.~Ranzato, A.~Beygelzimer, P.~Liang, J.~Vaughan\BCBL {}\ \BBA {} Y.~Dauphin\
  (\BEDS), \APACrefbtitle {35th Conference on Advances in Neural Information
  Processing Systems} {35th conference on advances in neural information
  processing systems}\ (\BVOL\ arXiv:2104.04646).
\PrintBackRefs{\CurrentBib}

\bibitem [\protect \citeauthoryear {%
Jacques%
, Tiganj%
, Sarkar%
, Howard%
\BCBL {}\ \BBA {} Sederberg%
}{%
Jacques%
\ \protect \BOthers {.}}{%
{\protect \APACyear {2022}}%
}]{%
JacqEtal22}
\APACinsertmetastar {%
JacqEtal22}%
\begin{APACrefauthors}%
Jacques, B.G.%
, Tiganj, Z.%
, Sarkar, A.%
, Howard, M.%
\BCBL {} Sederberg, P.%
\end{APACrefauthors}%
\unskip\
\newblock
\APACrefYearMonthDay{2022}{}{}.
\newblock
{\BBOQ}\APACrefatitle {A deep convolutional neural network that is invariant to
  time rescaling} {A deep convolutional neural network that is invariant to
  time rescaling}.{\BBCQ}
\newblock
 \APACrefbtitle {International Conference on Machine Learning} {International
  conference on machine learning}\ (\BPGS\ 9729--9738).
\PrintBackRefs{\CurrentBib}

\bibitem [\protect \citeauthoryear {%
Jansson%
\ \BBA {} Lindeberg%
}{%
Jansson%
\ \BBA {} Lindeberg%
}{%
{\protect \APACyear {2021}}%
}]{%
JansLind21}
\APACinsertmetastar {%
JansLind21}%
\begin{APACrefauthors}%
Jansson, Y.%
\BCBT {}\ \BBA {} Lindeberg, T.%
\end{APACrefauthors}%
\unskip\
\newblock
\APACrefYearMonthDay{2021}{}{}.
\newblock
{\BBOQ}\APACrefatitle {Scale-invariant scale-channel networks: Deep networks
  that generalise to previously unseen scales} {Scale-invariant scale-channel
  networks: Deep networks that generalise to previously unseen scales}.{\BBCQ}
\newblock
\APACjournalVolNumPages{arXiv preprint arXiv:2106.06418}{}{}{,}
\newblock

\newblock

\PrintBackRefs{\CurrentBib}

\bibitem [\protect \citeauthoryear {%
Jeong%
\ \protect \BOthers {.}}{%
Jeong%
\ \protect \BOthers {.}}{%
{\protect \APACyear {2022}}%
}]{%
JeonEtal22}
\APACinsertmetastar {%
JeonEtal22}%
\begin{APACrefauthors}%
Jeong, H.%
, Taylor, A.%
, Floeder, J.R.%
, Lohmann, M.%
, Mihalas, S.%
, Wu, B.%
\BDBL {}Namboodiri, V.M.K.%
\end{APACrefauthors}%
\unskip\
\newblock
\APACrefYearMonthDay{2022}{}{}.
\newblock
{\BBOQ}\APACrefatitle {Mesolimbic dopamine release conveys causal associations}
  {Mesolimbic dopamine release conveys causal associations}.{\BBCQ}
\newblock
\APACjournalVolNumPages{Science}{}{}{eabq6740,}
\newblock

\newblock

\PrintBackRefs{\CurrentBib}

\bibitem [\protect \citeauthoryear {%
Jin%
, Fujii%
\BCBL {}\ \BBA {} Graybiel%
}{%
Jin%
\ \protect \BOthers {.}}{%
{\protect \APACyear {2009}}%
}]{%
JinEtal09}
\APACinsertmetastar {%
JinEtal09}%
\begin{APACrefauthors}%
Jin, D.Z.%
, Fujii, N.%
\BCBL {} Graybiel, A.M.%
\end{APACrefauthors}%
\unskip\
\newblock
\APACrefYearMonthDay{2009}{}{}.
\newblock
{\BBOQ}\APACrefatitle {Neural representation of time in cortico-basal ganglia
  circuits} {Neural representation of time in cortico-basal ganglia
  circuits}.{\BBCQ}
\newblock
\APACjournalVolNumPages{Proceedings of the National Academy of
  Sciences}{106}{45}{19156--19161,}
\newblock

\newblock

\PrintBackRefs{\CurrentBib}

\bibitem [\protect \citeauthoryear {%
Jones%
\ \BBA {} Mewhort%
}{%
Jones%
\ \BBA {} Mewhort%
}{%
{\protect \APACyear {2007}}%
}]{%
JoneMewh07}
\APACinsertmetastar {%
JoneMewh07}%
\begin{APACrefauthors}%
Jones, M.N.%
\BCBT {}\ \BBA {} Mewhort, D.J.K.%
\end{APACrefauthors}%
\unskip\
\newblock
\APACrefYearMonthDay{2007}{}{}.
\newblock
{\BBOQ}\APACrefatitle {Representing Word Meaning and Order Information
  Composite Holographic Lexicon} {Representing word meaning and order
  information composite holographic lexicon}.{\BBCQ}
\newblock
\APACjournalVolNumPages{Psychological Review}{114}{}{1-32,}
\newblock

\newblock

\PrintBackRefs{\CurrentBib}

\bibitem [\protect \citeauthoryear {%
Kanter%
, Lykken%
, Moser%
\BCBL {}\ \BBA {} Moser%
}{%
Kanter%
\ \protect \BOthers {.}}{%
{\protect \APACyear {2024}}%
}]{%
KantEtal24}
\APACinsertmetastar {%
KantEtal24}%
\begin{APACrefauthors}%
Kanter, B.R.%
, Lykken, C.M.%
, Moser, M\BHBI B.%
\BCBL {} Moser, E.I.%
\end{APACrefauthors}%
\unskip\
\newblock
\APACrefYearMonthDay{2024}{}{}.
\newblock
{\BBOQ}\APACrefatitle {Event structure sculpts neural population dynamics in
  the lateral entorhinal cortex} {Event structure sculpts neural population
  dynamics in the lateral entorhinal cortex}.{\BBCQ}
\newblock
\APACjournalVolNumPages{bioRxiv}{}{}{2024--06,}
\newblock

\newblock

\PrintBackRefs{\CurrentBib}

\bibitem [\protect \citeauthoryear {%
Kato%
\ \BBA {} Caplan%
}{%
Kato%
\ \BBA {} Caplan%
}{%
{\protect \APACyear {2017}}%
}]{%
KatoCapl17}
\APACinsertmetastar {%
KatoCapl17}%
\begin{APACrefauthors}%
Kato, K.%
\BCBT {}\ \BBA {} Caplan, J.B.%
\end{APACrefauthors}%
\unskip\
\newblock
\APACrefYearMonthDay{2017}{}{}.
\newblock
{\BBOQ}\APACrefatitle {The brain's representations may be compatible with
  convolution-based memory models.} {The brain's representations may be
  compatible with convolution-based memory models.}{\BBCQ}
\newblock
\APACjournalVolNumPages{Canadian Journal of Experimental Psychology/Revue
  canadienne de psychologie exp{\'e}rimentale}{71}{4}{299,}
\newblock

\newblock

\PrintBackRefs{\CurrentBib}

\bibitem [\protect \citeauthoryear {%
Ke%
\ \protect \BOthers {.}}{%
Ke%
\ \protect \BOthers {.}}{%
{\protect \APACyear {2018}}%
}]{%
KeEtal18}
\APACinsertmetastar {%
KeEtal18}%
\begin{APACrefauthors}%
Ke, N.R.%
, Goyal, A.%
, Bilaniuk, O.%
, Binas, J.%
, Mozer, M.C.%
, Pal, C.%
\BCBL {} Bengio, Y.%
\end{APACrefauthors}%
\unskip\
\newblock
\APACrefYearMonthDay{2018}{}{}.
\newblock
{\BBOQ}\APACrefatitle {Sparse attentive backtracking: Temporal credit
  assignment through reminding} {Sparse attentive backtracking: Temporal credit
  assignment through reminding}.{\BBCQ}
\newblock
 \APACrefbtitle {Advances in Neural Information Processing Systems} {Advances
  in neural information processing systems}\ (\BPGS\ 7640--7651).
\PrintBackRefs{\CurrentBib}

\bibitem [\protect \citeauthoryear {%
Khona%
\ \BBA {} Fiete%
}{%
Khona%
\ \BBA {} Fiete%
}{%
{\protect \APACyear {2021}}%
}]{%
KhonFiet21}
\APACinsertmetastar {%
KhonFiet21}%
\begin{APACrefauthors}%
Khona, M.%
\BCBT {}\ \BBA {} Fiete, I.R.%
\end{APACrefauthors}%
\unskip\
\newblock
\APACrefYearMonthDay{2021}{}{}.
\newblock
{\BBOQ}\APACrefatitle {Attractor and integrator networks in the brain}
  {Attractor and integrator networks in the brain}.{\BBCQ}
\newblock
\APACjournalVolNumPages{arXiv preprint arXiv:2112.03978}{}{}{,}
\newblock

\newblock

\PrintBackRefs{\CurrentBib}

\bibitem [\protect \citeauthoryear {%
H.~Kim%
, Homann%
, Tank%
\BCBL {}\ \BBA {} Berry%
}{%
H.~Kim%
\ \protect \BOthers {.}}{%
{\protect \APACyear {2019}}%
}]{%
KimEtal19}
\APACinsertmetastar {%
KimEtal19}%
\begin{APACrefauthors}%
Kim, H.%
, Homann, J.%
, Tank, D.W.%
\BCBL {} Berry, M.J.%
\end{APACrefauthors}%
\unskip\
\newblock
\APACrefYearMonthDay{2019}{}{}.
\newblock
{\BBOQ}\APACrefatitle {A long timescale stimulus history effect in the primary
  visual cortex} {A long timescale stimulus history effect in the primary
  visual cortex}.{\BBCQ}
\newblock
\APACjournalVolNumPages{bioRxiv}{}{}{585539,}
\newblock

\newblock

\PrintBackRefs{\CurrentBib}

\bibitem [\protect \citeauthoryear {%
H.R.~Kim%
\ \protect \BOthers {.}}{%
H.R.~Kim%
\ \protect \BOthers {.}}{%
{\protect \APACyear {2020}}%
}]{%
KimEtal20}
\APACinsertmetastar {%
KimEtal20}%
\begin{APACrefauthors}%
Kim, H.R.%
, Malik, A.N.%
, Mikhael, J.G.%
, Bech, P.%
, Tsutsui-Kimura, I.%
, Sun, F.%
\BDBL {}others%
\end{APACrefauthors}%
\unskip\
\newblock
\APACrefYearMonthDay{2020}{}{}.
\newblock
{\BBOQ}\APACrefatitle {A unified framework for dopamine signals across
  timescales} {A unified framework for dopamine signals across
  timescales}.{\BBCQ}
\newblock
\APACjournalVolNumPages{Cell}{183}{6}{1600--1616,}
\newblock

\newblock

\PrintBackRefs{\CurrentBib}

\bibitem [\protect \citeauthoryear {%
Koay%
, Charles%
, Thiberge%
, Brody%
\BCBL {}\ \BBA {} Tank%
}{%
Koay%
\ \protect \BOthers {.}}{%
{\protect \APACyear {2022}}%
}]{%
KoayEtal22}
\APACinsertmetastar {%
KoayEtal22}%
\begin{APACrefauthors}%
Koay, S.A.%
, Charles, A.S.%
, Thiberge, S.Y.%
, Brody, C.D.%
\BCBL {} Tank, D.W.%
\end{APACrefauthors}%
\unskip\
\newblock
\APACrefYearMonthDay{2022}{}{}.
\newblock
{\BBOQ}\APACrefatitle {Sequential and efficient neural-population coding of
  complex task information} {Sequential and efficient neural-population coding
  of complex task information}.{\BBCQ}
\newblock
\APACjournalVolNumPages{Neuron}{110}{2}{328--349,}
\newblock

\newblock

\PrintBackRefs{\CurrentBib}

\bibitem [\protect \citeauthoryear {%
Komura%
\ \protect \BOthers {.}}{%
Komura%
\ \protect \BOthers {.}}{%
{\protect \APACyear {2001}}%
}]{%
KomuEtal01}
\APACinsertmetastar {%
KomuEtal01}%
\begin{APACrefauthors}%
Komura, Y.%
, Tamura, R.%
, Uwano, T.%
, Nishijo, H.%
, Kaga, K.%
\BCBL {} Ono, T.%
\end{APACrefauthors}%
\unskip\
\newblock
\APACrefYearMonthDay{2001}{}{}.
\newblock
{\BBOQ}\APACrefatitle {Retrospective and prospective coding for predicted
  reward in the sensory thalamus} {Retrospective and prospective coding for
  predicted reward in the sensory thalamus}.{\BBCQ}
\newblock
\APACjournalVolNumPages{Nature}{412}{6846}{546-9,}
\newblock
\begin{APACrefDOI} \doi{10.1038/35087595} \end{APACrefDOI}
\newblock

\newblock

\PrintBackRefs{\CurrentBib}

\bibitem [\protect \citeauthoryear {%
Kurth-Nelson%
\ \BBA {} Redish%
}{%
Kurth-Nelson%
\ \BBA {} Redish%
}{%
{\protect \APACyear {2009}}%
}]{%
KurtRedi09}
\APACinsertmetastar {%
KurtRedi09}%
\begin{APACrefauthors}%
Kurth-Nelson, Z.%
\BCBT {}\ \BBA {} Redish, A.D.%
\end{APACrefauthors}%
\unskip\
\newblock
\APACrefYearMonthDay{2009}{}{}.
\newblock
{\BBOQ}\APACrefatitle {Temporal-difference reinforcement learning with
  distributed representations.} {Temporal-difference reinforcement learning
  with distributed representations.}{\BBCQ}
\newblock
\APACjournalVolNumPages{{PLoS One}}{4}{10}{e7362,}
\newblock

\newblock

\PrintBackRefs{\CurrentBib}

\bibitem [\protect \citeauthoryear {%
Lee%
, Engelhard%
, Witten%
\BCBL {}\ \BBA {} Daw%
}{%
Lee%
\ \protect \BOthers {.}}{%
{\protect \APACyear {2022}}%
}]{%
LeeEtal22}
\APACinsertmetastar {%
LeeEtal22}%
\begin{APACrefauthors}%
Lee, R.S.%
, Engelhard, B.%
, Witten, I.B.%
\BCBL {} Daw, N.D.%
\end{APACrefauthors}%
\unskip\
\newblock
\APACrefYearMonthDay{2022}{}{}.
\newblock
{\BBOQ}\APACrefatitle {A vector reward prediction error model explains
  dopaminergic heterogeneity} {A vector reward prediction error model explains
  dopaminergic heterogeneity}.{\BBCQ}
\newblock
\APACjournalVolNumPages{bioRxiv}{}{}{2022--02,}
\newblock

\newblock

\PrintBackRefs{\CurrentBib}

\bibitem [\protect \citeauthoryear {%
Lerner%
, Honey%
, Katkov%
\BCBL {}\ \BBA {} Hasson%
}{%
Lerner%
\ \protect \BOthers {.}}{%
{\protect \APACyear {2014}}%
}]{%
LernEtal14}
\APACinsertmetastar {%
LernEtal14}%
\begin{APACrefauthors}%
Lerner, Y.%
, Honey, C.J.%
, Katkov, M.%
\BCBL {} Hasson, U.%
\end{APACrefauthors}%
\unskip\
\newblock
\APACrefYearMonthDay{2014}{}{}.
\newblock
{\BBOQ}\APACrefatitle {Temporal scaling of neural responses to compressed and
  dilated natural speech} {Temporal scaling of neural responses to compressed
  and dilated natural speech}.{\BBCQ}
\newblock
\APACjournalVolNumPages{Journal of neurophysiology}{111}{12}{2433--2444,}
\newblock

\newblock

\PrintBackRefs{\CurrentBib}

\bibitem [\protect \citeauthoryear {%
Lindeberg%
\ \BBA {} Fagerstr\"{o}m%
}{%
Lindeberg%
\ \BBA {} Fagerstr\"{o}m%
}{%
{\protect \APACyear {1996}}%
}]{%
LindFage96}
\APACinsertmetastar {%
LindFage96}%
\begin{APACrefauthors}%
Lindeberg, T.%
\BCBT {}\ \BBA {} Fagerstr\"{o}m, D.%
\end{APACrefauthors}%
\unskip\
\newblock
\APACrefYearMonthDay{1996}{}{}.
\newblock
{\BBOQ}\APACrefatitle {Scale-space with casual time direction} {Scale-space
  with casual time direction}.{\BBCQ}
\newblock
 \APACrefbtitle {European Conference on Computer Vision} {European conference
  on computer vision}\ (\BPGS\ 229--240).
\PrintBackRefs{\CurrentBib}

\bibitem [\protect \citeauthoryear {%
Lubenov%
\ \BBA {} Siapas%
}{%
Lubenov%
\ \BBA {} Siapas%
}{%
{\protect \APACyear {2009}}%
}]{%
LubeSiap09}
\APACinsertmetastar {%
LubeSiap09}%
\begin{APACrefauthors}%
Lubenov, E.V.%
\BCBT {}\ \BBA {} Siapas, A.G.%
\end{APACrefauthors}%
\unskip\
\newblock
\APACrefYearMonthDay{2009}{}{}.
\newblock
{\BBOQ}\APACrefatitle {Hippocampal theta oscillations are travelling waves.}
  {Hippocampal theta oscillations are travelling waves.}{\BBCQ}
\newblock
\APACjournalVolNumPages{Nature}{459}{7246}{534-9,}
\newblock

\newblock

\PrintBackRefs{\CurrentBib}

\bibitem [\protect \citeauthoryear {%
Mac{D}onald%
, Lepage%
, Eden%
\BCBL {}\ \BBA {} Eichenbaum%
}{%
Mac{D}onald%
\ \protect \BOthers {.}}{%
{\protect \APACyear {2011}}%
}]{%
MacDEtal11}
\APACinsertmetastar {%
MacDEtal11}%
\begin{APACrefauthors}%
Mac{D}onald, C.J.%
, Lepage, K.Q.%
, Eden, U.T.%
\BCBL {} Eichenbaum, H.%
\end{APACrefauthors}%
\unskip\
\newblock
\APACrefYearMonthDay{2011}{}{}.
\newblock
{\BBOQ}\APACrefatitle {Hippocampal ``Time Cells'' Bridge the Gap in Memory for
  Discontiguous Events} {Hippocampal ``time cells'' bridge the gap in memory
  for discontiguous events}.{\BBCQ}
\newblock
\APACjournalVolNumPages{Neuron}{71}{4}{737-749,}
\newblock

\newblock

\PrintBackRefs{\CurrentBib}

\bibitem [\protect \citeauthoryear {%
Mainen%
\ \BBA {} Kepecs%
}{%
Mainen%
\ \BBA {} Kepecs%
}{%
{\protect \APACyear {2009}}%
}]{%
MainKepe09}
\APACinsertmetastar {%
MainKepe09}%
\begin{APACrefauthors}%
Mainen, Z.F.%
\BCBT {}\ \BBA {} Kepecs, A.%
\end{APACrefauthors}%
\unskip\
\newblock
\APACrefYearMonthDay{2009}{}{}.
\newblock
{\BBOQ}\APACrefatitle {Neural representation of behavioral outcomes in the
  orbitofrontal cortex} {Neural representation of behavioral outcomes in the
  orbitofrontal cortex}.{\BBCQ}
\newblock
\APACjournalVolNumPages{Curr Opin Neurobiol}{19}{1}{84-91,}
\newblock
\begin{APACrefDOI} \doi{10.1016/j.conb.2009.03.010} \end{APACrefDOI}
\newblock

\newblock

\PrintBackRefs{\CurrentBib}

\bibitem [\protect \citeauthoryear {%
Marcus%
}{%
Marcus%
}{%
{\protect \APACyear {2018}}%
}]{%
Marc18a}
\APACinsertmetastar {%
Marc18a}%
\begin{APACrefauthors}%
Marcus, G.F.%
\end{APACrefauthors}%
\unskip\
\newblock
\APACrefYear{2018}.
\newblock
\APACrefbtitle {The algebraic mind: Integrating connectionism and cognitive
  science} {The algebraic mind: Integrating connectionism and cognitive
  science}.
\newblock
\APACaddressPublisher{}{MIT press}.
\PrintBackRefs{\CurrentBib}

\bibitem [\protect \citeauthoryear {%
Masset%
, Malik%
, Kim%
, {Bech~Vilaseca}%
\BCBL {}\ \BBA {} Uchida%
}{%
Masset%
\ \protect \BOthers {.}}{%
{\protect \APACyear {2022}}%
}]{%
MassEtal22}
\APACinsertmetastar {%
MassEtal22}%
\begin{APACrefauthors}%
Masset, P.%
, Malik, A.N.%
, Kim, H.R.%
, {Bech~Vilaseca}, P.%
\BCBL {} Uchida, N.%
\end{APACrefauthors}%
\unskip\
\newblock
\APACrefYearMonthDay{2022}{}{}.
\newblock
{\BBOQ}\APACrefatitle {A distributional code for learning across timescales in
  dopamine-based reinforcement learning} {A distributional code for learning
  across timescales in dopamine-based reinforcement learning}.{\BBCQ}
\newblock
 \APACrefbtitle {Society for Neuroscience Abstracts} {Society for neuroscience
  abstracts}\ (\BVOL\ 234.27).
\PrintBackRefs{\CurrentBib}

\bibitem [\protect \citeauthoryear {%
Masset%
\ \protect \BOthers {.}}{%
Masset%
\ \protect \BOthers {.}}{%
{\protect \APACyear {2023}}%
}]{%
MassEtal23}
\APACinsertmetastar {%
MassEtal23}%
\begin{APACrefauthors}%
Masset, P.%
, Tano, P.%
, Kim, H.R.%
, Malik, A.N.%
, Pouget, A.%
\BCBL {} Uchida, N.%
\end{APACrefauthors}%
\unskip\
\newblock
\APACrefYearMonthDay{2023}{}{}.
\newblock
{\BBOQ}\APACrefatitle {Multi-timescale reinforcement learning in the brain}
  {Multi-timescale reinforcement learning in the brain}.{\BBCQ}
\newblock
\APACjournalVolNumPages{bioRxiv}{}{}{2023--11,}
\newblock

\newblock

\PrintBackRefs{\CurrentBib}

\bibitem [\protect \citeauthoryear {%
Mello%
, Soares%
\BCBL {}\ \BBA {} Paton%
}{%
Mello%
\ \protect \BOthers {.}}{%
{\protect \APACyear {2015}}%
}]{%
MellEtal15}
\APACinsertmetastar {%
MellEtal15}%
\begin{APACrefauthors}%
Mello, G.B.%
, Soares, S.%
\BCBL {} Paton, J.J.%
\end{APACrefauthors}%
\unskip\
\newblock
\APACrefYearMonthDay{2015}{}{}.
\newblock
{\BBOQ}\APACrefatitle {A scalable population code for time in the striatum} {A
  scalable population code for time in the striatum}.{\BBCQ}
\newblock
\APACjournalVolNumPages{Current Biology}{25}{9}{1113--1122,}
\newblock

\newblock

\PrintBackRefs{\CurrentBib}

\bibitem [\protect \citeauthoryear {%
Momennejad%
\ \BBA {} Howard%
}{%
Momennejad%
\ \BBA {} Howard%
}{%
{\protect \APACyear {2018}}%
}]{%
MomeHowa18}
\APACinsertmetastar {%
MomeHowa18}%
\begin{APACrefauthors}%
Momennejad, I.%
\BCBT {}\ \BBA {} Howard, M.W.%
\end{APACrefauthors}%
\unskip\
\newblock
\APACrefYearMonthDay{2018}{}{}.
\newblock
{\BBOQ}\APACrefatitle {Predicting the future with multi-scale successor
  representations} {Predicting the future with multi-scale successor
  representations}.{\BBCQ}
\newblock
\APACjournalVolNumPages{bioRxiv}{}{}{449470,}
\newblock

\newblock

\PrintBackRefs{\CurrentBib}

\bibitem [\protect \citeauthoryear {%
Momennejad%
\ \protect \BOthers {.}}{%
Momennejad%
\ \protect \BOthers {.}}{%
{\protect \APACyear {2017}}%
}]{%
MomeEtal17}
\APACinsertmetastar {%
MomeEtal17}%
\begin{APACrefauthors}%
Momennejad, I.%
, Russek, E.M.%
, Cheong, J.H.%
, Botvinick, M.M.%
, Daw, N.%
\BCBL {} Gershman, S.J.%
\end{APACrefauthors}%
\unskip\
\newblock
\APACrefYearMonthDay{2017}{}{}.
\newblock
{\BBOQ}\APACrefatitle {The successor representation in human reinforcement
  learning} {The successor representation in human reinforcement
  learning}.{\BBCQ}
\newblock
\APACjournalVolNumPages{Nature Human Behaviour}{1}{9}{680,}
\newblock

\newblock

\PrintBackRefs{\CurrentBib}

\bibitem [\protect \citeauthoryear {%
Morcos%
\ \BBA {} Harvey%
}{%
Morcos%
\ \BBA {} Harvey%
}{%
{\protect \APACyear {2016}}%
}]{%
MorcHarv16}
\APACinsertmetastar {%
MorcHarv16}%
\begin{APACrefauthors}%
Morcos, A.S.%
\BCBT {}\ \BBA {} Harvey, C.D.%
\end{APACrefauthors}%
\unskip\
\newblock
\APACrefYearMonthDay{2016}{}{}.
\newblock
{\BBOQ}\APACrefatitle {History-dependent variability in population dynamics
  during evidence accumulation in cortex} {History-dependent variability in
  population dynamics during evidence accumulation in cortex}.{\BBCQ}
\newblock
\APACjournalVolNumPages{Nature Neuroscience}{19}{12}{1672--1681,}
\newblock

\newblock

\PrintBackRefs{\CurrentBib}

\bibitem [\protect \citeauthoryear {%
Murdock%
}{%
Murdock%
}{%
{\protect \APACyear {1982}}%
}]{%
Murd82}
\APACinsertmetastar {%
Murd82}%
\begin{APACrefauthors}%
Murdock, B.B.%
\end{APACrefauthors}%
\unskip\
\newblock
\APACrefYearMonthDay{1982}{}{}.
\newblock
{\BBOQ}\APACrefatitle {A theory for the storage and retrieval of item and
  associative information} {A theory for the storage and retrieval of item and
  associative information}.{\BBCQ}
\newblock
\APACjournalVolNumPages{Psychological Review}{89}{}{609-626,}
\newblock

\newblock

\PrintBackRefs{\CurrentBib}

\bibitem [\protect \citeauthoryear {%
Namboodiri%
}{%
Namboodiri%
}{%
{\protect \APACyear {2021}}%
}]{%
Namb21}
\APACinsertmetastar {%
Namb21}%
\begin{APACrefauthors}%
Namboodiri, V.M.K.%
\end{APACrefauthors}%
\unskip\
\newblock
\APACrefYearMonthDay{2021}{}{}.
\newblock
{\BBOQ}\APACrefatitle {What is the state space of the world for real animals?}
  {What is the state space of the world for real animals?}{\BBCQ}
\newblock
\APACjournalVolNumPages{BioRxiv}{}{}{,}
\newblock

\newblock

\PrintBackRefs{\CurrentBib}

\bibitem [\protect \citeauthoryear {%
Namboodiri%
\ \protect \BOthers {.}}{%
Namboodiri%
\ \protect \BOthers {.}}{%
{\protect \APACyear {2019}}%
}]{%
NambEtal19}
\APACinsertmetastar {%
NambEtal19}%
\begin{APACrefauthors}%
Namboodiri, V.M.K.%
, Otis, J.M.%
, van Heeswijk, K.%
, Voets, E.S.%
, Alghorazi, R.A.%
, Rodriguez-Romaguera, J.%
\BDBL {}Stuber, G.D.%
\end{APACrefauthors}%
\unskip\
\newblock
\APACrefYearMonthDay{2019}{}{}.
\newblock
{\BBOQ}\APACrefatitle {Single-cell activity tracking reveals that orbitofrontal
  neurons acquire and maintain a long-term memory to guide behavioral
  adaptation} {Single-cell activity tracking reveals that orbitofrontal neurons
  acquire and maintain a long-term memory to guide behavioral
  adaptation}.{\BBCQ}
\newblock
\APACjournalVolNumPages{Nature neuroscience}{22}{7}{1110--1121,}
\newblock

\newblock

\PrintBackRefs{\CurrentBib}

\bibitem [\protect \citeauthoryear {%
Namboodiri%
\ \BBA {} Stuber%
}{%
Namboodiri%
\ \BBA {} Stuber%
}{%
{\protect \APACyear {2021}}%
}]{%
NambStub21}
\APACinsertmetastar {%
NambStub21}%
\begin{APACrefauthors}%
Namboodiri, V.M.K.%
\BCBT {}\ \BBA {} Stuber, G.D.%
\end{APACrefauthors}%
\unskip\
\newblock
\APACrefYearMonthDay{2021}{}{}.
\newblock
{\BBOQ}\APACrefatitle {The learning of prospective and retrospective cognitive
  maps within neural circuits} {The learning of prospective and retrospective
  cognitive maps within neural circuits}.{\BBCQ}
\newblock
\APACjournalVolNumPages{Neuron}{109}{22}{3552--3575,}
\newblock

\newblock

\PrintBackRefs{\CurrentBib}

\bibitem [\protect \citeauthoryear {%
Ning%
, Bladon%
\BCBL {}\ \BBA {} Hasselmo%
}{%
Ning%
\ \protect \BOthers {.}}{%
{\protect \APACyear {2022}}%
}]{%
NingEtal22}
\APACinsertmetastar {%
NingEtal22}%
\begin{APACrefauthors}%
Ning, W.%
, Bladon, J.H.%
\BCBL {} Hasselmo, M.E.%
\end{APACrefauthors}%
\unskip\
\newblock
\APACrefYearMonthDay{2022}{}{}.
\newblock
{\BBOQ}\APACrefatitle {Complementary representations of time in the prefrontal
  cortex and hippocampus} {Complementary representations of time in the
  prefrontal cortex and hippocampus}.{\BBCQ}
\newblock
\APACjournalVolNumPages{Hippocampus}{}{}{,}
\newblock

\newblock

\PrintBackRefs{\CurrentBib}

\bibitem [\protect \citeauthoryear {%
Ogata%
}{%
Ogata%
}{%
{\protect \APACyear {1970}}%
}]{%
Ogat70}
\APACinsertmetastar {%
Ogat70}%
\begin{APACrefauthors}%
Ogata, K.%
\end{APACrefauthors}%
\unskip\
\newblock
\APACrefYearMonthDay{1970}{}{}.
\newblock
{\BBOQ}\APACrefatitle {Describing-Function Analysis of Nonlinear Control
  Systems} {Describing-function analysis of nonlinear control systems}.{\BBCQ}
\newblock
\APACjournalVolNumPages{Modern Control Engineering}{2}{}{645--676,}
\newblock

\newblock

\PrintBackRefs{\CurrentBib}

\bibitem [\protect \citeauthoryear {%
O'{K}eefe%
\ \BBA {} Nadel%
}{%
O'{K}eefe%
\ \BBA {} Nadel%
}{%
{\protect \APACyear {1978}}%
}]{%
OKeeNade78}
\APACinsertmetastar {%
OKeeNade78}%
\begin{APACrefauthors}%
O'{K}eefe, J.%
\BCBT {}\ \BBA {} Nadel, L.%
\end{APACrefauthors}%
\unskip\
\newblock
\APACrefYear{1978}.
\newblock
\APACrefbtitle {The hippocampus as a cognitive map} {The hippocampus as a
  cognitive map}.
\newblock
\APACaddressPublisher{New York}{Oxford University Press}.
\PrintBackRefs{\CurrentBib}

\bibitem [\protect \citeauthoryear {%
Palmer%
, Marre%
, Berry%
\BCBL {}\ \BBA {} Bialek%
}{%
Palmer%
\ \protect \BOthers {.}}{%
{\protect \APACyear {2015}}%
}]{%
PalmEtal15}
\APACinsertmetastar {%
PalmEtal15}%
\begin{APACrefauthors}%
Palmer, S.E.%
, Marre, O.%
, Berry, M.J., 2nd%
\BCBL {} Bialek, W.%
\end{APACrefauthors}%
\unskip\
\newblock
\APACrefYearMonthDay{2015}{}{}.
\newblock
{\BBOQ}\APACrefatitle {Predictive information in a sensory population}
  {Predictive information in a sensory population}.{\BBCQ}
\newblock
\APACjournalVolNumPages{Proceedings of the National Academy of Sciences
  USA}{112}{22}{6908-13,}
\newblock
\begin{APACrefDOI} \doi{10.1073/pnas.1506855112} \end{APACrefDOI}
\newblock

\newblock

\PrintBackRefs{\CurrentBib}

\bibitem [\protect \citeauthoryear {%
Pastalkova%
, Itskov%
, Amarasingham%
\BCBL {}\ \BBA {} Buzsaki%
}{%
Pastalkova%
\ \protect \BOthers {.}}{%
{\protect \APACyear {2008}}%
}]{%
PastEtal08}
\APACinsertmetastar {%
PastEtal08}%
\begin{APACrefauthors}%
Pastalkova, E.%
, Itskov, V.%
, Amarasingham, A.%
\BCBL {} Buzsaki, G.%
\end{APACrefauthors}%
\unskip\
\newblock
\APACrefYearMonthDay{2008}{}{}.
\newblock
{\BBOQ}\APACrefatitle {Internally generated cell assembly sequences in the rat
  hippocampus.} {Internally generated cell assembly sequences in the rat
  hippocampus.}{\BBCQ}
\newblock
\APACjournalVolNumPages{Science}{321}{5894}{1322-7,}
\newblock

\newblock

\PrintBackRefs{\CurrentBib}

\bibitem [\protect \citeauthoryear {%
Patel%
, Fujisawa%
, Ber{\'e}nyi%
, Royer%
\BCBL {}\ \BBA {} Buzs{\'a}ki%
}{%
Patel%
\ \protect \BOthers {.}}{%
{\protect \APACyear {2012}}%
}]{%
PateEtal12}
\APACinsertmetastar {%
PateEtal12}%
\begin{APACrefauthors}%
Patel, J.%
, Fujisawa, S.%
, Ber{\'e}nyi, A.%
, Royer, S.%
\BCBL {} Buzs{\'a}ki, G.%
\end{APACrefauthors}%
\unskip\
\newblock
\APACrefYearMonthDay{2012}{}{}.
\newblock
{\BBOQ}\APACrefatitle {Traveling theta waves along the entire septotemporal
  axis of the hippocampus} {Traveling theta waves along the entire
  septotemporal axis of the hippocampus}.{\BBCQ}
\newblock
\APACjournalVolNumPages{Neuron}{75}{3}{410-7,}
\newblock
\begin{APACrefDOI} \doi{10.1016/j.neuron.2012.07.015} \end{APACrefDOI}
\newblock

\newblock

\PrintBackRefs{\CurrentBib}

\bibitem [\protect \citeauthoryear {%
Phillips%
\ \protect \BOthers {.}}{%
Phillips%
\ \protect \BOthers {.}}{%
{\protect \APACyear {2019}}%
}]{%
PhilEtal19}
\APACinsertmetastar {%
PhilEtal19}%
\begin{APACrefauthors}%
Phillips, J.W.%
, Schulmann, A.%
, Hara, E.%
, Winnubst, J.%
, Liu, C.%
, Valakh, V.%
\BDBL {}others%
\end{APACrefauthors}%
\unskip\
\newblock
\APACrefYearMonthDay{2019}{}{}.
\newblock
{\BBOQ}\APACrefatitle {A repeated molecular architecture across thalamic
  pathways} {A repeated molecular architecture across thalamic
  pathways}.{\BBCQ}
\newblock
\APACjournalVolNumPages{Nature neuroscience}{22}{11}{1925--1935,}
\newblock

\newblock

\PrintBackRefs{\CurrentBib}

\bibitem [\protect \citeauthoryear {%
Piantadosi%
}{%
Piantadosi%
}{%
{\protect \APACyear {2016}}%
}]{%
Pian16}
\APACinsertmetastar {%
Pian16}%
\begin{APACrefauthors}%
Piantadosi, S.T.%
\end{APACrefauthors}%
\unskip\
\newblock
\APACrefYearMonthDay{2016}{}{}.
\newblock
{\BBOQ}\APACrefatitle {A rational analysis of the approximate number system} {A
  rational analysis of the approximate number system}.{\BBCQ}
\newblock
\APACjournalVolNumPages{Psychonomic Bulletin \& Review}{}{}{1--10,}
\newblock

\newblock

\PrintBackRefs{\CurrentBib}

\bibitem [\protect \citeauthoryear {%
Plate%
}{%
Plate%
}{%
{\protect \APACyear {1995}}%
}]{%
Plat95}
\APACinsertmetastar {%
Plat95}%
\begin{APACrefauthors}%
Plate, T.A.%
\end{APACrefauthors}%
\unskip\
\newblock
\APACrefYearMonthDay{1995}{}{}.
\newblock
{\BBOQ}\APACrefatitle {Holographic reduced representations} {Holographic
  reduced representations}.{\BBCQ}
\newblock
\APACjournalVolNumPages{IEEE Transactions on Neural networks}{6}{3}{623--641,}
\newblock

\newblock

\PrintBackRefs{\CurrentBib}

\bibitem [\protect \citeauthoryear {%
Rainer%
, Rao%
\BCBL {}\ \BBA {} Miller%
}{%
Rainer%
\ \protect \BOthers {.}}{%
{\protect \APACyear {1999}}%
}]{%
RainEtal99}
\APACinsertmetastar {%
RainEtal99}%
\begin{APACrefauthors}%
Rainer, G.%
, Rao, S.C.%
\BCBL {} Miller, E.K.%
\end{APACrefauthors}%
\unskip\
\newblock
\APACrefYearMonthDay{1999}{}{}.
\newblock
{\BBOQ}\APACrefatitle {Prospective coding for objects in primate prefrontal
  cortex} {Prospective coding for objects in primate prefrontal cortex}.{\BBCQ}
\newblock
\APACjournalVolNumPages{Journal of Neuroscience}{19}{13}{5493--5505,}
\newblock

\newblock

\PrintBackRefs{\CurrentBib}

\bibitem [\protect \citeauthoryear {%
Rao%
\ \BBA {} Ballard%
}{%
Rao%
\ \BBA {} Ballard%
}{%
{\protect \APACyear {1999}}%
}]{%
RaoBall99}
\APACinsertmetastar {%
RaoBall99}%
\begin{APACrefauthors}%
Rao, R.P.%
\BCBT {}\ \BBA {} Ballard, D.H.%
\end{APACrefauthors}%
\unskip\
\newblock
\APACrefYearMonthDay{1999}{}{}.
\newblock
{\BBOQ}\APACrefatitle {Predictive coding in the visual cortex: a functional
  interpretation of some extra-classical receptive-field effects} {Predictive
  coding in the visual cortex: a functional interpretation of some
  extra-classical receptive-field effects}.{\BBCQ}
\newblock
\APACjournalVolNumPages{Nature Neuroscience}{2}{1}{79-87,}
\newblock
\begin{APACrefDOI} \doi{10.1038/4580} \end{APACrefDOI}
\newblock

\newblock

\PrintBackRefs{\CurrentBib}

\bibitem [\protect \citeauthoryear {%
Rescorla%
\ \BBA {} Wagner%
}{%
Rescorla%
\ \BBA {} Wagner%
}{%
{\protect \APACyear {1972}}%
}]{%
RescWagn72}
\APACinsertmetastar {%
RescWagn72}%
\begin{APACrefauthors}%
Rescorla, R.A.%
\BCBT {}\ \BBA {} Wagner, A.R.%
\end{APACrefauthors}%
\unskip\
\newblock
\APACrefYearMonthDay{1972}{}{}.
\newblock
{\BBOQ}\APACrefatitle {A theory of {P}avlovian conditioning: Variations in the
  efectivenesss of reinforcement and nonreinforcement} {A theory of {P}avlovian
  conditioning: Variations in the efectivenesss of reinforcement and
  nonreinforcement}.{\BBCQ}
\newblock
 A.H.~Black\ \BBA {} W.F.~Prokasy\ (\BEDS), \APACrefbtitle {Classical
  conditioning {II}: Current research and theory.} {Classical conditioning
  {II}: Current research and theory.}
\newblock
\APACaddressPublisher{New York}{Appleton-Century-Crofts}.
\PrintBackRefs{\CurrentBib}

\bibitem [\protect \citeauthoryear {%
Roitman%
\ \BBA {} Shadlen%
}{%
Roitman%
\ \BBA {} Shadlen%
}{%
{\protect \APACyear {2002}}%
}]{%
RoitShad02}
\APACinsertmetastar {%
RoitShad02}%
\begin{APACrefauthors}%
Roitman, J.D.%
\BCBT {}\ \BBA {} Shadlen, M.N.%
\end{APACrefauthors}%
\unskip\
\newblock
\APACrefYearMonthDay{2002}{}{}.
\newblock
{\BBOQ}\APACrefatitle {Response of neurons in the lateral intraparietal area
  during a combined visual discrimination reaction time task} {Response of
  neurons in the lateral intraparietal area during a combined visual
  discrimination reaction time task}.{\BBCQ}
\newblock
\APACjournalVolNumPages{Journal of neuroscience}{22}{21}{9475--9489,}
\newblock

\newblock

\PrintBackRefs{\CurrentBib}

\bibitem [\protect \citeauthoryear {%
Roy%
, Zhang%
, Halassa%
\BCBL {}\ \BBA {} Feng%
}{%
Roy%
\ \protect \BOthers {.}}{%
{\protect \APACyear {2022}}%
}]{%
RoyEtal22}
\APACinsertmetastar {%
RoyEtal22}%
\begin{APACrefauthors}%
Roy, D.S.%
, Zhang, Y.%
, Halassa, M.M.%
\BCBL {} Feng, G.%
\end{APACrefauthors}%
\unskip\
\newblock
\APACrefYearMonthDay{2022}{}{}.
\newblock
{\BBOQ}\APACrefatitle {Thalamic subnetworks as units of function} {Thalamic
  subnetworks as units of function}.{\BBCQ}
\newblock
\APACjournalVolNumPages{Nature Neuroscience}{25}{2}{140--153,}
\newblock

\newblock

\PrintBackRefs{\CurrentBib}

\bibitem [\protect \citeauthoryear {%
Russ%
, Koyano%
, Day-Cooney%
, Perwez%
\BCBL {}\ \BBA {} Leopold%
}{%
Russ%
\ \protect \BOthers {.}}{%
{\protect \APACyear {2022}}%
}]{%
RussEtal22}
\APACinsertmetastar {%
RussEtal22}%
\begin{APACrefauthors}%
Russ, B.E.%
, Koyano, K.W.%
, Day-Cooney, J.%
, Perwez, N.%
\BCBL {} Leopold, D.A.%
\end{APACrefauthors}%
\unskip\
\newblock
\APACrefYearMonthDay{2022}{}{}.
\newblock
{\BBOQ}\APACrefatitle {Temporal continuity shapes visual responses of macaque
  face patch neurons} {Temporal continuity shapes visual responses of macaque
  face patch neurons}.{\BBCQ}
\newblock
\APACjournalVolNumPages{bioRxiv}{}{}{,}
\newblock

\newblock

\PrintBackRefs{\CurrentBib}

\bibitem [\protect \citeauthoryear {%
Salet%
, Kruijne%
, van Rijn%
, Los%
\BCBL {}\ \BBA {} Meeter%
}{%
Salet%
\ \protect \BOthers {.}}{%
{\protect \APACyear {2022}}%
}]{%
SaleEtal22}
\APACinsertmetastar {%
SaleEtal22}%
\begin{APACrefauthors}%
Salet, J.M.%
, Kruijne, W.%
, van Rijn, H.%
, Los, S.A.%
\BCBL {} Meeter, M.%
\end{APACrefauthors}%
\unskip\
\newblock
\APACrefYearMonthDay{2022}{}{}.
\newblock
{\BBOQ}\APACrefatitle {{FMTP}: A unifying computational framework of temporal
  preparation across time scales.} {{FMTP}: A unifying computational framework
  of temporal preparation across time scales.}{\BBCQ}
\newblock
\APACjournalVolNumPages{Psychological Review}{}{}{,}
\newblock

\newblock

\PrintBackRefs{\CurrentBib}

\bibitem [\protect \citeauthoryear {%
Sarel%
, Finkelstein%
, Las%
\BCBL {}\ \BBA {} Ulanovsky%
}{%
Sarel%
\ \protect \BOthers {.}}{%
{\protect \APACyear {2017}}%
}]{%
SareEtal17}
\APACinsertmetastar {%
SareEtal17}%
\begin{APACrefauthors}%
Sarel, A.%
, Finkelstein, A.%
, Las, L.%
\BCBL {} Ulanovsky, N.%
\end{APACrefauthors}%
\unskip\
\newblock
\APACrefYearMonthDay{2017}{}{}.
\newblock
{\BBOQ}\APACrefatitle {Vectorial representation of spatial goals in the
  hippocampus of bats} {Vectorial representation of spatial goals in the
  hippocampus of bats}.{\BBCQ}
\newblock
\APACjournalVolNumPages{Science}{355}{6321}{176--180,}
\newblock

\newblock

\PrintBackRefs{\CurrentBib}

\bibitem [\protect \citeauthoryear {%
Schlegel%
, Neubert%
\BCBL {}\ \BBA {} Protzel%
}{%
Schlegel%
\ \protect \BOthers {.}}{%
{\protect \APACyear {2022}}%
}]{%
SchlEtal22}
\APACinsertmetastar {%
SchlEtal22}%
\begin{APACrefauthors}%
Schlegel, K.%
, Neubert, P.%
\BCBL {} Protzel, P.%
\end{APACrefauthors}%
\unskip\
\newblock
\APACrefYearMonthDay{2022}{}{}.
\newblock
{\BBOQ}\APACrefatitle {A comparison of vector symbolic architectures} {A
  comparison of vector symbolic architectures}.{\BBCQ}
\newblock
\APACjournalVolNumPages{Artificial Intelligence Review}{55}{6}{4523--4555,}
\newblock

\newblock

\PrintBackRefs{\CurrentBib}

\bibitem [\protect \citeauthoryear {%
Schoenbaum%
, Chiba%
\BCBL {}\ \BBA {} Gallagher%
}{%
Schoenbaum%
\ \protect \BOthers {.}}{%
{\protect \APACyear {1998}}%
}]{%
SchoEtal98}
\APACinsertmetastar {%
SchoEtal98}%
\begin{APACrefauthors}%
Schoenbaum, G.%
, Chiba, A.A.%
\BCBL {} Gallagher, M.%
\end{APACrefauthors}%
\unskip\
\newblock
\APACrefYearMonthDay{1998}{}{}.
\newblock
{\BBOQ}\APACrefatitle {Orbitofrontal cortex and basolateral amygdala encode
  expected outcomes during learning} {Orbitofrontal cortex and basolateral
  amygdala encode expected outcomes during learning}.{\BBCQ}
\newblock
\APACjournalVolNumPages{Nature neuroscience}{1}{2}{155--159,}
\newblock

\newblock

\PrintBackRefs{\CurrentBib}

\bibitem [\protect \citeauthoryear {%
Schoenbaum%
\ \BBA {} Roesch%
}{%
Schoenbaum%
\ \BBA {} Roesch%
}{%
{\protect \APACyear {2005}}%
}]{%
SchoRoes05}
\APACinsertmetastar {%
SchoRoes05}%
\begin{APACrefauthors}%
Schoenbaum, G.%
\BCBT {}\ \BBA {} Roesch, M.%
\end{APACrefauthors}%
\unskip\
\newblock
\APACrefYearMonthDay{2005}{}{}.
\newblock
{\BBOQ}\APACrefatitle {Orbitofrontal cortex, associative learning, and
  expectancies} {Orbitofrontal cortex, associative learning, and
  expectancies}.{\BBCQ}
\newblock
\APACjournalVolNumPages{Neuron}{47}{5}{633-6,}
\newblock
\begin{APACrefDOI} \doi{10.1016/j.neuron.2005.07.018} \end{APACrefDOI}
\newblock

\newblock

\PrintBackRefs{\CurrentBib}

\bibitem [\protect \citeauthoryear {%
Schonhaut%
, Aghajan%
, Kahana%
\BCBL {}\ \BBA {} Fried%
}{%
Schonhaut%
\ \protect \BOthers {.}}{%
{\protect \APACyear {2022}}%
}]{%
SchoEtal22}
\APACinsertmetastar {%
SchoEtal22}%
\begin{APACrefauthors}%
Schonhaut, D.R.%
, Aghajan, Z.M.%
, Kahana, M.J.%
\BCBL {} Fried, I.%
\end{APACrefauthors}%
\unskip\
\newblock
\APACrefYearMonthDay{2022}{}{}.
\newblock
{\BBOQ}\APACrefatitle {A neural code for spatiotemporal context} {A neural code
  for spatiotemporal context}.{\BBCQ}
\newblock
\APACjournalVolNumPages{bioRxiv}{}{}{,}
\newblock

\newblock

\PrintBackRefs{\CurrentBib}

\bibitem [\protect \citeauthoryear {%
Schonhaut%
, Aghajan%
, Kahana%
\BCBL {}\ \BBA {} Fried%
}{%
Schonhaut%
\ \protect \BOthers {.}}{%
{\protect \APACyear {2023}}%
}]{%
SchoEtal23}
\APACinsertmetastar {%
SchoEtal23}%
\begin{APACrefauthors}%
Schonhaut, D.R.%
, Aghajan, Z.M.%
, Kahana, M.J.%
\BCBL {} Fried, I.%
\end{APACrefauthors}%
\unskip\
\newblock
\APACrefYearMonthDay{2023}{}{}.
\newblock
{\BBOQ}\APACrefatitle {A neural code for time and space in the human brain} {A
  neural code for time and space in the human brain}.{\BBCQ}
\newblock
\APACjournalVolNumPages{Cell Reports}{42}{11}{,}
\newblock

\newblock

\PrintBackRefs{\CurrentBib}

\bibitem [\protect \citeauthoryear {%
Schultz%
, Dayan%
\BCBL {}\ \BBA {} Montague%
}{%
Schultz%
\ \protect \BOthers {.}}{%
{\protect \APACyear {1997}}%
}]{%
SchuEtal97}
\APACinsertmetastar {%
SchuEtal97}%
\begin{APACrefauthors}%
Schultz, W.%
, Dayan, P.%
\BCBL {} Montague, P.R.%
\end{APACrefauthors}%
\unskip\
\newblock
\APACrefYearMonthDay{1997}{}{}.
\newblock
{\BBOQ}\APACrefatitle {A neural substrate of prediction and reward} {A neural
  substrate of prediction and reward}.{\BBCQ}
\newblock
\APACjournalVolNumPages{Science}{275}{}{1593-1599,}
\newblock

\newblock

\PrintBackRefs{\CurrentBib}

\bibitem [\protect \citeauthoryear {%
Scott%
\ \protect \BOthers {.}}{%
Scott%
\ \protect \BOthers {.}}{%
{\protect \APACyear {2017}}%
}]{%
ScotEtal17}
\APACinsertmetastar {%
ScotEtal17}%
\begin{APACrefauthors}%
Scott, B.B.%
, Constantinople, C.M.%
, Akrami, A.%
, Hanks, T.D.%
, Brody, C.D.%
\BCBL {} Tank, D.W.%
\end{APACrefauthors}%
\unskip\
\newblock
\APACrefYearMonthDay{2017}{}{}.
\newblock
{\BBOQ}\APACrefatitle {Fronto-parietal Cortical Circuits Encode Accumulated
  Evidence with a Diversity of Timescales} {Fronto-parietal cortical circuits
  encode accumulated evidence with a diversity of timescales}.{\BBCQ}
\newblock
\APACjournalVolNumPages{Neuron}{95}{2}{385--398,}
\newblock

\newblock

\PrintBackRefs{\CurrentBib}

\bibitem [\protect \citeauthoryear {%
Shahbaba%
\ \protect \BOthers {.}}{%
Shahbaba%
\ \protect \BOthers {.}}{%
{\protect \APACyear {2022}}%
}]{%
ShahEtal22}
\APACinsertmetastar {%
ShahEtal22}%
\begin{APACrefauthors}%
Shahbaba, B.%
, Li, L.%
, Agostinelli, F.%
, Saraf, M.%
, Cooper, K.W.%
, Haghverdian, D.%
\BDBL {}Fortin, N.J.%
\end{APACrefauthors}%
\unskip\
\newblock
\APACrefYearMonthDay{2022}{}{}.
\newblock
{\BBOQ}\APACrefatitle {Hippocampal ensembles represent sequential relationships
  among an extended sequence of nonspatial events} {Hippocampal ensembles
  represent sequential relationships among an extended sequence of nonspatial
  events}.{\BBCQ}
\newblock
\APACjournalVolNumPages{Nature communications}{13}{1}{1--17,}
\newblock

\newblock

\PrintBackRefs{\CurrentBib}

\bibitem [\protect \citeauthoryear {%
Shankar%
\ \BBA {} Howard%
}{%
Shankar%
\ \BBA {} Howard%
}{%
{\protect \APACyear {2013}}%
}]{%
ShanHowa13}
\APACinsertmetastar {%
ShanHowa13}%
\begin{APACrefauthors}%
Shankar, K.H.%
\BCBT {}\ \BBA {} Howard, M.W.%
\end{APACrefauthors}%
\unskip\
\newblock
\APACrefYearMonthDay{2013}{}{}.
\newblock
{\BBOQ}\APACrefatitle {Optimally fuzzy temporal memory} {Optimally fuzzy
  temporal memory}.{\BBCQ}
\newblock
\APACjournalVolNumPages{Journal of Machine Learning Research}{14}{}{3753-3780,}
\newblock

\newblock

\PrintBackRefs{\CurrentBib}

\bibitem [\protect \citeauthoryear {%
Shankar%
, Singh%
\BCBL {}\ \BBA {} Howard%
}{%
Shankar%
\ \protect \BOthers {.}}{%
{\protect \APACyear {2016}}%
}]{%
ShanEtal16}
\APACinsertmetastar {%
ShanEtal16}%
\begin{APACrefauthors}%
Shankar, K.H.%
, Singh, I.%
\BCBL {} Howard, M.W.%
\end{APACrefauthors}%
\unskip\
\newblock
\APACrefYearMonthDay{2016}{}{}.
\newblock
{\BBOQ}\APACrefatitle {Neural mechanism to simulate a scale-invariant future}
  {Neural mechanism to simulate a scale-invariant future}.{\BBCQ}
\newblock
\APACjournalVolNumPages{Neural Computation}{28}{}{2594--2627,}
\newblock

\newblock

\PrintBackRefs{\CurrentBib}

\bibitem [\protect \citeauthoryear {%
Shikano%
, Ikegaya%
\BCBL {}\ \BBA {} Sasaki%
}{%
Shikano%
\ \protect \BOthers {.}}{%
{\protect \APACyear {2021}}%
}]{%
ShikEtal21}
\APACinsertmetastar {%
ShikEtal21}%
\begin{APACrefauthors}%
Shikano, Y.%
, Ikegaya, Y.%
\BCBL {} Sasaki, T.%
\end{APACrefauthors}%
\unskip\
\newblock
\APACrefYearMonthDay{2021}{}{}.
\newblock
{\BBOQ}\APACrefatitle {Minute-encoding neurons in hippocampal-striatal
  circuits} {Minute-encoding neurons in hippocampal-striatal circuits}.{\BBCQ}
\newblock
\APACjournalVolNumPages{Current Biology}{31}{7}{1438--1449,}
\newblock

\newblock

\PrintBackRefs{\CurrentBib}

\bibitem [\protect \citeauthoryear {%
Solstad%
, Boccara%
, Kropff%
, Moser%
\BCBL {}\ \BBA {} Moser%
}{%
Solstad%
\ \protect \BOthers {.}}{%
{\protect \APACyear {2008}}%
}]{%
SolsEtal08}
\APACinsertmetastar {%
SolsEtal08}%
\begin{APACrefauthors}%
Solstad, T.%
, Boccara, C.N.%
, Kropff, E.%
, Moser, M.B.%
\BCBL {} Moser, E.I.%
\end{APACrefauthors}%
\unskip\
\newblock
\APACrefYearMonthDay{2008}{}{}.
\newblock
{\BBOQ}\APACrefatitle {Representation of geometric borders in the entorhinal
  cortex.} {Representation of geometric borders in the entorhinal
  cortex.}{\BBCQ}
\newblock
\APACjournalVolNumPages{Science}{322}{5909}{1865-8,}
\newblock

\newblock

\PrintBackRefs{\CurrentBib}

\bibitem [\protect \citeauthoryear {%
Sousa%
\ \protect \BOthers {.}}{%
Sousa%
\ \protect \BOthers {.}}{%
{\protect \APACyear {2023}}%
}]{%
SousEtal23}
\APACinsertmetastar {%
SousEtal23}%
\begin{APACrefauthors}%
Sousa, M.%
, Bujalski, P.%
, Cruz, B.F.%
, Louie, K.%
, McNamee, D.%
\BCBL {} Paton, J.J.%
\end{APACrefauthors}%
\unskip\
\newblock
\APACrefYearMonthDay{2023}{}{}.
\newblock
{\BBOQ}\APACrefatitle {Dopamine neurons encode a multidimensional probabilistic
  map of future reward} {Dopamine neurons encode a multidimensional
  probabilistic map of future reward}.{\BBCQ}
\newblock
\APACjournalVolNumPages{bioRxiv}{}{}{2023--11,}
\newblock

\newblock

\PrintBackRefs{\CurrentBib}

\bibitem [\protect \citeauthoryear {%
Spitmaan%
, Seo%
, Lee%
\BCBL {}\ \BBA {} Soltani%
}{%
Spitmaan%
\ \protect \BOthers {.}}{%
{\protect \APACyear {2020}}%
}]{%
SpitEtal20}
\APACinsertmetastar {%
SpitEtal20}%
\begin{APACrefauthors}%
Spitmaan, M.%
, Seo, H.%
, Lee, D.%
\BCBL {} Soltani, A.%
\end{APACrefauthors}%
\unskip\
\newblock
\APACrefYearMonthDay{2020}{}{}.
\newblock
{\BBOQ}\APACrefatitle {Multiple timescales of neural dynamics and integration
  of task-relevant signals across cortex} {Multiple timescales of neural
  dynamics and integration of task-relevant signals across cortex}.{\BBCQ}
\newblock
\APACjournalVolNumPages{Proceedings of the National Academy of
  Sciences}{117}{36}{22522--22531,}
\newblock

\newblock

\PrintBackRefs{\CurrentBib}

\bibitem [\protect \citeauthoryear {%
Stachenfeld%
, Botvinick%
\BCBL {}\ \BBA {} Gershman%
}{%
Stachenfeld%
\ \protect \BOthers {.}}{%
{\protect \APACyear {2017}}%
}]{%
StacEtal17}
\APACinsertmetastar {%
StacEtal17}%
\begin{APACrefauthors}%
Stachenfeld, K.L.%
, Botvinick, M.M.%
\BCBL {} Gershman, S.J.%
\end{APACrefauthors}%
\unskip\
\newblock
\APACrefYearMonthDay{2017}{}{}.
\newblock
{\BBOQ}\APACrefatitle {The hippocampus as a predictive map} {The hippocampus as
  a predictive map}.{\BBCQ}
\newblock
\APACjournalVolNumPages{Nature Neuroscience}{20}{11}{1643-1653,}
\newblock
\begin{APACrefDOI} \doi{10.1038/nn.4650} \end{APACrefDOI}
\newblock

\newblock

\PrintBackRefs{\CurrentBib}

\bibitem [\protect \citeauthoryear {%
Steinberg%
\ \BBA {} Sompolinsky%
}{%
Steinberg%
\ \BBA {} Sompolinsky%
}{%
{\protect \APACyear {2022}}%
}]{%
SteiSomp22}
\APACinsertmetastar {%
SteiSomp22}%
\begin{APACrefauthors}%
Steinberg, J.%
\BCBT {}\ \BBA {} Sompolinsky, H.%
\end{APACrefauthors}%
\unskip\
\newblock
\APACrefYearMonthDay{2022}{}{}.
\newblock
{\BBOQ}\APACrefatitle {Associative memory of structured knowledge} {Associative
  memory of structured knowledge}.{\BBCQ}
\newblock
\APACjournalVolNumPages{Scientific Reports}{12}{1}{21808,}
\newblock

\newblock

\PrintBackRefs{\CurrentBib}

\bibitem [\protect \citeauthoryear {%
Subramanian%
\ \BBA {} Smith%
}{%
Subramanian%
\ \BBA {} Smith%
}{%
{\protect \APACyear {2024}}%
}]{%
SubrSmit24}
\APACinsertmetastar {%
SubrSmit24}%
\begin{APACrefauthors}%
Subramanian, D.L.%
\BCBT {}\ \BBA {} Smith, D.M.%
\end{APACrefauthors}%
\unskip\
\newblock
\APACrefYearMonthDay{2024}{}{}.
\newblock
{\BBOQ}\APACrefatitle {Time Cells in the Retrosplenial Cortex} {Time cells in
  the retrosplenial cortex}.{\BBCQ}
\newblock
\APACjournalVolNumPages{bioRxiv}{}{}{,}
\newblock

\newblock

\PrintBackRefs{\CurrentBib}

\bibitem [\protect \citeauthoryear {%
Sutton%
\ \BBA {} Barto%
}{%
Sutton%
\ \BBA {} Barto%
}{%
{\protect \APACyear {1981}}%
}]{%
SuttBart81}
\APACinsertmetastar {%
SuttBart81}%
\begin{APACrefauthors}%
Sutton, R.S.%
\BCBT {}\ \BBA {} Barto, A.G.%
\end{APACrefauthors}%
\unskip\
\newblock
\APACrefYearMonthDay{1981}{}{}.
\newblock
{\BBOQ}\APACrefatitle {Toward a modern theory of adaptive networks: Expectation
  and Prediction} {Toward a modern theory of adaptive networks: Expectation and
  prediction}.{\BBCQ}
\newblock
\APACjournalVolNumPages{Psychological Review}{88}{}{135-171,}
\newblock

\newblock

\PrintBackRefs{\CurrentBib}

\bibitem [\protect \citeauthoryear {%
Tano%
, Dayan%
\BCBL {}\ \BBA {} Pouget%
}{%
Tano%
\ \protect \BOthers {.}}{%
{\protect \APACyear {2020}}%
}]{%
TanoEtal20}
\APACinsertmetastar {%
TanoEtal20}%
\begin{APACrefauthors}%
Tano, P.%
, Dayan, P.%
\BCBL {} Pouget, A.%
\end{APACrefauthors}%
\unskip\
\newblock
\APACrefYearMonthDay{2020}{}{}.
\newblock
{\BBOQ}\APACrefatitle {A Local Temporal Difference Code for Distributional
  Reinforcement Learning} {A local temporal difference code for distributional
  reinforcement learning}.{\BBCQ}
\newblock
\APACjournalVolNumPages{Advances in Neural Information Processing
  Systems}{33}{}{13662--13673,}
\newblock

\newblock

\PrintBackRefs{\CurrentBib}

\bibitem [\protect \citeauthoryear {%
Taxidis%
\ \protect \BOthers {.}}{%
Taxidis%
\ \protect \BOthers {.}}{%
{\protect \APACyear {2020}}%
}]{%
TaxiEtal20}
\APACinsertmetastar {%
TaxiEtal20}%
\begin{APACrefauthors}%
Taxidis, J.%
, Pnevmatikakis, E.A.%
, Dorian, C.C.%
, Mylavarapu, A.L.%
, Arora, J.S.%
, Samadian, K.D.%
\BDBL {}Golshani, P.%
\end{APACrefauthors}%
\unskip\
\newblock
\APACrefYearMonthDay{2020}{}{}.
\newblock
{\BBOQ}\APACrefatitle {Differential Emergence and Stability of Sensory and
  Temporal Representations in Context-Specific Hippocampal Sequences}
  {Differential emergence and stability of sensory and temporal representations
  in context-specific hippocampal sequences}.{\BBCQ}
\newblock
\APACjournalVolNumPages{Neuron}{108}{5}{984--998.e9,}
\newblock

\newblock

\PrintBackRefs{\CurrentBib}

\bibitem [\protect \citeauthoryear {%
Tiganj%
, Cromer%
, Roy%
, Miller%
\BCBL {}\ \BBA {} Howard%
}{%
Tiganj%
\ \protect \BOthers {.}}{%
{\protect \APACyear {2018}}%
}]{%
TigaEtal18a}
\APACinsertmetastar {%
TigaEtal18a}%
\begin{APACrefauthors}%
Tiganj, Z.%
, Cromer, J.A.%
, Roy, J.E.%
, Miller, E.K.%
\BCBL {} Howard, M.W.%
\end{APACrefauthors}%
\unskip\
\newblock
\APACrefYearMonthDay{2018}{}{}.
\newblock
{\BBOQ}\APACrefatitle {Compressed timeline of recent experience in monkey
  {lPFC}} {Compressed timeline of recent experience in monkey {lPFC}}.{\BBCQ}
\newblock
\APACjournalVolNumPages{Journal of Cognitive Neuroscience}{30}{}{935-950,}
\newblock

\newblock

\PrintBackRefs{\CurrentBib}

\bibitem [\protect \citeauthoryear {%
{Tiganj}%
, Gershman%
, Sederberg%
\BCBL {}\ \BBA {} Howard%
}{%
{Tiganj}%
\ \protect \BOthers {.}}{%
{\protect \APACyear {2019}}%
}]{%
TigaEtal19a}
\APACinsertmetastar {%
TigaEtal19a}%
\begin{APACrefauthors}%
{Tiganj}, Z.%
, Gershman, S.J.%
, Sederberg, P.B.%
\BCBL {} Howard, M.W.%
\end{APACrefauthors}%
\unskip\
\newblock
\APACrefYearMonthDay{2019}{}{}.
\newblock
{\BBOQ}\APACrefatitle {Estimating scale-invariant future in continuous time}
  {Estimating scale-invariant future in continuous time}.{\BBCQ}
\newblock
\APACjournalVolNumPages{Neural Computation}{31}{4}{681--709,}
\newblock

\newblock

\PrintBackRefs{\CurrentBib}

\bibitem [\protect \citeauthoryear {%
Tiganj%
, Kim%
, Jung%
\BCBL {}\ \BBA {} Howard%
}{%
Tiganj%
\ \protect \BOthers {.}}{%
{\protect \APACyear {2017}}%
}]{%
TigaEtal17b}
\APACinsertmetastar {%
TigaEtal17b}%
\begin{APACrefauthors}%
Tiganj, Z.%
, Kim, J.%
, Jung, M.W.%
\BCBL {} Howard, M.W.%
\end{APACrefauthors}%
\unskip\
\newblock
\APACrefYearMonthDay{2017}{}{}.
\newblock
{\BBOQ}\APACrefatitle {Sequential firing codes for time in rodent {mPFC}}
  {Sequential firing codes for time in rodent {mPFC}}.{\BBCQ}
\newblock
\APACjournalVolNumPages{Cerebral Cortex}{27}{}{5663-5671,}
\newblock

\newblock

\PrintBackRefs{\CurrentBib}

\bibitem [\protect \citeauthoryear {%
Tiganj%
, Singh%
, Esfahani%
\BCBL {}\ \BBA {} Howard%
}{%
Tiganj%
\ \protect \BOthers {.}}{%
{\protect \APACyear {2022}}%
}]{%
TigaEtal22}
\APACinsertmetastar {%
TigaEtal22}%
\begin{APACrefauthors}%
Tiganj, Z.%
, Singh, I.%
, Esfahani, Z.G.%
\BCBL {} Howard, M.W.%
\end{APACrefauthors}%
\unskip\
\newblock
\APACrefYearMonthDay{2022}{}{}.
\newblock
{\BBOQ}\APACrefatitle {Scanning a compressed ordered representation of the
  future} {Scanning a compressed ordered representation of the future}.{\BBCQ}
\newblock
\APACjournalVolNumPages{Journal of Experimental Psychology:
  General}{151}{3082--3096}{,}
\newblock

\newblock

\PrintBackRefs{\CurrentBib}

\bibitem [\protect \citeauthoryear {%
Tolman%
}{%
Tolman%
}{%
{\protect \APACyear {1948}}%
}]{%
Tolm48}
\APACinsertmetastar {%
Tolm48}%
\begin{APACrefauthors}%
Tolman, E.C.%
\end{APACrefauthors}%
\unskip\
\newblock
\APACrefYearMonthDay{1948}{}{}.
\newblock
{\BBOQ}\APACrefatitle {Cognitive maps in rats and men} {Cognitive maps in rats
  and men}.{\BBCQ}
\newblock
\APACjournalVolNumPages{Psychological Review}{55}{4}{189-208,}
\newblock

\newblock

\PrintBackRefs{\CurrentBib}

\bibitem [\protect \citeauthoryear {%
Tsao%
\ \protect \BOthers {.}}{%
Tsao%
\ \protect \BOthers {.}}{%
{\protect \APACyear {2018}}%
}]{%
TsaoEtal18}
\APACinsertmetastar {%
TsaoEtal18}%
\begin{APACrefauthors}%
Tsao, A.%
, Sugar, J.%
, Lu, L.%
, Wang, C.%
, Knierim, J.J.%
, Moser, M\BHBI B.%
\BCBL {} Moser, E.I.%
\end{APACrefauthors}%
\unskip\
\newblock
\APACrefYearMonthDay{2018}{}{}.
\newblock
{\BBOQ}\APACrefatitle {Integrating time from experience in the lateral
  entorhinal cortex} {Integrating time from experience in the lateral
  entorhinal cortex}.{\BBCQ}
\newblock
\APACjournalVolNumPages{Nature}{561}{}{57-62,}
\newblock

\newblock

\PrintBackRefs{\CurrentBib}

\bibitem [\protect \citeauthoryear {%
van~der Meer%
\ \BBA {} Redish%
}{%
van~der Meer%
\ \BBA {} Redish%
}{%
{\protect \APACyear {2011}}%
}]{%
MeerEtal11}
\APACinsertmetastar {%
MeerEtal11}%
\begin{APACrefauthors}%
van~der Meer, M.A.A.%
\BCBT {}\ \BBA {} Redish, A.D.%
\end{APACrefauthors}%
\unskip\
\newblock
\APACrefYearMonthDay{2011}{}{}.
\newblock
{\BBOQ}\APACrefatitle {Theta phase precession in rat ventral striatum links
  place and reward information} {Theta phase precession in rat ventral striatum
  links place and reward information}.{\BBCQ}
\newblock
\APACjournalVolNumPages{Journal of Neuroscience}{31}{8}{2843-54,}
\newblock
\begin{APACrefDOI} \doi{10.1523/JNEUROSCI.4869-10.2011} \end{APACrefDOI}
\newblock

\newblock

\PrintBackRefs{\CurrentBib}

\bibitem [\protect \citeauthoryear {%
Waelti%
, Dickinson%
\BCBL {}\ \BBA {} Schultz%
}{%
Waelti%
\ \protect \BOthers {.}}{%
{\protect \APACyear {2001}}%
}]{%
WaelEtal01}
\APACinsertmetastar {%
WaelEtal01}%
\begin{APACrefauthors}%
Waelti, P.%
, Dickinson, A.%
\BCBL {} Schultz, W.%
\end{APACrefauthors}%
\unskip\
\newblock
\APACrefYearMonthDay{2001}{}{}.
\newblock
{\BBOQ}\APACrefatitle {Dopamine responses comply with basic assumptions of
  formal learning theory.} {Dopamine responses comply with basic assumptions of
  formal learning theory.}{\BBCQ}
\newblock
\APACjournalVolNumPages{Nature}{412}{6842}{43-8,}
\newblock

\newblock

\PrintBackRefs{\CurrentBib}

\bibitem [\protect \citeauthoryear {%
Wagner%
, Kim%
, Savall%
, Schnitzer%
\BCBL {}\ \BBA {} Luo%
}{%
Wagner%
\ \protect \BOthers {.}}{%
{\protect \APACyear {2017}}%
}]{%
WagnEtal17}
\APACinsertmetastar {%
WagnEtal17}%
\begin{APACrefauthors}%
Wagner, M.J.%
, Kim, T.H.%
, Savall, J.%
, Schnitzer, M.J.%
\BCBL {} Luo, L.%
\end{APACrefauthors}%
\unskip\
\newblock
\APACrefYearMonthDay{2017}{}{}.
\newblock
{\BBOQ}\APACrefatitle {Cerebellar granule cells encode the expectation of
  reward} {Cerebellar granule cells encode the expectation of reward}.{\BBCQ}
\newblock
\APACjournalVolNumPages{Nature}{544}{7648}{96--100,}
\newblock

\newblock

\PrintBackRefs{\CurrentBib}

\bibitem [\protect \citeauthoryear {%
Wagner%
\ \BBA {} Luo%
}{%
Wagner%
\ \BBA {} Luo%
}{%
{\protect \APACyear {2020}}%
}]{%
WagnLuo20}
\APACinsertmetastar {%
WagnLuo20}%
\begin{APACrefauthors}%
Wagner, M.J.%
\BCBT {}\ \BBA {} Luo, L.%
\end{APACrefauthors}%
\unskip\
\newblock
\APACrefYearMonthDay{2020}{}{}.
\newblock
{\BBOQ}\APACrefatitle {Neocortex--Cerebellum Circuits for Cognitive Processing}
  {Neocortex--cerebellum circuits for cognitive processing}.{\BBCQ}
\newblock
\APACjournalVolNumPages{Trends in neurosciences}{43}{1}{42--54,}
\newblock

\newblock

\PrintBackRefs{\CurrentBib}

\bibitem [\protect \citeauthoryear {%
Watanabe%
, Kadohisa%
, Kusunoki%
, Buckley%
\BCBL {}\ \BBA {} Duncan%
}{%
Watanabe%
\ \protect \BOthers {.}}{%
{\protect \APACyear {2023}}%
}]{%
WataEtal23}
\APACinsertmetastar {%
WataEtal23}%
\begin{APACrefauthors}%
Watanabe, K.%
, Kadohisa, M.%
, Kusunoki, M.%
, Buckley, M.J.%
\BCBL {} Duncan, J.%
\end{APACrefauthors}%
\unskip\
\newblock
\APACrefYearMonthDay{2023}{}{}.
\newblock
{\BBOQ}\APACrefatitle {Cycles of goal silencing and reactivation underlie
  complex problem-solving in primate frontal and parietal cortex} {Cycles of
  goal silencing and reactivation underlie complex problem-solving in primate
  frontal and parietal cortex}.{\BBCQ}
\newblock
\APACjournalVolNumPages{Nature Communications}{14}{1}{5054,}
\newblock

\newblock

\PrintBackRefs{\CurrentBib}

\bibitem [\protect \citeauthoryear {%
W.~Wei%
, Mohebi%
\BCBL {}\ \BBA {} Berke%
}{%
W.~Wei%
\ \protect \BOthers {.}}{%
{\protect \APACyear {2021}}%
}]{%
WeiEtal21}
\APACinsertmetastar {%
WeiEtal21}%
\begin{APACrefauthors}%
Wei, W.%
, Mohebi, A.%
\BCBL {} Berke, J.D.%
\end{APACrefauthors}%
\unskip\
\newblock
\APACrefYearMonthDay{2021}{}{}.
\newblock
{\BBOQ}\APACrefatitle {A spectrum of time horizons for dopamine signals} {A
  spectrum of time horizons for dopamine signals}.{\BBCQ}
\newblock
\APACjournalVolNumPages{bioRxiv}{}{}{2021--10,}
\newblock

\newblock

\PrintBackRefs{\CurrentBib}

\bibitem [\protect \citeauthoryear {%
X\BHBI X.~Wei%
\ \BBA {} Stocker%
}{%
X\BHBI X.~Wei%
\ \BBA {} Stocker%
}{%
{\protect \APACyear {2012}}%
}]{%
WeiStoc12a}
\APACinsertmetastar {%
WeiStoc12a}%
\begin{APACrefauthors}%
Wei, X\BHBI X.%
\BCBT {}\ \BBA {} Stocker, A.A.%
\end{APACrefauthors}%
\unskip\
\newblock
\APACrefYearMonthDay{2012}{}{}.
\newblock
{\BBOQ}\APACrefatitle {Efficient coding provides a direct link between prior
  and likelihood in perceptual Bayesian inference} {Efficient coding provides a
  direct link between prior and likelihood in perceptual bayesian
  inference}.{\BBCQ}
\newblock
\APACjournalVolNumPages{Advances in neural information processing
  systems}{25}{}{,}
\newblock

\newblock

\PrintBackRefs{\CurrentBib}

\bibitem [\protect \citeauthoryear {%
Whittington%
\ \protect \BOthers {.}}{%
Whittington%
\ \protect \BOthers {.}}{%
{\protect \APACyear {2020}}%
}]{%
WhitEtal20}
\APACinsertmetastar {%
WhitEtal20}%
\begin{APACrefauthors}%
Whittington, J.C.%
, Muller, T.H.%
, Mark, S.%
, Chen, G.%
, Barry, C.%
, Burgess, N.%
\BCBL {} Behrens, T.E.%
\end{APACrefauthors}%
\unskip\
\newblock
\APACrefYearMonthDay{2020}{}{}.
\newblock
{\BBOQ}\APACrefatitle {The Tolman-Eichenbaum machine: unifying space and
  relational memory through generalization in the hippocampal formation} {The
  tolman-eichenbaum machine: unifying space and relational memory through
  generalization in the hippocampal formation}.{\BBCQ}
\newblock
\APACjournalVolNumPages{Cell}{183}{5}{1249--1263,}
\newblock

\newblock

\PrintBackRefs{\CurrentBib}

\bibitem [\protect \citeauthoryear {%
Wilson%
\ \BBA {} McNaughton%
}{%
Wilson%
\ \BBA {} McNaughton%
}{%
{\protect \APACyear {1993}}%
}]{%
WilsMcNa93}
\APACinsertmetastar {%
WilsMcNa93}%
\begin{APACrefauthors}%
Wilson, M.A.%
\BCBT {}\ \BBA {} McNaughton, B.L.%
\end{APACrefauthors}%
\unskip\
\newblock
\APACrefYearMonthDay{1993}{}{}.
\newblock
{\BBOQ}\APACrefatitle {Dynamics of the hippocampal ensemble code for space}
  {Dynamics of the hippocampal ensemble code for space}.{\BBCQ}
\newblock
\APACjournalVolNumPages{Science}{261}{}{1055-8,}
\newblock

\newblock

\PrintBackRefs{\CurrentBib}

\bibitem [\protect \citeauthoryear {%
Young%
\ \BBA {} Shapiro%
}{%
Young%
\ \BBA {} Shapiro%
}{%
{\protect \APACyear {2011}}%
}]{%
YounShap11}
\APACinsertmetastar {%
YounShap11}%
\begin{APACrefauthors}%
Young, J.J.%
\BCBT {}\ \BBA {} Shapiro, M.L.%
\end{APACrefauthors}%
\unskip\
\newblock
\APACrefYearMonthDay{2011}{}{}.
\newblock
{\BBOQ}\APACrefatitle {The orbitofrontal cortex and response selection} {The
  orbitofrontal cortex and response selection}.{\BBCQ}
\newblock
\APACjournalVolNumPages{Ann N Y Acad Sci}{1239}{}{25-32,}
\newblock
\begin{APACrefDOI} \doi{10.1111/j.1749-6632.2011.06279.x} \end{APACrefDOI}
\newblock

\newblock

\PrintBackRefs{\CurrentBib}

\bibitem [\protect \citeauthoryear {%
Yu%
\ \protect \BOthers {.}}{%
Yu%
\ \protect \BOthers {.}}{%
{\protect \APACyear {2022}}%
}]{%
YuEtal22}
\APACinsertmetastar {%
YuEtal22}%
\begin{APACrefauthors}%
Yu, Q.%
, Bi, Z.%
, Jiang, S.%
, Yan, B.%
, Chen, H.%
, Wang, Y.%
\BDBL {}Zhang, J.%
\end{APACrefauthors}%
\unskip\
\newblock
\APACrefYearMonthDay{2022}{}{}.
\newblock
{\BBOQ}\APACrefatitle {Visual cortex encodes timing information in humans and
  mice} {Visual cortex encodes timing information in humans and mice}.{\BBCQ}
\newblock
\APACjournalVolNumPages{Neuron}{110}{}{4194-4211,}
\newblock

\newblock

\PrintBackRefs{\CurrentBib}

\bibitem [\protect \citeauthoryear {%
Zhang%
\ \BBA {} Jacobs%
}{%
Zhang%
\ \BBA {} Jacobs%
}{%
{\protect \APACyear {2015}}%
}]{%
ZhanJaco15}
\APACinsertmetastar {%
ZhanJaco15}%
\begin{APACrefauthors}%
Zhang, H.%
\BCBT {}\ \BBA {} Jacobs, J.%
\end{APACrefauthors}%
\unskip\
\newblock
\APACrefYearMonthDay{2015}{}{}.
\newblock
{\BBOQ}\APACrefatitle {Traveling Theta Waves in the Human Hippocampus}
  {Traveling theta waves in the human hippocampus}.{\BBCQ}
\newblock
\APACjournalVolNumPages{Journal of Neuroscience}{35}{36}{12477-87,}
\newblock
\begin{APACrefDOI} \doi{10.1523/JNEUROSCI.5102-14.2015} \end{APACrefDOI}
\newblock

\newblock

\PrintBackRefs{\CurrentBib}

\bibitem [\protect \citeauthoryear {%
Zuo%
\ \protect \BOthers {.}}{%
Zuo%
\ \protect \BOthers {.}}{%
{\protect \APACyear {2023}}%
}]{%
ZuoEtal23}
\APACinsertmetastar {%
ZuoEtal23}%
\begin{APACrefauthors}%
Zuo, S.%
, Wang, C.%
, Wang, L.%
, Jin, Z.%
, Kusunoki, M.%
\BCBL {} Kwok, S.C.%
\end{APACrefauthors}%
\unskip\
\newblock
\APACrefYearMonthDay{2023}{}{}.
\newblock
{\BBOQ}\APACrefatitle {Neural signatures for temporal-order memory in the
  medial posterior parietal cortex} {Neural signatures for temporal-order
  memory in the medial posterior parietal cortex}.{\BBCQ}
\newblock
\APACjournalVolNumPages{bioRxiv}{}{}{2023--08,}
\newblock

\newblock

\PrintBackRefs{\CurrentBib}

\end{thebibliography}

\end{document}